\documentclass{cernrep} 

\usepackage{mathptmx}

\usepackage{texnames}
\usepackage[T1]{fontenc}
\usepackage[bookmarks, colorlinks=true, linktoc=page, pdftex, linkcolor=red, citecolor=red, urlcolor=red]{hyperref}
\usepackage{wrapfig}
\usepackage{parskip}

\usepackage{xcolor}
\hypersetup{
    colorlinks = true,
    linkcolor = [rgb]{0,0.0,0.7},
    urlcolor = [rgb]{0,0.0,0.7},
    citecolor = [rgb]{0,0.0,0.7},
}

\newcommand{\bei}{\begin{itemize}}
\newcommand{\eei}{\end{itemize}}
\newcommand{\beq}{\begin{equation}}
\newcommand{\eeq}{\end{equation}}
\newcommand{\beqn}{\begin{eqnarray}}
\newcommand{\eeqn}{\end{eqnarray}}
\newcommand{\beqns}{\begin{eqnarray*}}
\newcommand{\eeqns}{\end{eqnarray*}}

\newcommand{\e}{\varepsilon}

\newcommand{\ifb}{\;fb$^{-1}$\xspace}
\newcommand{\ipb}{\;pb$^{-1}$\xspace}
\newcommand{\MET}{E_T^{\rm miss}}
\newcommand{\ttbar}{t\overline t}
\newcommand{\seffsf}[1]{\sin\!^2\theta^{#1}_{{\rm eff}}}

\newcommand{\sinfeff}{\ensuremath{\seffsf{f}}\xspace}
\newcommand{\sinleff}{\ensuremath{\seffsf{\ell}}\xspace}

\newcommand{\as}{\ensuremath{\alpha_{\scriptscriptstyle S}}\xspace}
\newcommand\allFontSize{\small}
\newenvironment{myquote}
               {\list{}{\leftmargin0cm}%
                \item\relax}
               {\endlist}
\newcommand\detailsSize{\allFontSize}
\newenvironment{details}%
{\begin{myquote}\vspace{-0.2cm}\detailsSize}{\end{myquote}\vspace{-0.2cm}}

\begin{document}

\title{Physics at the LHC Run-2 and Beyond}
 
\author{Andreas Hoecker}

\institute{CERN, Geneva, Switzerland}

\maketitle 

\begin{abstract}
After an astounding Run-1 with 8\;TeV proton--proton collisions featuring 
among others the discovery of the Higgs boson, Run-2 of the Large Hadron Collider (LHC) 
has started in 2015 colliding protons with unprecedented 13\;TeV centre-of-mass energy. The 
higher energy and large expected integrated luminosity significantly increases the discovery 
potential for new physics, and allows for more detailed Higgs boson studies as well as improved 
Standard Model measurements. The  lecture discusses methods, recent results and future 
prospects in proton--proton physics at the LHC.\footnote{As I am a member of the ATLAS experiment, 
for practical reasons, this lecture writeup leans somewhat towards ATLAS results. In the 
majority of the cases, the plots shown can be interchanged against those from CMS (and vice versa)
without altering the message.}

\end{abstract}

 
\section{Introduction}
 
The Large Hadron Collider (LHC) at CERN probes nature at the smallest distances ever
explored on Earth to study and improve our current knowledge of space and time, matter 
and force as it is encoded in the Standard Model (SM) of particle physics.
The SM is {\em the} legacy of 20$^{\rm th}$ century particle physics:
it unifies quantum mechanics, special relativity and field theory; it unifies electromagnetic 
and weak interactions; it describes (about) all laboratory data. 
Does the SM deliver a complete answer to the complexity of the world
generated by the simultaneous existence of very small as well as very large, seemingly 
fundamental numbers? We have reasons to believe that this is not the case. 

The SM is made of spin one-half matter particles consisting of three generations of massive 
quarks and leptons, and force carriers in the form of partially massive spin one gauge bosons. 
An additional doublet of complex scalar fields, the Brout-Englert-Higgs (BEH) field $\phi$, is 
dictated by the requirement of local gauge symmetry~\cite{BEHmechanism}. 
Its condensation after spontaneous symmetry breaking at low temperature is responsible 
for the masses of the SM gauge bosons and (chiral Dirac-) fermions, leaving the electromagnetic force 
with infinite range, but making the weak force short-ranged (about $10^{-15}$\;cm).
The new field is not only a constant background field, but it has its own 
massive quantum, the scalar Higgs boson. Being a boson, 
we might want to call it a fifth force. However, unlike the other forces, the new force is not
a gauge force. Its non-universal coupling to masses of fermions and gauge bosons reminds 
us of classical gravitation, but the BEH force is much stronger than gravity and short-ranged. 

The potential of the scalar double field $\phi$ consists in its simplest form at 
low temperature of three terms: a quadratic term with negative coefficient $\mu$, 
a quartic term with positive coefficient $\lambda$ realising the ``Mexican hat'' shape, 
and a Yukawa term describing the helicity-changing couplings between the BEH field and 
the fermions. The discovery of the Higgs boson and measurement of its mass fixes the 
coefficients to $\lambda=m_H^2/2\upsilon^2\approx0.13$ and 
$|\mu|=\sqrt{\lambda}\cdot \upsilon=m_H/\sqrt{2}\approx89$\;GeV, 
where $\upsilon=|\mu|/\sqrt{\lambda}=(\sqrt{2}\cdot G_F)^{-1/2}\approx246$\;GeV 
is the vacuum expectation value of $\phi$.  

We may wonder how the potential evolves
with the decreasing temperature of the expanding universe. Above the critical temperature 
$T_{\rm EW}$ of approximately 100\;GeV, during the earliest $10^{-11}$ second after the 
big bang where the universe covered a causal domain of a few cm, the evolution of the 
potential is such that its 
minimum is $\langle0|\phi|0\rangle_{T>T_{\rm EW}}=0$. Gauge symmetry is respected, ie, 
all matter particles are massless and weak interaction is long-ranged. A spontaneous 
phase transition at $T_{\rm EW}$ (within the SM expected to be continuous, that is of second order) 
displaces the ground state of the BEH field to $\langle0|\phi|0\rangle_{T<T_{\rm EW}}=\upsilon$, 
breaking gauge symmetry. This spontaneous symmetry breaking corresponds to choosing 
a direction in the $SU(2)_L\times U(1)_Y$ group space. The condensed field fills all 
space-time, but without orientation as it has no spin. (One could imagine it as a 
Lorentz-invariant ether~\cite{Bernreuther}). The massive gauge bosons and fermions 
interact with the condensate which effectively reduces their velocity. The acquired mass is 
proportional to the strength of that interaction. The action of the BEH field thus creates a 
``vacuum viscosity''.

There are many questions about the structure of the SM in particular related to the 
matter sector: a large mass hierarchy is observed; CP violation has been observed in the 
quark sector, consistent with a single CP-violating phase in the quark mixing matrix. In effect,
three quark generations allow for exactly one such phase. The neutrino sector still bears many 
unknowns, among which the origin and values of neutrino masses, the neutrino nature, CP 
violation and whether or not there are sterile neutrinos that are singlets under the SM 
interactions (but possibly not under the new force). 

\begin{wrapfigure}{R}{0.52\textwidth}
\centering
\vspace{-0.43cm}
\includegraphics[width=0.52\textwidth]{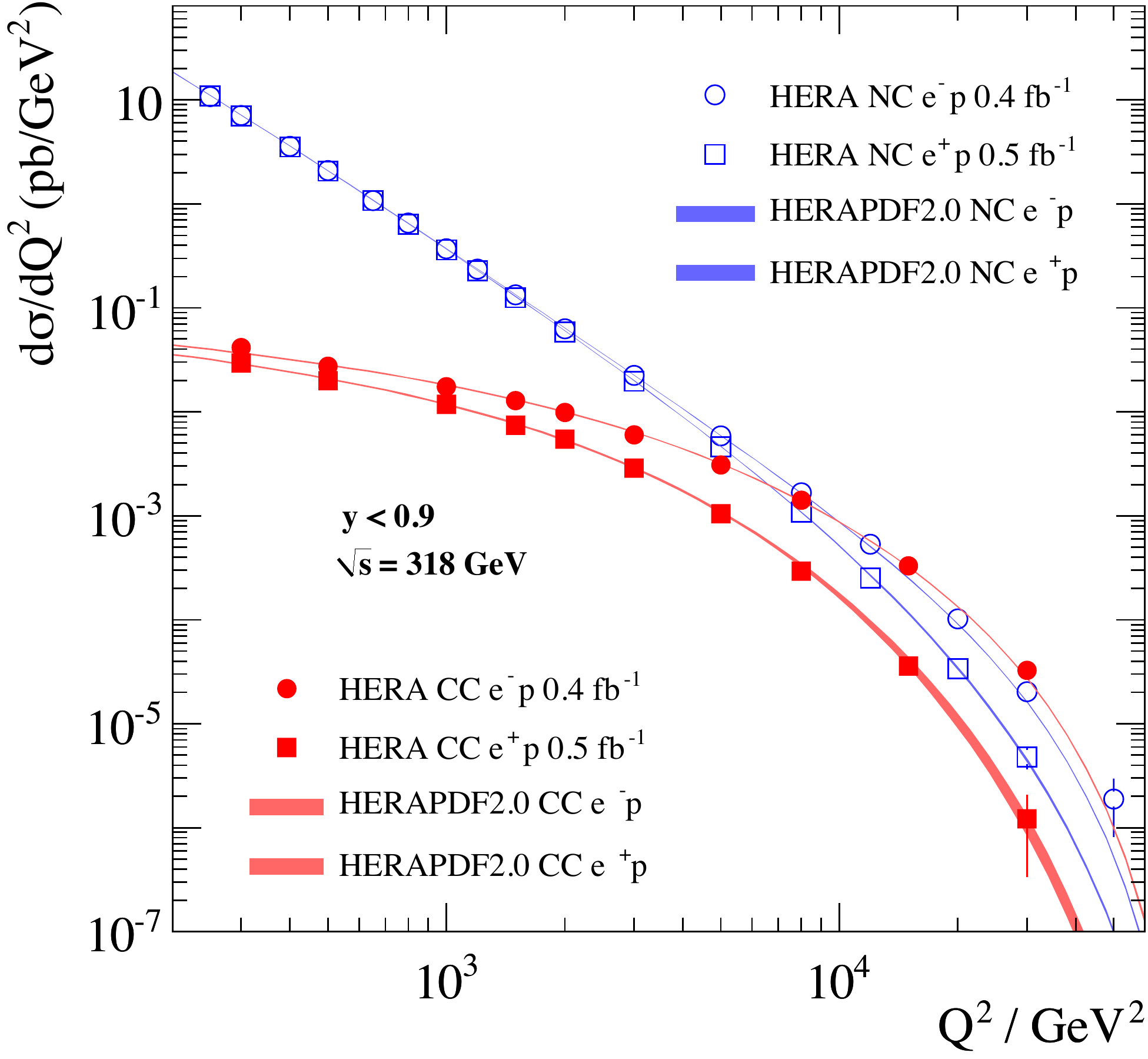}
\caption[.]{Differential cross-section versus momentum 
                 transfer-squared measured at the HERA collider for neutral (blue) and 
                 charged (red) current deep inelastic scattering processes~\cite{Hera}. The data points are 
                 integrated over the Bjorken-$x$ variable.
  \label{fig:Hera}}
\vspace{-0.5cm}
\end{wrapfigure}
Electroweak unification reduces the number of SM parameters from 20 to 19 (including 
the strong-CP parameter and neglecting the massive neutrino sector, which is irrelevant for LHC
physics unless there are new right-handed neutrinos in reach of the LHC). Figure~\ref{fig:Hera}
demonstrates beautifully electroweak unification at work at the HERA collider in electron--proton
and positron--proton scattering~\cite{Hera}. At low momentum transfer, neutral current processes 
with photon exchange producing an electron/positron in the final state dominate over charged 
current processes mediated via $W$ bosons. Above 100\;GeV, however, neutral and charged 
current processes are of similar size: electromagnetic and weak interactions are 
unified.\footnote{Looking in more detail into Fig.~\ref{fig:Hera}, 
the neutral current  cross sections 
for $e^-p$ and $e^+p$ are almost identical at small $Q^2$ but start to diverge as $Q^2$ grows. 
This is due to $\gamma$--$Z$ interference, which has the opposite effect on the $e^-p$ and $e^+p$ 
cross sections. The charged current cross sections also differ between $e^-p$ and $e^+p$ scattering, 
with two effects contributing: the helicity structure of the $W^\pm$ exchange and the fact that 
charged current $e^-p$ ($e^+p$) scattering probes the $u$-valence ($d$-valence) quarks~\cite{Hera}.}
Electroweak unification relates  the electromagnetic and weak coupling strengths to each other 
(the latter coupling given at lowest order by the  ratio-squared of weak gauge boson masses).
This relation has been tested experimentally to high precision~\cite{sin2thW}. 

The scales of particle physics reach from zero mass (for the photon and gluons) to as much 
as (and beyond) the Planck scale set by the  strength of gravity. Within these two 
extremes lies the range of sensitivity of the LHC covering three orders of magnitude 
between roughly 10\;GeV and 10\;TeV, probing length scales down to an attometre. That 
range comprises the study of high-energetic 
radiation such as jets, charm and bottom flavour physics, top quarks, $Z$, $W$ 
bosons and the Higgs boson, and any new physics that may reside therein. The physics 
at scales above that of the LHC is highly speculative. 
There could be right-handed neutrinos of mass above $10^{10}$\;GeV, as predicted by the 
(type 1) seesaw mechanism, the Peccei-Quinn axion scale above $10^{10}$\;GeV 
to suppress strong CP violation, grand unification of the electroweak and strong forces 
at roughly $10^{15}$\;GeV, quantum gravity at roughly $10^{18}$\;GeV and the 
hypercharge Landau pole well above the Planck scale.

\section{Particle physics at the dawn of the LHC}

The Higgs boson --- last of the particles? The SM predicts all properties, except for 
its mass. But before coming to the Higgs boson let us briefly recall the status of particle 
physics at the dawn of the LHC.
\begin{itemize}\setlength{\itemsep}{0.5\baselineskip}

\item LEP and SLC had ended their experimental programmes, with among their 
        main results the proof of three light active neutrino flavours, and direct Higgs boson
        searches that excluded $m_H < 114$\;GeV. Moreover, SM tests to unprecedented
        precision were performed with no direct or indirect hint for beyond the SM (BSM) physics.
        Among these, asymptotic freedom of strong interactions was tested       
        to the percent level through measurements of the strong coupling strength
        $\as(\mu)$ at scales $\mu=m_\tau$ and $\mu=m_Z$, respectively, and comparison
        with the accurately predicted evolution from the QCD renormalisation group. In both 
        cases the extraction occurred by comparing  experimental results 
        for the inclusive $\tau$ or $Z$ hadronic widths (among other $Z$ pole observables) 
        with NNNLO (3NLO) perturbative QCD predictions. 

\item Precision measurements at the 
        $Z$ pole and of the top-quark and $W$-boson masses allowed to exclude an 
        SM Higgs boson heavier than about 160\;GeV at 95\% confidence level. There are also 
        theoretical arguments in favour of a not too heavy Higgs boson, which is required to moderate 
        longitudinal  weak-boson scattering at large momentum transfer. 
        The evolution of the quartic coupling in the scalar
        potential with the energy scale $\Lambda$ representing the SM cut-off scale where new 
        physics occurs leads to constraints on $m_H$ in terms of upper perturbativity and lower 
        (meta)stability bounds. Indeed, the SM Higgs boson must steer a narrow course between 
        two disastrous situations if the SM is to survive up to the Planck scale 
        $\Lambda = M_{\rm Planck}$.

\item The Tevatron collider at Fermilab, USA  still continued Run-2. That collider led
         to the discovery of the top quark and the measurement of its mass to better than 
         1\%. The $W$ boson mass was measured more precisely than at LEP and by today
         Tevatron dominates the world average. 
         The mixing frequency of neutral $B_s$ mesons was measured for the first time, and found
         in agreement with the SM prediction. The Higgs boson was beyond Tevatron's sensitivity 
         except for masses around 160\;GeV, which could be excluded. No hint for BSM physics 
         was seen. 

\item The $B$ factory experiments BABAR at SLAC, USA and Belle at KEK, Japan were about 
         to end with a precise confirmation of the Kobayashi-Maskawa paradigm of a phase in 
         the three-generation CKM quark matrix being the sole responsible of the observed CP 
         violation in the quark sector. Ambiguous initial hints about a possible difference in the unitarity 
         triangle angle $\beta$ extracted from tree and loop (``penguin'') processes disappeared with 
         increasing statistics. 
         The $B$-factory experiments measured many rare processes and observed for the first time 
         CP violation in the charm sector. 
         
\item There was (and still is) no hint for charged-lepton flavour violation in spite of 
         ever increasing experimental sensitivity. 
         Any non-zero measurement would indicate new physics as the SM predictions via the 
         massive neutrino sector
         are immeasurably small. Also, no sign of a CP-violating electric dipole moment (EDM) was seen 
         in atoms or neutrons. The absence of a neutron EDM strongly constrains QCD induced 
         CP violation that would be expected in the SM. Only the anomalous magnetic moment of the muon 
         exhibits a long-standing $>$$3\sigma$ discrepancy between data and the SM prediction. 

\item The neutrino sector has seen a revolution after the discovery of neutrino oscillation
         and the measurement of all three angles of the neutrino mixing matrix. These 
         measurements establish that neutrinos have mass, but their nature (Dirac versus 
         Majorana), mass hierarchy (normal versus inverted), as well as CP violating mixing phase 
         remain unknown and are the subject of intense experimental activity.

\item Finally, there has been no signal other than gravitational effects for dark matter, 
         no signs of axions or of proton decay.\footnote{It is not possible to  reach energies 
           in the laboratory that would allow to directly study the physics at the expected 
           grand unification scale. Even Enrico Fermi's ``Globatron'' (that was to be built in 
           1994) would with current LHC magnet technology ``only'' reach insufficient 20 PeV 
           proton--proton centre-of-mass energy. 
           Proton decay is among the greatest mysteries in elementary particle physics. 
           It is required for baryogenesis and predicted by grand unified theories (GUT).
           Its discovery could therefore provide a probe of GUT-scale physics. The best current  
           limit on the partial proton lifetime from Super-Kamiokande, combining all its data from 1996 
           until now, is $\tau(p \to e^+\pi^0) > 1.7\cdot10^{34}$\;years.}

\end{itemize}

\section{Experimental setup}

Producing the Higgs boson and searching for new physics at the TeV scale requires a 
huge machine. Particle accelerators exploit three principles: $(i)$ they look deep into matter
requiring high energy to resolve small de Broglie wave lengths (particle accelerators are 
powerful microscopes), $(ii)$ Einstein's relation between energy and mass allows to 
produce potentially new heavy particles at high energy, and $(iii)$ accelerators probe the conditions 
of the early universe through Boltzmann's relation between energy and temperature. 

\begin{figure}[t]
\centerline{\includegraphics[width=0.9\linewidth]{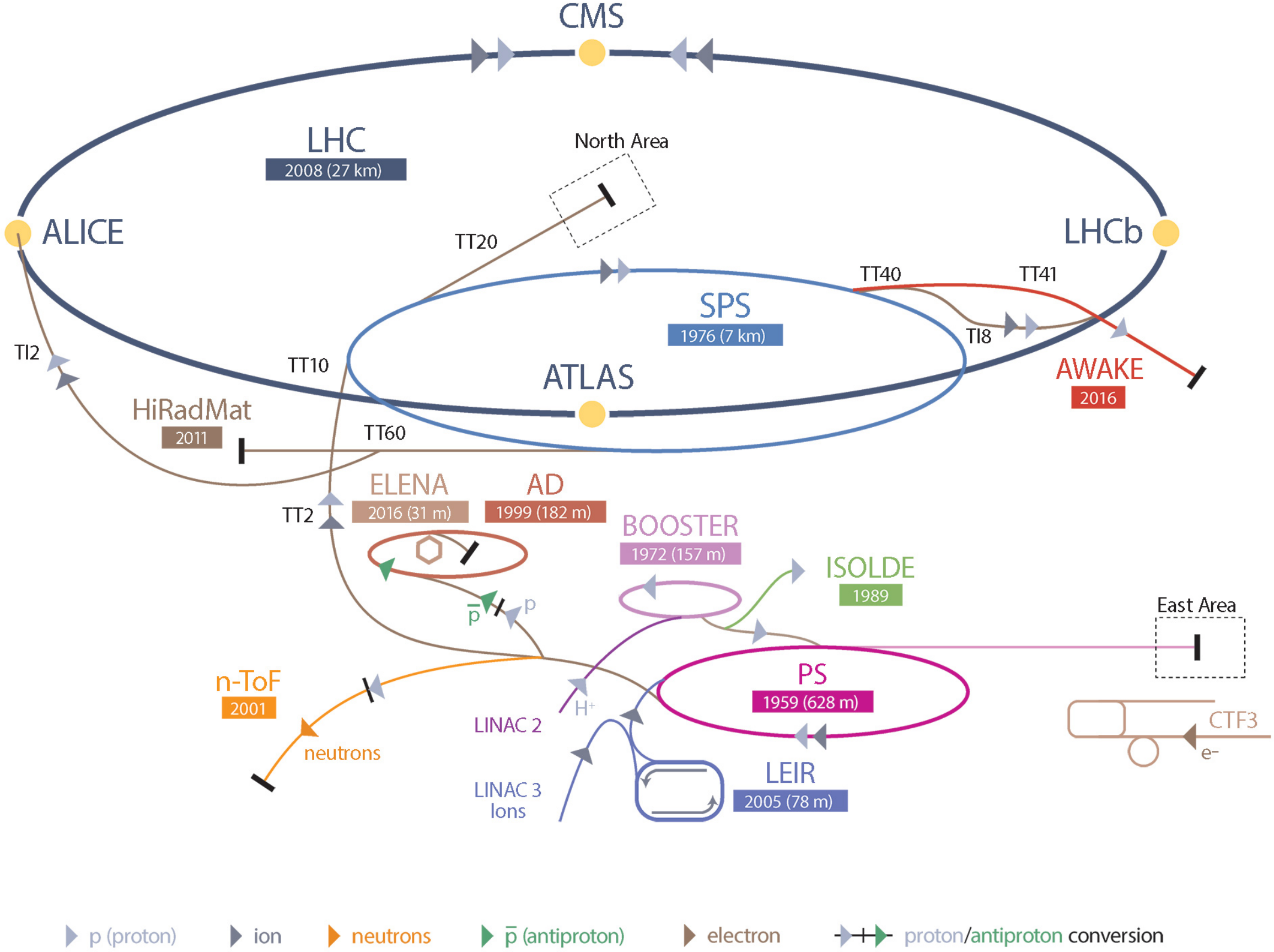}}
\vspace{-0.0cm}
\caption[]{Schematic view of CERN's accelerator complex.
  \label{fig:CERNacc}}
\end{figure}
Figure~\ref{fig:CERNacc} gives a schematic view of CERN's accelerator complex. It 
consists of a succession of machines that accelerate particles to increasingly higher 
energies and condition the particle beams. Each machine boosts the energy of a 
beam before injecting it into the next machine in the sequence. 
Protons are accelerated to 50\;MeV in the Linac 2, to 1.4\;GeV in the PS Booster, 
to 26\;GeV in the PS (Proton Synchrotron) where also spatial proton bunches with  25\;ns 
(7.5\;m) distance and bunch trains
are formed, 450\;GeV in the SPS (Super Proton Synchrotron) before being injected 
in opposite directions into the LHC.  Booster, PS and SPS have their own experimental halls 
where fractions of the beams are used for fixed target experiments at lower energies.
The LHC~\cite{LHC} is a superconducting proton/ion accelerator and collider installed 
in a 26.7\;km circumference, 70--140\;m underground tunnel with 4\;m cross-section 
diameter.  Up to 2\,800 bunches containing each more than 100 billion protons
are accelerated within roughly 20 minutes from 450\;GeV up to the design
energy of 7\;TeV per beam. So far, proton--proton collisions with centre-of-mass 
energies of 0.9, 2.8, 5, 7, 8, and 13\;TeV were delivered by the LHC. In 
addition, data with 5 and 8\;TeV proton--lead and 2.8 and 5\;TeV lead--lead 
collisions were taken. 

The most challenging component of the LHC are the 1\,232 superconducting dipole 
magnets realised in a novel ``2-in-1'' design that guide the protons along their circular trajectory 
around the ring. The dipoles have a length of 14.3\;m each, and are cooled to 1.9$^\circ$\;K 
by means of a closed circuit of 120 tonnes liquid-helium. The LHC also features almost 
400 focusing quadrupole magnets and 3\,700 multipole corrector magnets. 
The maximum dipole field strength of 8.3\;T, achieved with a current of almost 12\;kA, 
limits the energy to which the protons can be accelerated. The proton's energy is
given by 
$E_{p}[{\rm TeV}]=\sqrt{4\pi\alpha}\cdot B[{\rm T}]\cdot r[{\rm km}]$ so that, with the
radius $r=4.3$\;km,  one finds $E_{p}\sim7$\;TeV  taking into account that only roughly
two-third of the ring are equipped with dipoles. Following the scale-energy relation 
$\mu\approx 200$\;GeV\,am$\,/E[{\rm TeV}]$, does the LHC thus probe length scales of 
$\mu\sim10^{-20}$\;m at 14\;TeV centre-of-mass energy? It is not quite that small
as the protons are composite particles whose energy is distributed among its 
constituents (partons).  

The LHC hosts four large, ultra-sophisticated experiments among which the general 
purpose detectors ATLAS and CMS, as well as ALICE and LHCb dedicated mainly 
(but not only) to heavy-ion and flavour physics studies, respectively, requiring 
optimisation for low transverse momentum physics. There are additional smaller-scale 
experiments dedicated to forward physics. The design of the ATLAS detector
emphasises excellent jet and missing transverse momentum resolution, 
particle identification, flavour tagging, and standalone muon measurement. CMS 
features excellent electron/photon energy and track (muon) momentum resolution, 
and flavour tagging. Both detectors are highly hermetic with very few 
acceptance holes allowing to precisely determine missing transverse momentum. 
ALICE has highly efficient track reconstruction in busy heavy-ion 
environment and particle identification. LHCb is a forward spectrometer with a trigger 
for fully hadronic $B$ and $D$ hadron events, excellent low-momentum track resolution, 
and particle identification (pion/kaon separation). 

The particle detectors measure particles produced as debris from the proton--proton collisions 
through interaction with active material. Different concentric detector layers measure different
properties. The innermost parts of the detectors measure tracks of charged particles 
in layers of semi-conductors or straw tubes that respond to traversing charges. The 
momentum and charge of these particles is measured through their immersion in a 
homogeneous magnetic field. Outside the tracking volume are thick calorimeters
that absorb most particles and measure their energy. Additional tracking chambers 
behind the calorimeters identify and measure muons, which are minimum ionising 
in most of their momentum range and thus penetrate the calorimeter layers. Neutrinos
do not interact with the detector and therefore induce missing energy. 
Due to the non-zero longitudinal momentum of the collisions in the laboratory frame
and the missing acceptance coverage along the beam line, only the transverse missing
momentum in a collision is conserved. Hence, the reconstructed missing transverse momentum
is used to detect neutrinos or any unknown non or weakly interacting particles.
In addition to stable particles, the experiments also reconstruct jets from near-by 
calorimeter depositions or tracks. Jets are narrow cones of hadrons and other particles 
produced by the hadronisation of a quark or gluon due to QCD confinement. 
Reconstruction of long-lived states in a jet allows to tag
jets originating from $b$ or $c$ quarks. The ensemble of measurements 
in a given proton--proton bunch crossing makes up an event. It contains
more than the original hard parton scattering due to underlying event
interaction and additional soft proton--proton interactions (dubbed ``pileup''). 

LHC computing represents ``big data'': the LHC experiments started more 
than a decade ago with large scale  computing, which  is now present everywhere. The 
ATLAS managed data volume of roughly 150\;petabyte (dominated by simulated data) 
is of similar order as the Google search index or the content uploaded to Facebook 
every year. Unlike these companies, however, the LHC has to manage its data 
volume with a public science budget.

\section{Experimental methods}

We will review in this chapter a few (basic) experimental concepts at the LHC.

\subsection{Luminosity}

Besides energy, luminosity is the single most important quantity in collider physics.
The instantaneous luminosity of the beam collision, expressed in units of s$^{-1}$cm$^{-2}$, 
is a function of the LHC beam parameters as follows
\beq
\label{eq:luminosity}
       L  = \frac{f_{\rm rev} \cdot n_{\rm b} \cdot N^2_p}{4\pi\cdot \sigma_x\cdot \sigma_y}
                   \cdot F(\theta_c,\sigma_x,\sigma_z)
                \;,
\eeq
where $f_{\rm rev} =11\,245.5$\;Hz is the bunch revolution frequency determined by the 
size of the LHC (27\;km) and the speed of light, $n_{\rm b}=1,\dots,2\,808$ is the number
of proton bunches in the machine (2\,808 is the maximum number of possible 25\;ns slots; 
the theoretical maximum of 3\,564 bunches cannot be reached due to space needed between 
bunch trains and for the beam dump kicker magnets (abort gap)), $N_p\approx 1.15\cdot10^{11}$
is the number of protons per bunch (the bunch intensity), and 
$\sigma_{x,y}=12,\dots,50\;\mu$m is the transverse beam width characterising the 
beam optics. The factor $F(\theta_c,\sigma_x,\sigma_z)$ accounts 
for  luminosity reduction due to the beam crossing angle $\theta_c$, roughly given by 
$(1+(\sigma_z/\sigma_x)^2\cdot(\theta_c/2)^2)^{-1/2}$), the hourglass effect leading
to a varying transverse bunch size in the collision point because of the several cm longitudinal 
bunch extension, and other effects. 

Luminosity drives the statistical precision of any measurement and our ability to observe 
low cross section processes as
\beq
       N_{\rm events}^{\rm obs} = {\rm cross~section}\,\times\,{\rm efficiency}\,\times\,\int\! L\cdot dt\;,
\eeq
where the cross section is given by Nature, the efficiency of detection is optimised by 
the experimentalist, and the integrated luminosity is delivered by the LHC. There are several 
options to maximise the luminosity of the machine as outlined below.
\begin{itemize}\setlength{\itemsep}{0.5\baselineskip}

\item {\bf Maximise the total beam current.} 
        The cryogenic system limits the maximum beam 
        current leading to an anticorrelation between $N_p$ and $n_{\rm b}$.
        Improvements in beam collimation, cryogenics vacuum, and background 
        protection allow to extend that limit. 

\item {\bf\boldmath Maximise brightness and energy, minimise $\beta^*$.} 
         The transverse beam size is given by $\sigma(s)=\sqrt{\beta(s)\varepsilon_n/\gamma}$,
         where $\sigma^*=\sigma(s=0)\approx17\;\mu$m at the collision point.
         The value $\beta^*\approx60\;$cm is the longitudinal distance from the focus 
         point where the transverse beam size grows twice as wide.
         The emittance $\varepsilon\cdot\pi$ is the area in phase space occupied 
         by the beam, and $\varepsilon_n\approx3.8\;\mu$m is the normalised emittance,
         where the Lorentz $\gamma$ factor is taken out. 
         To reduce (``squeeze'') $\beta^*$ one needs to respect the quadrupole aperture limit.
         The beam brightness, $N_p/\varepsilon_n$, is limited by beam--beam interactions 
         which have a quadrupole de-focusing effect, and by space-charge tune shift and spread 
         (the tune spread is limited by resonances).

\item {\bf Compensate reduction factor.} 
         The crossing angle is required to avoid  parasitic long-range beam encounters.
         The hour glass effect may be reduced by shorter bunches, at the expense of a higher
         longitudinal pileup density.

\end{itemize} 
The LHC group offers an excellent tool~\cite{LHCtool} to study the dependence of the 
expected LHC luminosity under various parameter settings. 

The instantaneous luminosity is measured by the 
experiments~\cite{ATLAS-lumi,CMS-lumi,LHCb-lumi} with dedicated detector systems 
that are calibrated with the use of so-called van-der-Meer beam-separation scans~\cite{vdMeer}. Such 
scans are performed in specific low-intensity LHC fills during which the beams are separated 
from each other by an increasing distance in both $x$ and $y$ directions. The dedicated luminosity 
detectors count for each scan point the hits they receive from inelastic minimum-bias events. 
From the Gaussian profile of the hit counts versus the beam separation one can determine the 
transverse beam profiles entering Eq.~(\ref{eq:luminosity}) via the convolution  relation 
$\sigma_{x,y}=\Sigma_{x,y}/\sqrt{2}$, where it is assumed that both LHC beams have the 
same width. The knowledge of $L$ from the measured beam currents and beam widths in the 
specific LHC fill allows to extract the visible cross section, $\sigma_{\rm vis}$, 
for any dedicated luminosity detector (this may include, for example, cluster counting in the 
Pixel detector). During normal (ie, high-luminosity) data taking, the counts measured in that 
detector together with the calibrated visible cross section from the van-der-Meer scan
allow to extract the luminosity via
$L = N_{\rm counts}/\sigma_{\rm vis}$. Note that this method does not require to know the 
acceptance nor the efficiency of the luminosity detector, which are included in $\sigma_{\rm vis}$.

The precision of the luminosity calibration depends on many factors. Systematic uncertainties
arise from correlations between the $x$ and $y$ transverse beam positions during the scan, 
beam--beam corrections, beam orbit drifts and position jitters, stability (reproducibility) of the results,
and instrumental effects such as the absolute length scale calibration for the separated beams,
beam backgrounds and noise, the reference specific luminosity, the measurement of the beam 
currents, and  the extrapolation of the calibration from the low to  high-luminosity 
regimes as well as the run-by-run stability of the luminosity detectors. The best precision on the 
integrated luminosity achieved by the experiments undercuts 2\%.

During Run-1 of the LHC, spanning the years 2010--2012 of data taking, the peak luminosity achieved was 
$L_{\rm peak}=7.7\cdot10^{33}\;{\rm cm}^{-2}{\rm s}^{-1}$ and an integrated luminosity during the year
2012 of 23\ifb at 8\;TeV proton--proton centre-of-mass energy was delivered to the experiments. With 50\;ns
bunch distance a maximum number of 1380 colliding bunches was reached. At $L_{\rm peak}$
the LHC produced every 45 minutes a $H\to\gamma\gamma$ event, and typically 
two 160\ipb fills were needed to produce one $H\to4\ell$ ($\ell=e,\mu$) event.

The high luminosity of the LHC comes to the price of additional inelastic proton--proton pileup interactions
within a bunch crossing. An average of $\langle\mu\rangle=21$ (maximum $\langle\mu\rangle=40$) interactions per 
crossing occurred during 2012, with a similar or slightly higher rate in 2016. The LHC design 
pileup value at 14\;TeV is obtained as follows
\beq
\label{eq:pileup}
      \langle\mu\rangle = \frac{\sigma_{\rm inel}\cdot L}{f_{\rm rev}\cdot n_{\rm b}}
             \approx \frac{80\;{\rm mb}\cdot10\;{\rm nb}^{-1}{\rm s}^{-1}}{11\,245\;{\rm s}^{-1}\cdot2\,808}
             \approx 25\,,
\eeq
where we used $10^{34}\;{\rm cm}^{-2}{\rm s}^{-1}=10\;{\rm nb}^{-1}{\rm s}^{-1}$.
When the detector response integrates over several bunch crossings,  as for example 
the calorimeter pulse shape, pileup occurring in the recorded proton collision  (in-time pileup) as well
as that in neighbouring collisions (out-of-time pileup) affect the event reconstruction.
Most analyses are fairly insensitive to pileup at the  rates experienced so far.  Mitigation 
methods have been developed to further improve the robustness of the physics object reconstruction 
and analyses. Pileup does, hoever,  affect the trigger requiring higher thresholds,
which impacts the low transverse momentum physics programme of the experiments. It 
also increases the stored event size and CPU time needed for track reconstruction.

\subsection{Proton--proton collisions}

\begin{figure}[t]
\centerline{\includegraphics[width=0.9\linewidth]{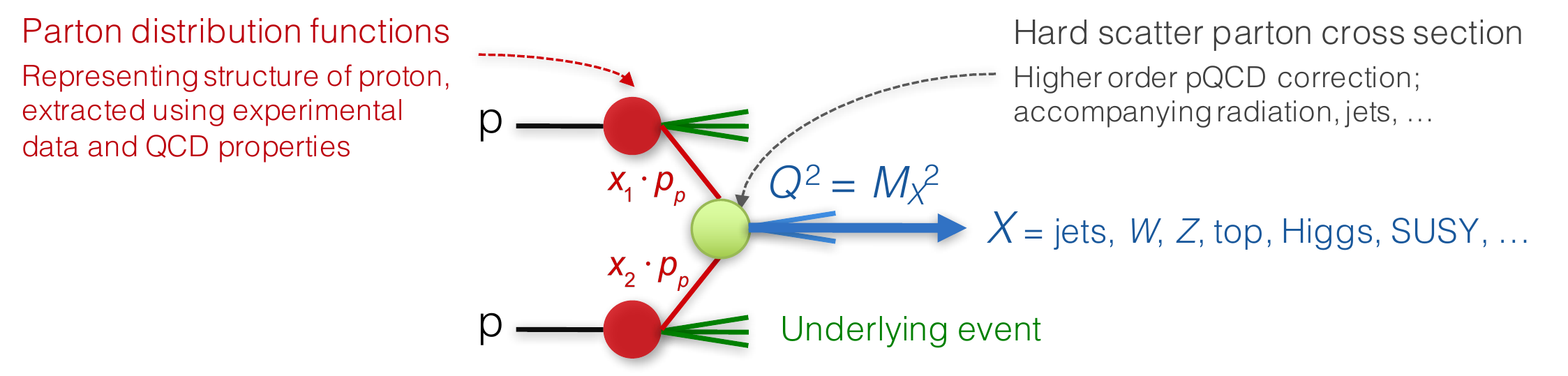}}
\vspace{-0.1cm}
\caption[]{Simplified view of a proton--proton collision. 
  \label{fig:ProtonCollision}}
\end{figure}
\begin{wrapfigure}{R}{0.53\textwidth}
\centering
\vspace{-0.52cm}
\includegraphics[width=0.51\textwidth]{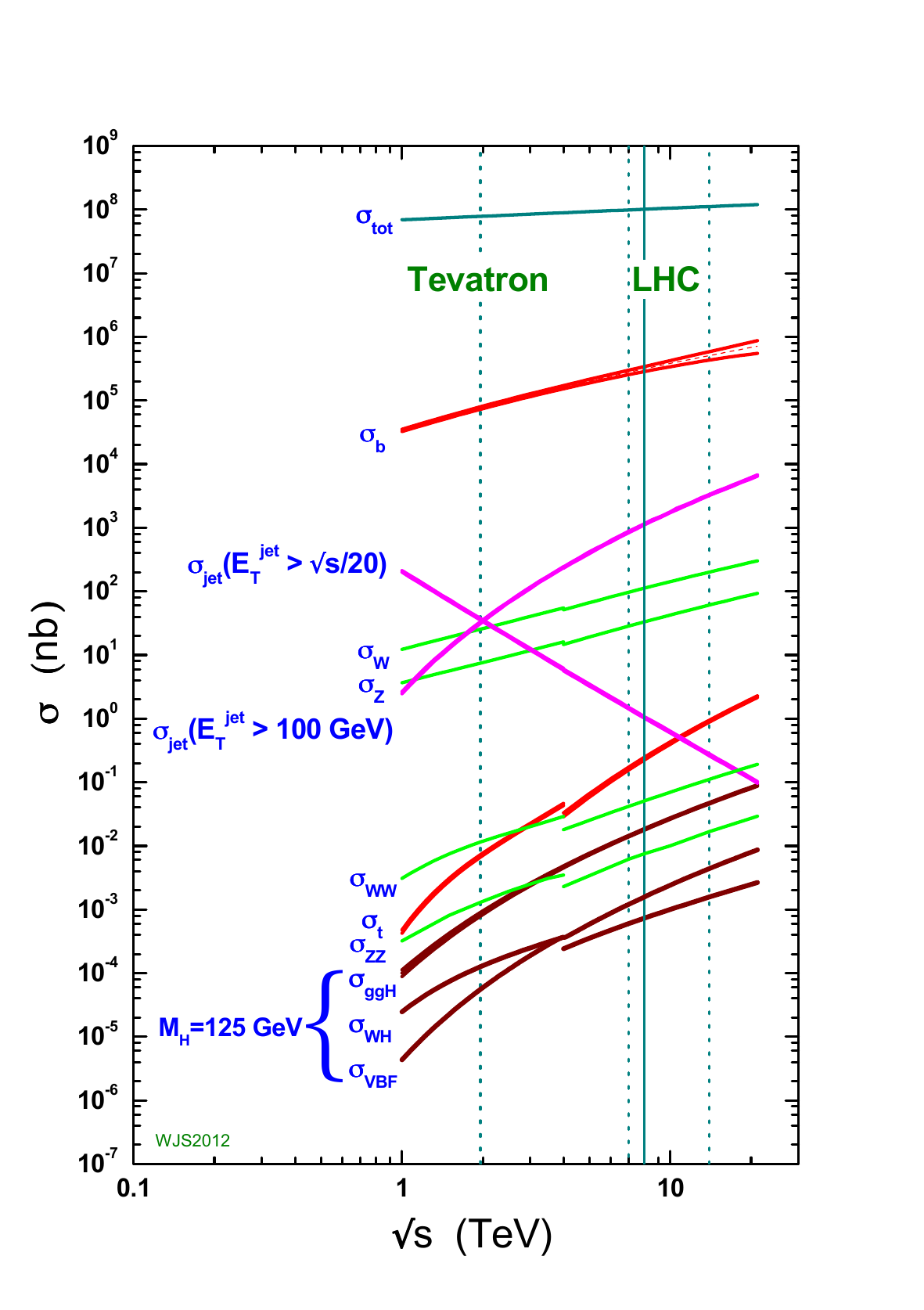}
\caption[.]{Cross sections of selected proton--(anti-)proton processes versus centre-of-mass
               energy~\cite{StirlingPartonLuminosities}. 
  \label{fig:ppXS}}
\vspace{-1.0cm}
\end{wrapfigure}
Owing to factorisation (see for example~\cite{Factorisation}), 
the cross section of a proton--proton collision  can be computed 
as the convolution of parton density functions\footnote{Parton density functions were 
  introduced 1969 by Feynman  in the parton model to explain Bjorken scaling in 
  deep inelastic scattering data.} (PDF) with the parton scattering matrix 
element (cf. Fig.~\ref{fig:ProtonCollision}). 
The PDFs are universal distributions containing the long-distance structure 
of the proton (or hadrons in general) in terms of valence and sea quarks and gluons.
They are related to parton model distributions at leading order, but with logarithmic 
scaling violations (DGLAP\footnote{In analogy with a running coupling strength, one 
  can vary the factorisation scale and obtain the renormalisation group equation for PDFs. 
  The DGLAP equations~\cite{DGLAP} describe the $Q^2$ dependence of 
  the PDFs.}).
Since precise Lattice QCD predictions are not yet available, 
the PDFs are extracted versus the parton momentum 
fraction $x$ and the momentum transfer $Q^2$ using experimental 
data and exploiting QCD evolution properties. The centre-of-mass 
energy-squared of the parton collision, $\hat s$, is given by the product of the 
momentum fractions of the colliding partons, $x_{1,2}$, times the proton centre-of-mass 
energy: $\hat s = x_1\cdot x_2\cdot s$. The production of the 125\;GeV Higgs boson 
thus occurs at an average momentum fraction$\langle x\rangle \sim  0.01$ at 
$\sqrt{s}=13\;$TeV.

The parton density functions rise dramatically towards low $x$ in particular at high $Q^2$
and most notably for the gluon density. The consequences are: the cross section of a given 
process increases with increasing proton--proton collision energy, more luminosity allows to 
reach higher parton collision energy, and the low-$x$ regime is dominated by gluon--gluon 
collisions. Hence, although the
parton-level cross sections falls  with centre-of-mass energy, the cross-sections of 
proton--proton processes rise due to the convolution with increasing PDFs. This is depicted
in Fig.~\ref{fig:ppXS}: all cross sections rise with centre-of-mass energy, where gluon 
initiated processes have a steeper slope than quark initiated ones because of the strong
enhancement of the gluon PDF towards lower $x$. One notes that when requiring a 
centre-of-mass-energy dependent minimum transverse momentum for jets, the jet cross section
decreases with centre-of-mass energy as expected for a parton--parton cross section. The inclusive
cross section is dominated by inelastic scattering (also denoted minimum bias events),
the interesting jet and boson physics processes have many orders of magnitude lower rates. 

\subsection{The experimental data path in a nutshell}

The LHC detectors cannot record events at the filled proton bunch crossings rate of 
approximately 30\;MHz\footnote{The LHCb phase-1 upgrade is preparing for exactly that!} as 
each event has an approximate raw size of 2\;MB, requiring to store
60\;TB per second. This is not only impractical given the computing resources, but 
also unnecessary in view of the LHC physics goals as most events contain only soft minimum 
bias interactions. Instead, online custom hardware 
and software triggers reduce that rate by filtering out events with a million and
more times smaller cross sections than minimum bias events. 

The data path of an LHC experiment can be described in a nutshell as follows.
\begin{enumerate}\setlength{\itemsep}{0.5\baselineskip}

\item LHC bunches collide every 25\;n (50\;ns during Run-1), but not all bunches are filled
        with protons. 

\item LHC detectors record the detector response in pipelined on-detector memories
         that are time-stamped (synchronised) to the LHC collisions they belong to. 
         The events are kept during the latency of the first level trigger decision 
         (2--3\;$\mu$s).

\item Level-1 hardware and high-level software triggers filter the interesting events 
         that are written to disk. The level-1 trigger system reduces the initial bunch 
         crossing rate to up to 100\;kHz. The high-level trigger further reduces this rate 
         to about 1\;kHz that are kept. A trigger menu is a large collection of 
         physics and monitoring triggers. Among these are low-threshold single lepton 
         triggers, single missing transverse momentum and jet triggers, and lower threshold 
         di-object and topological triggers. The online system also provides detailed 
         data quality monitoring.

\item The recorded data are subject to prompt offline calibration and refined monitoring, 
         followed by the prompt reconstruction (mostly) at CERN. 

\item The reconstructed data are distributed to computing centres world-wide from where 
         standardised derived datasets are produced for physics and performance analysis.

\item Large amounts of Monte Carlo events using the same reconstruction software as 
         used for data are also produced and distributed for analysis. 

\item Performance groups provide standard physics objects with calibrations and uncertainties, 
         unified in analysis releases. Analysis groups build physics analyses on top of this ground work.

\end{enumerate}

\subsection{Monte Carlo event simulation}

A crucial ingredient to any physics and performance analysis is Monte Carlo (MC) event simulation. 
MC events mimic the physics processes, which allows to isolate specific processes by 
subtracting simulated background processes, to evaluate the acceptance and efficiency of 
signal processes, to optimise signal selection, and to evaluate systematic uncertainties by 
varying MC parameters.

The MC generation path is sketched in Fig.~\ref{fig:MCpath}. A matrix element generator 
calculates the hard  parton--parton scattering event and stores it in a common 
data format. The event is passed through the detector simulation which simulates 
the interactions of the stable particles with the active and passive detector material. 
The simulation may use Geant4~\cite{Geant4} or a parametrised fast simulation. The output
of that process is subject to the digitisation step during which the detector response 
and readout is mimicked. After this step the simulated data have the same format as real 
detector data except for the so-called truth information which records the information 
about the generated particle types, decay chains and four-momenta. The following event 
reconstruction is identical to that of real data (also format wise).

\begin{figure}[t]
\centerline{\includegraphics[width=1\linewidth]{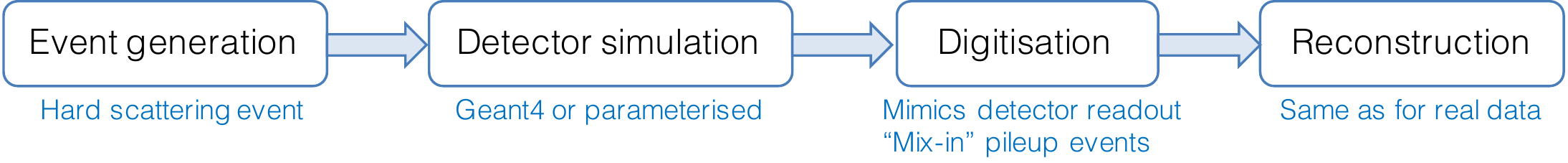}}
\vspace{-0.0cm}
\caption[]{Monte Carlo event simulation chain. 
  \label{fig:MCpath}}
\end{figure}
The physics modelling with event generators proceeds as follows. The hard scattering matrix 
element calculation including initial and final-state radiation (ISR/FSR) is convolved with the 
parton density functions. Decays of the hard subprocesses, and multiple parton interactions (and 
their ISR/FSR) are also generated. Matrix elements are used as much as possible, but one cannot 
fully avoid phenomenological description of nonperturbative effects such as parton showers, 
hadronisation, and the underlying event. State-of-the-art event generation includes next-to-leading
order (NLO) matrix elements up to two partons, leading order (LO) matrix elements up to 5 partons, 
parton shower matching. Nonperturbative and electroweak corrections are sometimes applied.
Fixed-order calculations are known to higher order (NNLO or even 3NLO). The physics modelling, 
including PDFs, has the largest systematic uncertainty in many analyses and is therefore a 
critical step that requires care, extensive validation and sometimes correction 
with data. 

There has been significant progress in both generator developments and fixed-order calculations.
Fixed-order predictions have seen an NNLO revolution with about 20 new results during the last two
years. Some of the resulting NNLO to NLO K-factors, in particular for diboson production, were not 
covered by the  uncertainties assigned to the NLO calculations as obtained from the canonical factors 
two and one-half variations of the renormalisation scale~\cite{SalamLHCP}. 

\subsection{Cross section measurement and data unfolding}

The measurement of the cross section of a given process requires to isolate this process 
via selection of its final state, followed by the subtraction of contributing
background processes. The remaining number of events needs to be corrected for resolution 
effects, the selection efficiency and, in case of the total cross section, the acceptance:
\beq
\label{eq:XS}
       \sigma^{\rm tot}_{pp\to X} = \frac{1}{A_X} \sigma^{\rm fid}_{pp\to X} 
                                            = \frac{1}{A_X} 
                                               \left(\frac{N_{\rm obs}-N_{\rm bkg}}{L \cdot C_X}\right)\;.
\eeq
Here, $N_{\rm obs}$ and $N_{\rm bkg}$ are the total observed and the estimated number 
of background events, $L$ the integrated luminosity,
$A_X=N_{\rm gen,fid}/N_{\rm gen}$ is the acceptance factor, given by the fraction of 
generated $pp\to X$ events falling into the fiducial acceptance defined close to the 
final state selection, $C_X=N_{\rm reco,sel}/N_{\rm gen,fid}$ is a correction factor that
corrects for the detector resolution and inefficiency of events generated in the fiducial region,
and $\sigma^{\rm fid}_{pp\to X}$ is the fiducial cross section. 

The acceptance factor is computed entirely from theory using the best available fixed order 
calculation. The correction factor depends on the particle detector and is partly or fully 
determined from MC simulation. It is therefore the main interest of the experimentalist 
to determine the fiducial cross section, which is corrected for experimental effects and has 
minimal theory dependence. The acceptance correction to obtain the total inclusive cross 
section is left for theory. 

The definition of the fiducial cross section should facilitate the comparison between 
theoretical predictions and experimental results and thus should have the least possible 
dependence on the MC event generators available at the time of the measurement.
A suitable definition of the observables is based on the physical particles that enter 
the detector. This  includes the stable particles which account for the majority of 
interactions with the detector material, and from which the measurements are 
ultimately made. A detailed discussion about the definition of particle level objects
is provided in a dedicated ATLAS note~\cite{TruthATLASNote}.\footnote{The
recommendations in Ref.~\cite{TruthATLASNote} about how to suitably define an event topology 
at particle level on MC are: 1. Select the stable particles. 2. Select prompt leptons ($e$, $\nu_e$, 
$\mu$, $\nu_\mu$) and associate photons (not from hadrons) to electrons and muons 
to define dressed-level charged leptons. 3. Define particle-level jets by clustering all stable 
particles excluding the particles found in step (2). 4. Assign the jet flavour based on 
heavy-flavour hadrons ghost-matched to jets. 5. Sum all prompt neutrinos defined in 
(2) to form the missing transverse momentum. 6. Resolve lepton-lepton and jet-lepton
overlap following a procedure close to that used at the detector level. 7. Define other 
particle-level observables in complex event topologies based on the particle-level 
objects defined above.}

A differential cross section corresponds to a binned fiducial cross section. It requires the application
of unfolding due to bin-to-bin correlations. Unfolding is a mathematically unstable inversion 
problem that requires careful regularisation. 

\subsection{Background determination} 

\begin{wrapfigure}{R}{0.48\textwidth}
\centering
\vspace{-0.45cm}
\includegraphics[width=0.46\textwidth]{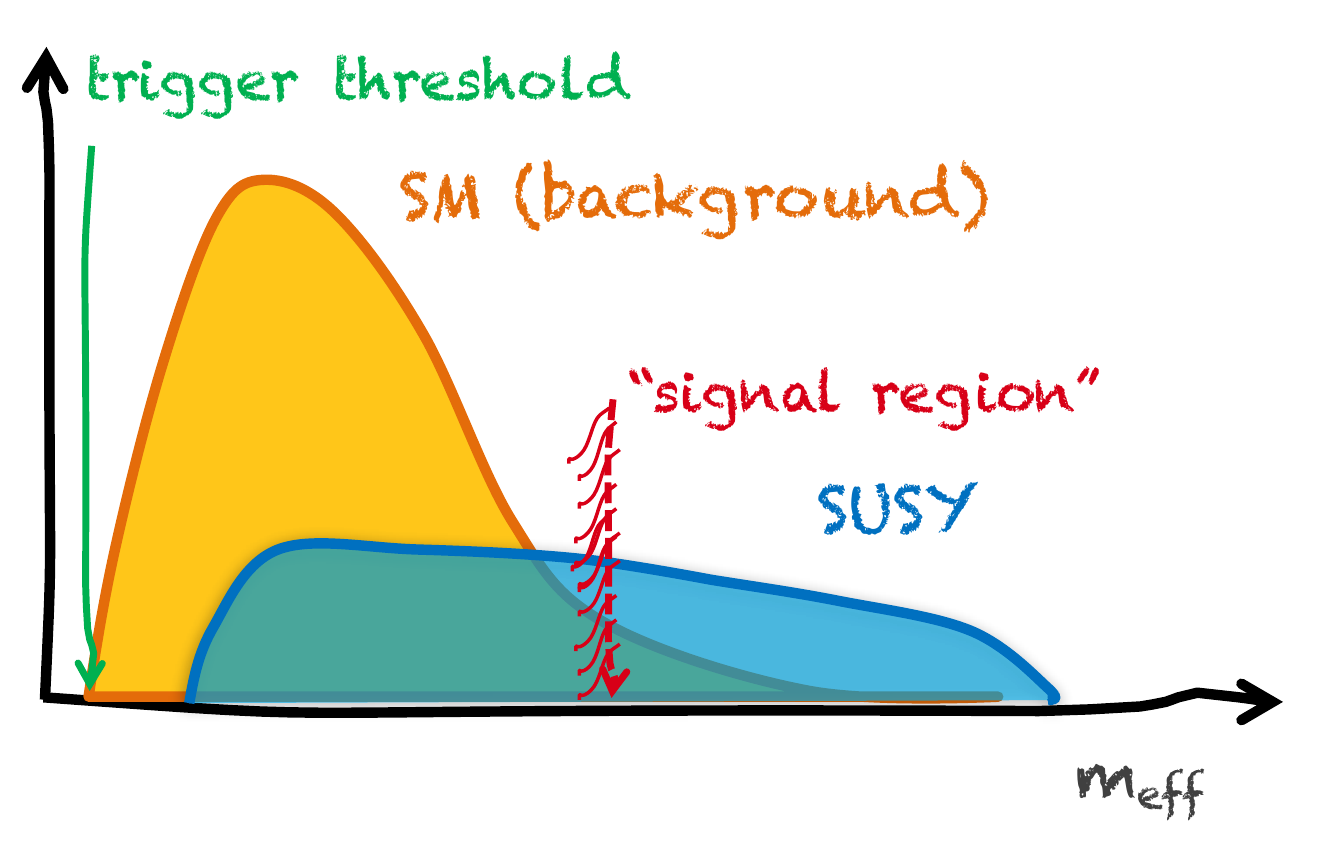}
\vspace{-0.1cm}
\caption[.]{Sketch illustrating signal and background distributions of a discriminating variable.
                A requirement is applied to enrich the signal. 
  \label{fig:Bkg}}
\vspace{-0.9cm}
\end{wrapfigure}
The subtraction of backgrounds as in Eq.~(\ref{eq:XS}) usually relies on MC simulation. 
In case of new physics searches, which often select extreme phase space regions such as a very 
large effective mass\footnote{The effective mass, $m_{\rm eff}$,  is defined by the scalar 
sum of the transverse momenta of all selected objects and the missing transverse momentum.
\label{ftn:effectivemass}}
as illustrated in Fig.~\ref{fig:Bkg}, the MC predictions may suffer from large and difficult
to estimate modelling uncertainties. More robust and reliable background estimates normalise
the MC predictions of the main backgrounds in phase space regions (dubbed ``control regions'') 
close to the signal region. Sometimes further data driven corrections for the transfer of the MC
normalisation from the control region to a signal region are required. 

For a typical search for supersymmetry looking for jets, possibly $b$-jets and leptons, and missing 
transverse momentum, the main background sources stem from top production, $W+{\rm jets}$ and 
$Z(\to\nu\nu)+{\rm jets}$ events, $WW$, $WZ$, $ZZ$ diboson production, and rare processes 
such as $tt+W$ or $tt+Z$. All these backgrounds may produce true missing transverse momentum 
due to decays to neutrinos. Additional backgrounds may arise from misreconstruction of 
multijet events, or the misidentification of non-prompt leptons or jets as prompt leptons.
Such backgrounds are usually determined from data using control regions or 
so-called ABCD sideband methods exploiting two or more none or only weakly correlated 
variables. The former backgrounds are often called irreducible, and the latter
due to misreconstruction are denoted reducible backgrounds. MC simulation is 
mainly used to predict irreducible backgrounds, in particular if they are sub-dominant. 

\subsection{Basic physics objects}

All ATLAS and CMS physics analyses are built upon basic physics objects. These correspond 
to single stable particles, ensembles of particles or event properties. 
\begin{itemize}\setlength{\itemsep}{0.5\baselineskip}

\item[] {\bf Tracks and vertices} are measured in the inner tracking systems. Their precise 
           measurement requires an accurate detector alignment which is obtained from data by 
           minimising hit residuals with respect to fitted tracks. Also important is a precise mapping
           of the inner tracker geometry and material, which is made with the help of survey data, 
           and from collision data using reconstructed vertices from hadronic interactions, photon 
           conversions to electron--positron pairs, track extensions, and long-lived hadrons.

\item {\bf Electrons and photons} are reconstructed as energetic clusters in the electromagnetic 
         calorimeter associated or not with an inner detector track. Due to significant amount of 
         active and passive tracker material (between 0.4 and 2.4 radiation lengths depending 
         on $|\eta|$),\footnote{The ``rapidity'' of a particle is defined by
           $y = \frac{1}{2}\ln[(E + p_z )/(E - p_z )]$, where $E$ denotes the particle's energy 
           and $p_z$ the  particle's momentum along the beam direction ($z$). 
           Differences in rapidity are Lorentz invariant under a boost along $z$.
           The ``pseudorapidity'' is defined 
           by $\eta = \frac{1}{2}\ln[(p + p_z )/(p - p_z )] = -\ln(\tan\!\frac{\theta}{2})$. 
           The azimuthal angle 
           $\phi$ is measured in the plane transverse to the beam direction and the polar angle 
           $\theta$ is measured with respect to the beam direction. Rapidity 
           and pseudorapidity are equal for massless particles. } 
         roughly 40\% of the 
         photons convert to electron--positron pairs and hence are reconstructed as one or two displaced 
         electron tracks.
         The electron efficiency, energy scale and resolution are precisely calibrated in data using
         $Z\to ee$,  $J/\psi\to ee$ and $W\to e\nu$ events. Photons are calibrated using MC and 
         radiative  $Z\to ee(\mu\mu)+\gamma$ events, as well as $\pi^0\to\gamma\gamma$ decays
         at low energy.

\item {\bf Muons} are reconstructed in the inner tracker and the outer muon systems. Combined
         tracking improves the momentum resolution for high transverse momentum ($p_T$) 
         muons, which are dominated in precision by alignment uncertainties.
         Muons are calibrated using $Z\to  \mu\mu$, $J/\psi\to \mu\mu$ and  
         $\Upsilon(1S)\to \mu\mu$ events.

\item {\bf\boldmath Hadronic $\tau$ decays ($\tau_h$)} to a narrow jet of charged and neutral pions or 
         kaons are reconstructed in the inner tracker and the electromagnetic and hadronic calorimeters. 
         Multivariate analysers (and particle flow in CMS, see next item) are used to combine 
         the available detector information and improve the efficiency and purity of the selection as well as
         the energy measurement (using multivariate regression). Taus are calibrated using $Z\to \tau_h\tau_h$
         decays and $E/p$ for the hadronic tracks. 

\item {\bf Jets} are formed by clustering  particles using the infrared and 
         collinear safe\footnote{Inrared saftey requires that a jet remains unaffected when adding a particle 
           with $|p_T| \to 0$ to it. Collinear safety requires that a jet remains unaffected 
           when replacing a particle $i$ with four-momentum $p_i$ by
           two particles $j$ and $k$ with four-momenta $p_j +p_k = p_i$ 
           such that $|\vec\rho_i - \vec\rho_j| = 0$, where $\vec\rho=(y,\phi)$.} 
         anti-$k_t$ algorithm~\cite{anti-ktJets} (for which the distance between clustered 
         particles is defined using negative $p_T$ power) via a pairwise successive 
         aggregation of proto-jets. Jet particles
         are reconstructed in the electromagnetic and hadronic calorimeters in ATLAS, and 
         with the use of a particle flow algorithm in CMS.  The particle flow algorithm aims at 
         identifying and reconstructing all the particles from the collision by combining the 
         information from the tracking and calorimeter devices. The algorithm results in a list of particles, 
         namely charged hadrons, neutral hadrons, electrons, photons and muons, which are
         used to reconstruct jets and missing transverse momentum (see next item), and to 
         reconstruct and identify hadronic $\tau$ decays. In ATLAS, tracks are 
         used via a multivariate algorithm to identify low transverse momentum jets  
         from pileup interactions. Neutral energy contributions from pileup are corrected 
         by subtracting from the calorimeter jet energy a contribution equal to the product 
         of the jet area and the median energy density of the event. 
         The jet energy scale and resolution are calibrated using the constraint
         from transversely balanced dijet and multijet events, photon plus jet and $Z$ plus jet events,
         and $E/p$ together with test beam results to extrapolate the absolute calibration 
         to  large transverse momenta with insufficient data coverage.

\item {\bf Missing transverse momentum} is computed as the negative vector sum of 
         the transverse momenta of all identified objects (leptons, photons, jets, \dots), and a
         contribution denoted soft term from objects originating from the primary 
         event vertex that are not associated to any identified object. ATLAS uses a track-based 
         soft term and CMS uses the particle flow algorithm. The missing transverse momentum 
         magnetitude is denoted $\MET$.

\item {\bf Flavoured jets} containing a $b$ or $c$ hadron are identified in the inner tracking 
         detector as a property of a reconstructed jet. The characteristics of (long-lived) weakly decaying 
         heavy flavour hadrons include a displaced secondary vertex, large impact parameter, 
         a large hadron mass, and semi-leptonic decays in 30--40\% of the cases. A multivariate 
         algorithm combines the available information to tag jets containing a heavy-flavour 
         hadron (and hence originating from a heavy-flavour quark). The efficiency 
         of $b$-tagging is calibrated from data using $\ttbar$ events, muons from heavy 
         flavour decays in dijet events, and using MC simulation to extrapolate the calibration 
         to high transverse momenta. Charm tagging is calibrated using $W+c$ or 
         $D^*\to D^0(\to K\pi)\pi$ events. Mistag rates are obtained from tracks with negative 
         impact parameters or secondary vertices with negative decay lengths.

\end{itemize}

\subsection{Boosted objects}

The high centre-of-mass energy of the LHC can produce highly boosted $W$, $Z$, $H$ bosons 
or top quarks so that their hadronic decays are merged into a single jet. This would occur in 
particular in presence of hypothetical heavy states that decay to massive bosons or top quarks. 
The identification and reconstruction of such merged objects requires a jet substructure analysis. Boosted 
signatures can also be used to enhance the signal-to-background ratio in some analyses such as 
 $H\to\tau\tau$ and $H\to bb$. Boosted signatures originating from 
a very hard ISR jet can be used to render visible to the trigger and data analysis collisions 
with soft final state activity (eg., WIMPs, compressed spectra in supersymmetry or other 
new physics models).

The averge transverse distance between two bodies originating from the decay of a resonance 
with mass $m$ and transverse momentum $p_T$ can be approximated
by $\Delta R=\sqrt{\Delta\phi^2+\Delta\eta^2}\approx2m/p_T$. A $W$ boson with 
$p_T =200$\;GeV (400\;GeV) has $\Delta R=0.8$ (0.4).
To ensure that all final state objects are fully contained the experiments usually employ 
so-called ``fat jets'', which are jets with radius parameter $R=1$ or 1.2, compared to 
standard anti-$k_t$ jets~\cite{anti-ktJets} of $R=0.4$. 
There exist many strategies to reconstruct the substructure in a fat jet (eg., jet mass), and 
to correct for pileup effects (jet grooming), see, eg. \cite{ATLAS-jetsubstr,CMS-jetsubstr}
and references therein.

\subsection{Systematic uncertainties}

\begin{wrapfigure}{R}{0.25\textwidth}
\centering
\vspace{-0.41cm}
\includegraphics[width=0.25\textwidth]{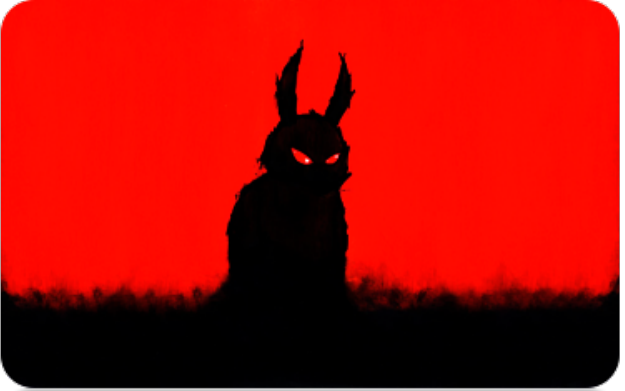}
\vspace{-0.6cm}
\end{wrapfigure}
Systematic uncertainties are the evil (see figure on the right) 
in every measurement. Well designed experiments minimise 
systematic uncertainties by achieving maximum phase space coverage, high measurement precision, 
response homogeneity and linearity, high calorimeter depth, sufficient longevity of the detector components 
including resistance against irradiation, etc. The understanding, evaluation and reduction of systematic 
uncertainties is often the main analysis challenge. A high quality analysis stands out by its 
thoroughness on all relevant sources of systematic uncertainty. It is thereby important to distinguish 
relevant from irrelevant sources, where in doubt a source should be considered relevant. For many 
uncertainty sources, in particular theoretical ones, estimating a ``one-sigma'' error is very difficult
or simply impossible. In such cases conservative uncertainties should be chosen where possible.

(Reasonably) conservative uncertainty estimates are a must! It is of no use to the scientific 
endeavour  to make over-aggressive statements that one cannot fully trust.

\section{Physics highlights from the LHC Run-1}

The LHC Run-1 featured proton--proton collisions at 7 and 8\;TeV with datasets corresponding
to approximately 5 and 20\ifb integrated luminosity for ATLAS and CMS, and 
a total of 3\ifb for LHCb. There are numerous physics highlights published in altogether 
more than a thousand physics papers. Only a small subset of these are recollected here. 

\subsection{Standard Model and top-quark physics}

\begin{figure}[p]
\centerline{\includegraphics[width=0.65\linewidth]{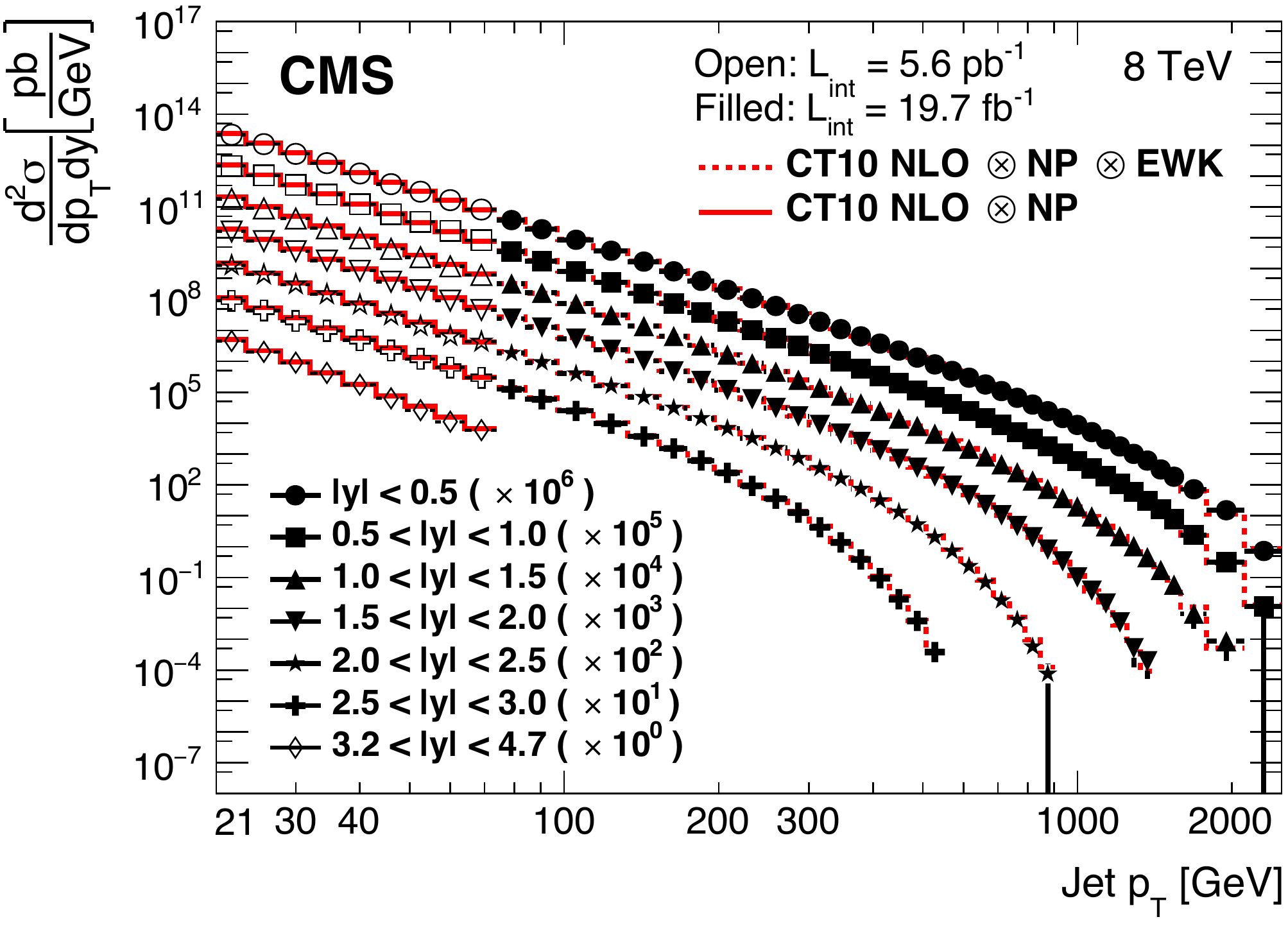}}
\vspace{-0.0cm}
\caption[]{Double-differential inclusive jet cross sections as function of jet transverse momentum 
                measured by CMS (dots with error bars) in 8\;TeV data~\cite{CMS-jetXS-Run1}. The 
                red lines indicate the SM predictions using NLO perturbative QCD and applying 
                nonperturbative (low-$p_T$) and electroweak corrections.                 
  \label{fig:CMS-jetXS-Run1}}
\vspace{1cm}
\centerline{\includegraphics[width=1\linewidth]{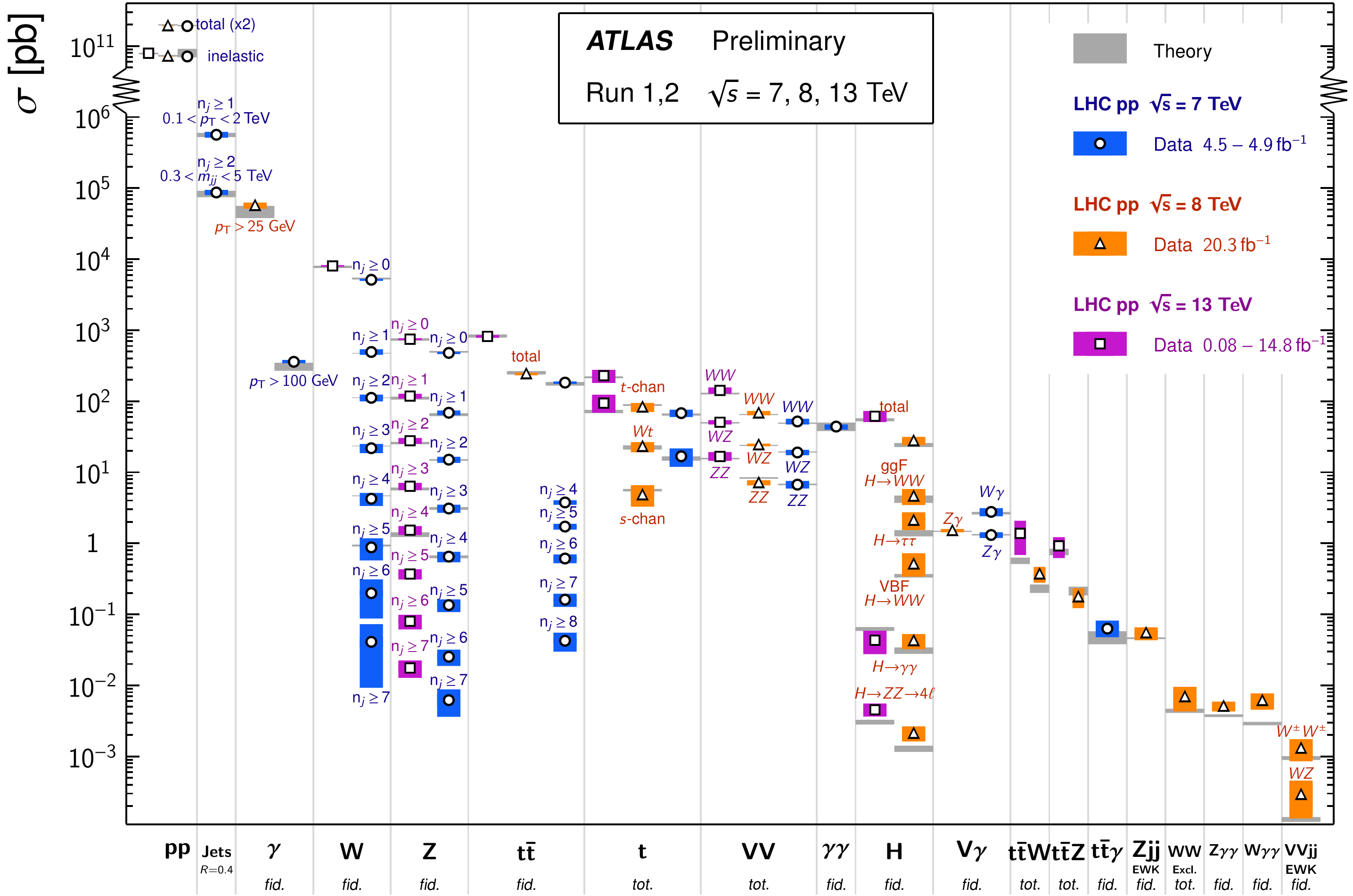}}
\vspace{0.1cm}
\caption[]{Summary of  ATLAS Run-1 and Run-2 cross-section measurements 
               (status August 2016).                  
  \label{fig:ATLAS-XSs-Run1}}
\end{figure}
We should praise the extraordinary match between a plethora of total, fiducial and differential
cross-section measurements of all known proton--proton scattering  processes and their 
theoretical predictions, confirming the predictive power of the SM. An example for the
measurement of double-differential jet cross-sections by CMS compared to theory prediction
is shown in Fig.~\ref{fig:CMS-jetXS-Run1}~\cite{CMS-jetXS-Run1}.
Figure~\ref{fig:ATLAS-XSs-Run1} gives a summary of ATLAS Run-1 and Run-2 cross-section 
measurements witnessing the large variety of channels and cross section magnitude, as well as 
the agreement with the SM predictions. There are many subtleties in this comparison that are 
not represented in such a summary plot. For example, diboson cross sections exhibit some 
discrepancy with the NLO SM predictions, which are resolved by moving to NNLO and by 
taking into account soft-gluon resummation corrections that are needed in case of phase space 
cuts sensitive to such effects (as, eg., a low-$p_T$ jet veto). A compilation like Fig.~\ref{fig:ATLAS-XSs-Run1} 
delivers a strong statement about the depth of the understanding of hadron collider physics at the 
highest centre-of-mass energies. It gives confidence that new physics searches, which depend
on a good understanding of SM processes, can be reliably performed. We should stress that 
Run-1 analysis is not over yet: it  represents a high-quality, extremely well understood data 
sample for precision measurements.

\begin{figure}[t]
\centerline{\includegraphics[width=0.9\linewidth]{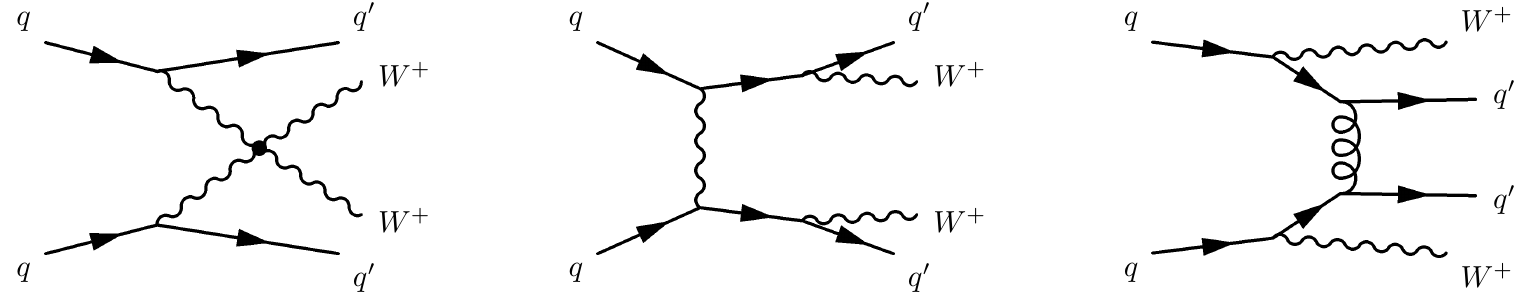}}
\vspace{0.0cm}
\caption[]{Feynman graphs for same-charge $W^\pm W^\pm+2\,{\rm jets}$ production. 
                Top: electroweak VBS; middle: electroweak non-VBS; bottom: gluon exchange. 
  \label{fig:VBS-Feyn}}
\end{figure}
The analysis of Run-1 data 
allowed first critical electroweak studies of vector-boson scattering (VBS). In electroweak theory 
the Higgs boson acts as ``moderator'' to unitarise high-energy longitudinal vector boson 
scattering. Indeed, if only $Z$ and $W$ bosons are exchanged, the amplitude of (longitudinal) 
$W_L W_L$ scattering, $A_{Z,\gamma}\sim v^{-2}\cdot(s+t)$, rises with the centre-of-mass energy
and violates unitarity. Higgs-boson exchange regularises this amplitude via the negative term
$A_{H}\sim -(m_H^2/v^2)\cdot(s/(s-m_H^2)+t/(t-m_H^2))$, if $m_H ^2/v^2$ is not too large,
which is the case for $m_H=125$\;GeV. That mechanism can be tested by, eg., measuring 
same-charge $W^\pm W^\pm+2{\rm jets}$ production at the LHC (see graphs in Fig.~\ref{fig:VBS-Feyn}).
Requiring same-sign 
$W^\pm W^\pm$ 
production greatly reduces strong production (see right-hand graph in Fig.~\ref{fig:VBS-Feyn})
due to the lack of contributions from two initial gluons or one quark and one gluon.
It also suppresses the s-channel Higgs amplitude, but moderation through t-channel Higgs exchange 
remains. The two electroweak processes in Fig.~\ref{fig:VBS-Feyn} cannot be separated in 
a gauge-invariant way.
Contributions from electroweak VBS to this process can be separated from non-VBS electroweak 
and strong processes by requiring a large dijet invariant mass and a rapidity gap for hadronic 
activity. Evidence for electroweak production at the 3.6$\sigma$ (2.0$\sigma$) level was found
by ATLAS (CMS)~\cite{ATLAS-VBS-Run1,CMS-VBS-Run1}.

\begin{wrapfigure}{R}{0.50\textwidth}
\centering
\vspace{-0.42cm}
\includegraphics[width=0.50\textwidth]{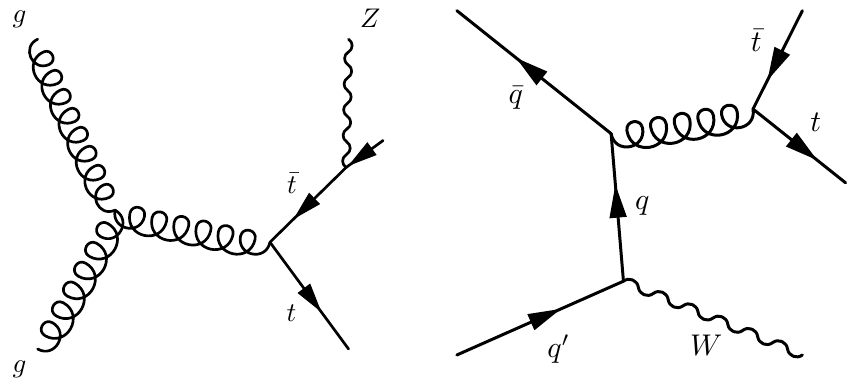}
\caption[.]{Feynman graphs for leading order $ttZ$ (left) and $ttW$ production. 
  \label{fig:ttV-Feyn}}
\vspace{-0.4cm}
\end{wrapfigure}
Strong top-quark pair production has been studies with unprecedented experimental precision
at the LHC. Inclusive cross sections are best measured in the dilepton $e\mu$ final state
that is very pure and can be isolated with a minimal set of selection requirements. The measurements
provide precise tests of NNLO QCD including leading-logarithmic resummation.\footnote{The cross
section of soft gluon emission is infrared divergent (eikonal factor). The divergence is 
cancelled by virtual corrections up to logarithmic leftover terms 
$\sigma(a\to b) \to \sigma\ln^2(1-m_b^2/s)$, which need to be resummed. Several resummation
strategies denoted ``threshold resummation'', ``transverse momentum resummation'', or 
``high-energy resummation'' exist in the literature.} In addition, many top properties 
(mass, charge, charge asymmetry, polarisation, spin correlations, suppressed flavour-changing 
neutral currents (FCNC), etc.) were measured or probed. 

The large luminosity and high centre-of-mass energy also allowed to observe the rare $tt+W$ and 
$tt+Z$ production (see Feynman graphs in Fig.~\ref{fig:ttV-Feyn}) with more than 7.1$\sigma$ 
combined significance for both modes~\cite{ATLAS-ttV-Run1,CMS-ttV-Run1}. 
The neutral-current $tZ$ coupling is directly probed in $tt+Z$. 

Run-1 also allowed detailed studies of electroweak single top production and property 
measurements (see Fig.~\ref{fig:single-top} for representative leading-order diagrams).
Single top cross sections are  enhanced at the LHC compared to the Tevatron: at 8\;TeV
LHC centre-of-mass energy, factors of 42 (t-channel), 31 ($Wt$), but only 5 for s-channel production
so that the signal to background ratio is worse for the latter channel at the LHC. Production
of t-channel single top has been studied in great differential detail 
already~\cite{ATLAS-tchan-Run1,CMS-tchan-Run1}. 
The separate measurement of $tq$ and $\overline t q$ production provides sensitivity to 
$u$ and $d$ quark  PDFs. Inclusive 
$Wt$ channel production was clearly observed by both ATLAS and 
CMS~\cite{ATLAS-Wt-Run1,CMS-Wt-Run1}.
Production via an s-channel process (see bottom diagram 
in Fig.~\ref{fig:single-top} was recently observed by the Tevatron experiments with 
6.3$\sigma$ combined significance in agreement with the SM 
prediction~\cite{Tevatron-s-channel}. ATLAS reported 
an observed evidence of 3.2$\sigma$ (for 3.9$\sigma$ expected significance), also 
in agreement with the SM prediction~\cite{ATLAS-s-chan-Run1}. 
\begin{figure}[t]
\centerline{\includegraphics[width=1\linewidth]{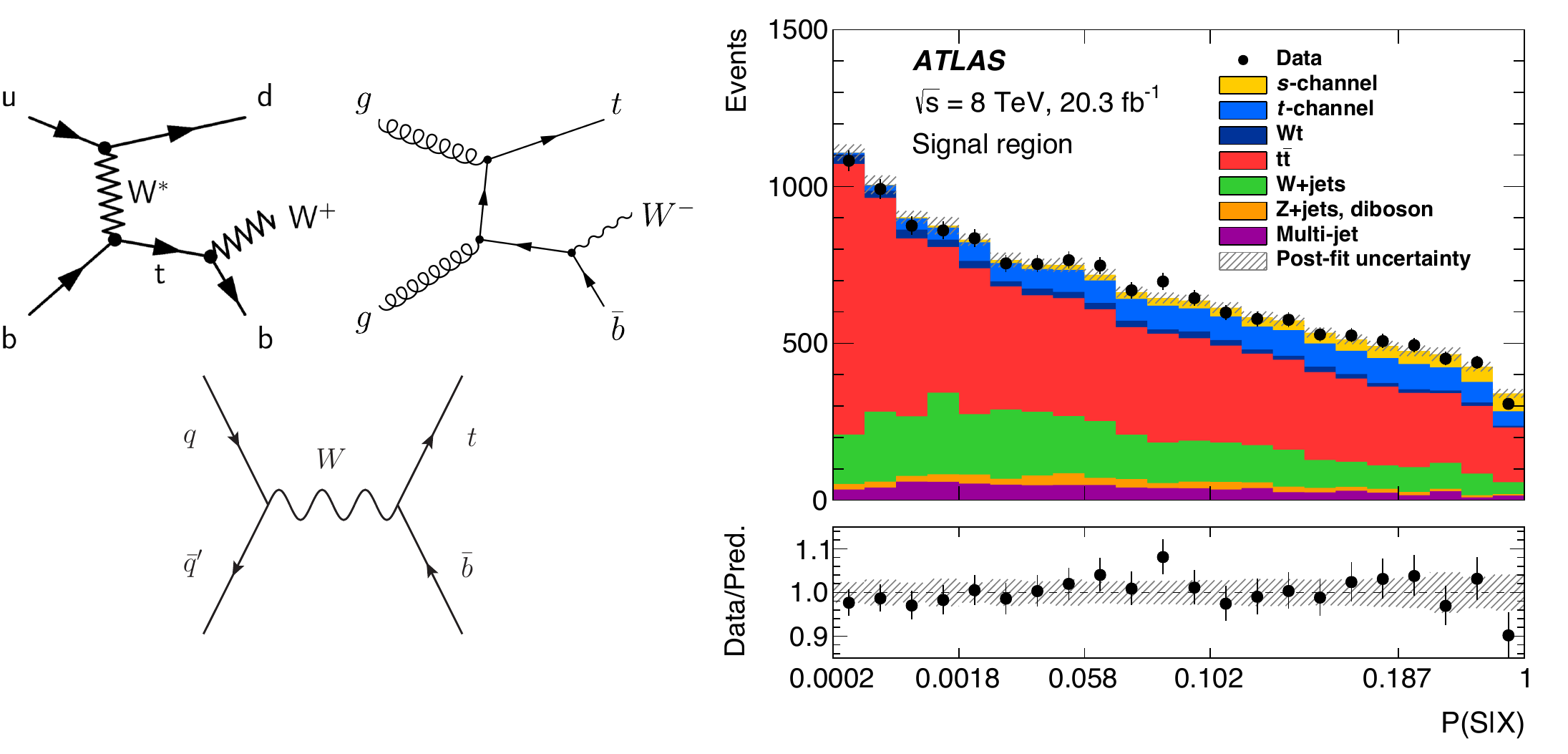}}
\vspace{-0.0cm}
\caption[]{Left: leading-order Feynman graphs for t-channel (top left), $Wt$-channel (top right),
               and s-channel (bottom) single top production at the LHC. Right: 
               distribution of a discriminant variable in an ATLAS search for s-channel single top 
               production~\cite{ATLAS-s-chan-Run1}. 
  \label{fig:single-top}}
\end{figure}

\vfill\pagebreak
\subsection{Higgs boson physics}

Among all the Run-1 physics results, the discovery of the Higgs boson is  (so far) the 
magnum opus~\cite{ATLAS-HiggsDiscovery,CMS-HiggsDiscovery} (articles that each have 
collected about 6\,700 citations to date). 
The Higgs boson had been vainly searched for at many accelerators. The 
most stringent non-LHC  limits came from the Large Electron--Positron Collider (LEP at
CERN, 1989--2000) and the proton--antiproton collider Tevatron (Fermilab, 1990--2011)
excluding at 95\% confidence level $m_H < 114$\;GeV~\cite{LEP-Higgs}
and $149<m_H<182\;$GeV~\cite{Tevatron-Higgs}, 
respectively. Global fits to electroweak precision data (see Section~\ref{sec:EWfit}) 
constrained the Higgs boson mass via logarithmic corrections and excluded about $m_H > 160$\;GeV
at 95\% confidence level~\cite{EW-globalfits}. 

At the LHC the Higgs boson is dominantly produced via gluon fusion with a cross
section of 19.3\;pb at 8\;TeV for $m_H = 125$\;GeV~\cite{YR4} (see also~\cite{Higgs-PDG} 
for a recent review on Higgs boson physics). 
The cross section steeply falls with the Higgs boson  mass. Additional production modes
are weak boson fusion (VBF) with 1.6\;pb, associated production with a weak
boson (also denoted Higgs-strahlung) with 0.70\;pb (0.42\;pb) for $WH$ ($ZH$), and 
associated production with a $\ttbar$ or $b\overline b$ pair ($ttH$, $bbH$) 
with 0.13\;pb and 0.20\;pb, respectively (cf. 
Fig.~\ref{fig:HiggsProdAndBRs} for the corresponding Feynman diagrams). The 
uncertainties in the predictions are larger (7$\sim$14\%) for the gluon initiated 
processes than for the quark initiated ones ($\sim$3\%, dominated by PDF uncertainties).
The inclusive 8\;TeV Higgs  cross section amounts to 22\;pb. In total, 
about 470 thousand SM Higgs bosons of 125 GeV were produced in 2012 at 
8\;TeV in each ATLAS and CMS. 

Because of the coupling to the mass of the decay particles ($\propto m_V^2,m_f$)
the Higgs boson decays with preference to the heaviest 
particles allowed. It does not couple directly to photons and gluons
but proceeds via loops involving preferentially heavy particles (eg., top, $W$ boson).
The branching fractions predicted for an SM Higgs boson of mass 125\;GeV
are shown on the right panel of Fig.~\ref{fig:HiggsProdAndBRs}. The theoretical 
uncertainty in these predictions ranges from 3\% to about 12\%. The leptonic 
($\ell=e,\mu$) and photon final states provide the best discovery sensitivity.
The decays $H\to\gamma\gamma$ and $H\to ZZ^{(*)}\to 4\ell$ provide the 
best mass resolution (1--2\% for $m_H=125$\;GeV). The decay 
$H\to WW^{(*)}\to2\ell2\nu$ ($\sim$20\% mass resolution due to the neutrinos
in the final state)  has a good trigger, a sustainable background 
level, and large branching fraction. The fermionic modes $H\to\tau\tau$ and 
$H\to bb$ have mass resolutions of about 10\% and 15\%, respectively, and 
are more challenging to detect due to large backgrounds. The decays $H\to \mu\mu$
and $H\to Z(\to\ell\ell)\gamma$ have excellent mass resolution but  too low branching fractions
to be in reach with the current datasets.

\begin{figure}[t]
\centerline{\includegraphics[width=0.99\linewidth]{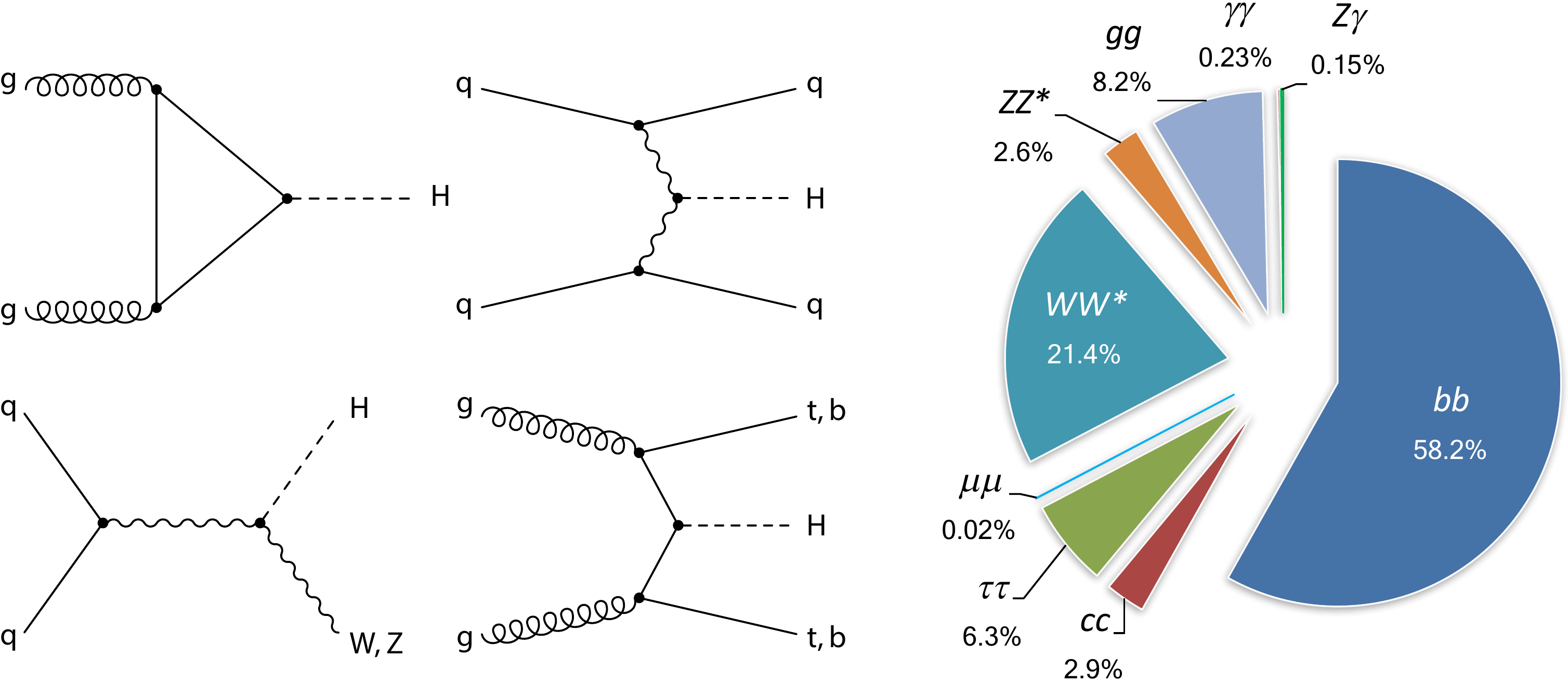}}
\vspace{0.2cm}
\caption[]{Left: Feynman graphs for the dominant Higgs  production channels:
               gluon fusion via (dominantly) a top quark triangle (top left), 
               weak boson fusion (top right), associated production with a weak boson (bottom left)
               and with a heavy quark pair (bottom right).
               Right: branching fractions predicted for an SM Higgs boson of mass 125\;GeV~\cite{YR4}.
               Considering only leptonic decays to $e,\mu$, the exploitable branching fractions
               to $WW^*$ and $ZZ^*$ are 1.1\% and 0.012\%, respectively. 
  \label{fig:HiggsProdAndBRs}}
\end{figure}
It is fortunate  that at $m_H = 125$\;GeV many decays of the Higgs boson
are experimentally accessible. The phenomenological aspects of that mass might 
appear less appealing as we will see later. The dominant $H\to bb$ mode is only 
exploitable in association with $W/Z$ or $tt$. 
Their  leptonic decays provide a trigger signal and help to reduce the overwhelming
background from strong interaction $bb$ continuum production, 
$\sigma(bb)\sim {\cal O}(100\;\mu$b). A boost of the Higgs boson helps to improve 
the signal purity at the expense of reduced efficiency. 

\begin{figure}[p]
\centerline{\includegraphics[width=1\linewidth]{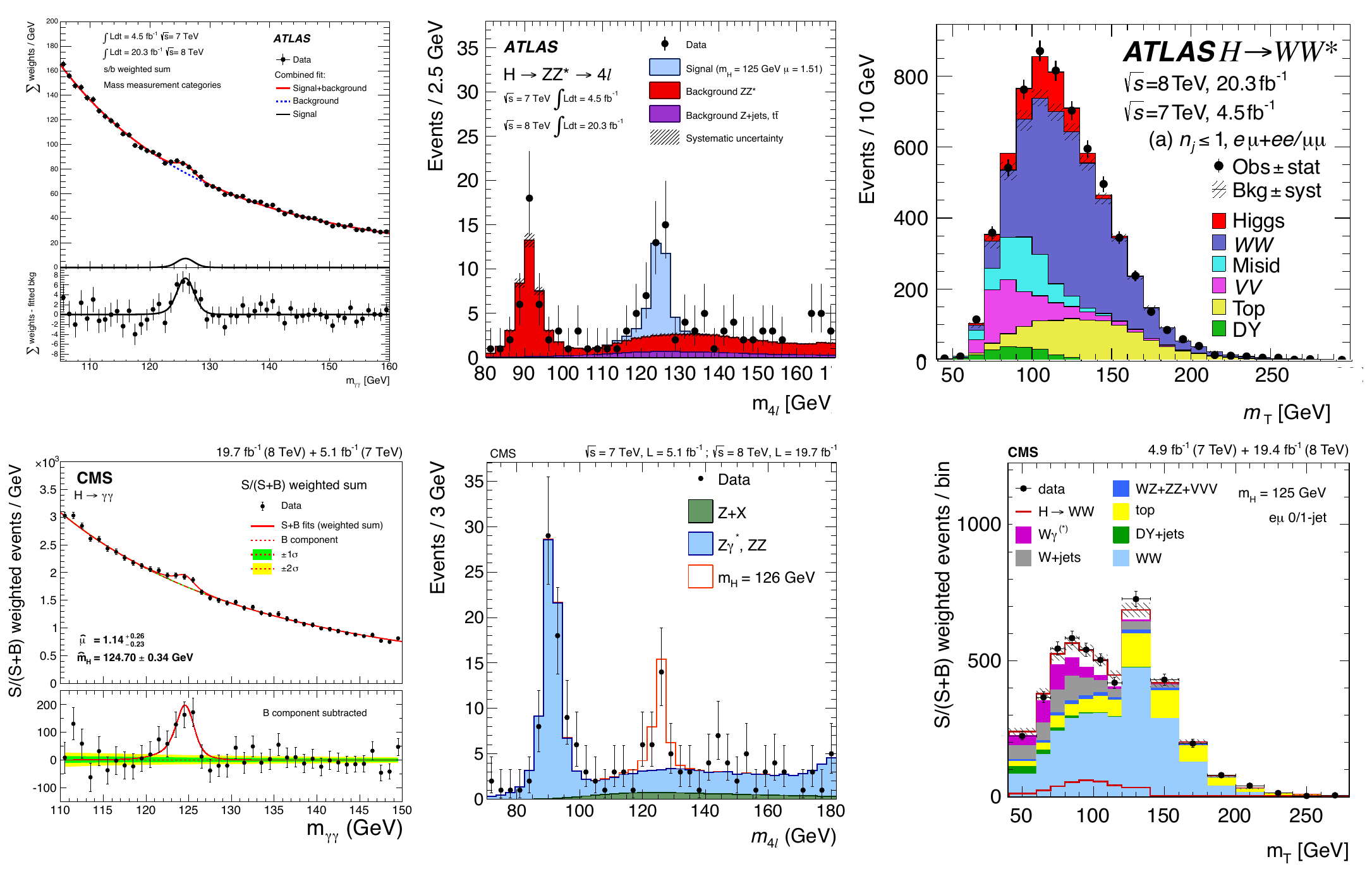}}
\vspace{0.0cm}
\caption[]{Reconstructed inclusive Higgs candidate masses in the bosonic decay channels 
              $H\to\gamma\gamma$ (left), $H\to 4\ell$ (middle), and
              $H\to 2\ell 2\nu$ (right, shown is the transverse mass)
              for the ATLAS (top row) and CMS (bottom row) Run-1 
              analyses~\cite{ATLAS-Hyy-Run1,ATLAS-H4l-Run1,ATLAS-HWW-Run1,CMS-Hyy-Run1,CMS-H4l-Run1,CMS-HWW-Run1}.
  \label{fig:ATLASCMSHiggs-Run1}}
\vspace{1cm}
\centerline{\includegraphics[width=1\linewidth]{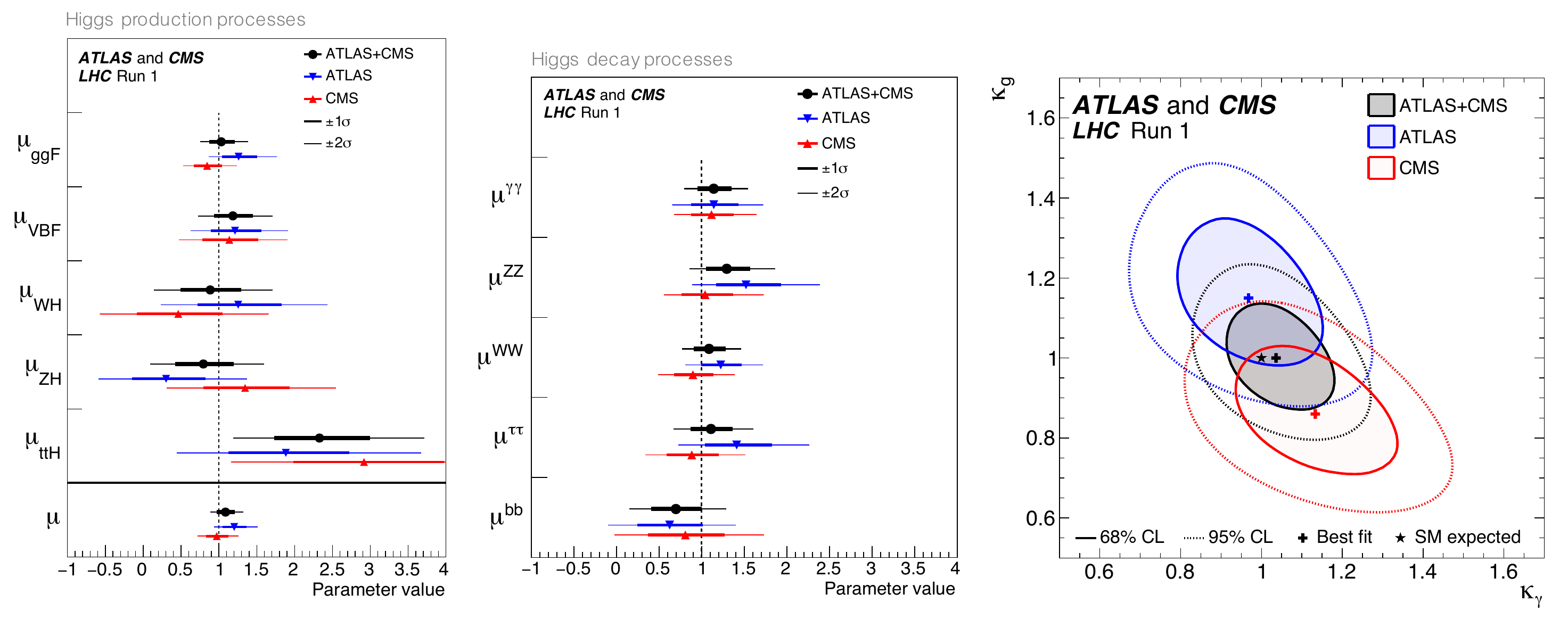}}
\vspace{0.0cm}
\caption[]{Left: Higgs production signal strengths for  ATLAS and CMS 
                and their combination. Also shown is the measurement of the global signal 
                strength. Middle: Higgs decay signal strengths for ATLAS, CMS 
                and their combination. Right: 68\% and 95\% confidence level contours 
                in the Higgs-to-gluon versus Higgs-to-photon coupling modifiers for ATLAS,
                CMS and their combination. The SM prediction is $\kappa_g=\kappa_\gamma=1$.
                See text for the assumptions underlying these plots. The figures are taken 
                from~\cite{ATLASCMSHcouplings-Run1}.
  \label{fig:ATLASCMSHiggsCouplings-Run1}}
\end{figure}
There is no doubt about the discovery of the Higgs boson. 
Each of the most sensitive bosonic channels  $H\to\gamma\gamma$, $H\to 4\ell$ 
and $H\to 2\ell 2\nu$ from ATLAS and CMS have achieved an independent 
observation 
(cf. Fig.~\ref{fig:ATLASCMSHiggs-Run1})~\cite{ATLAS-Hyy-Run1,ATLAS-H4l-Run1,ATLAS-HWW-Run1,CMS-Hyy-Run1,CMS-H4l-Run1,CMS-HWW-Run1}.
The combination of ATLAS and CMS mass measurements gives
$m_H=125.09 \pm 0.21_{\rm stat} \pm 0.11_{\rm syst}$\;GeV~\cite{ATLASCMSHmass-Run1}.
There are very different experimental challenges in each Higgs channel.  
All analyses have constantly increased their sensitivity during Run-1 owing to improved 
understanding of lepton reconstruction and calibration, as well as improved background modelling 
and signal against background discrimination. 

In addition to sophisticated individual analyses, ATLAS and CMS have joined forces and 
combined their Higgs  mass and coupling 
measurements~\cite{ATLASCMSHmass-Run1,ATLASCMSHcouplings-Run1}. 
These combinations represent the full picture of what the experiments have learned in 
a framework that consistently treats all processes in terms of production 
mechanism and decay. Figure~\ref{fig:ATLASCMSHiggsCouplings-Run1} shows 
as an example the ratios of measured to predicted signal strengths per production 
process (left panel, assuming the Higgs  decays to proceed according to the SM), 
and vice versa per decay channel (middle panel,  assuming SM Higgs  
production)~\cite{ATLASCMSHcouplings-Run1}. The overall signal strength, assuming 
an overall scale for all individual signal strengths, is measured to be $\mu=1.09 \pm 0.11$.
The right hand panel of Fig.~\ref{fig:ATLASCMSHiggsCouplings-Run1} shows the 
results of a fit of leading order coupling modifiers~\cite{HiggsCouplingModifiers} to
the combined Higgs boson data, where for a given production process or decay mode, 
denoted $j$, the coupling modifier $\kappa_j$ is defined such that 
$\kappa_j^2=\sigma_j/\sigma_j^{\rm SM}$. Shown in the figure are the coupling 
modifiers $\kappa_g$ versus $\kappa_\gamma$ of the Higgs-to-gluon and 
Higgs-to-photon couplings, respectively. The fit was performed by constraining all the other 
coupling modifiers to their SM values and assuming no non-SM decays of the Higgs boson. 
The resulting agreement with the SM is remarkable as these couplings proceed through 
loops involving heavy fermions and also, in the photon case, bosons ($W$). 
It is  a powerful probe for new heavy degrees of freedom. For example, the result
allows to reject a theory with heavy fermions with SM-like Yukawa couplings.\footnote{Naively
an additional heavy fermion generation would increase the gluon fusion Higgs cross section 
by a factor of nine with respect to the SM prediction due to the quadratic fermion form-factor 
dependence of the cross section.} The ATLAS
and CMS Higgs  coupling combination exhibits agreement among the two experiments. 
It yields sufficient significance for the observation of the Higgs decay to fermions, 
$H\to\tau\tau$~\cite{ATLAS-Htautau-Run1,CMS-Htautau-Run1}, 
and of VBF production. The $ttH$ process~\cite{ATLAS-ttH-Run1,CMS-ttH-Run1}
 comes out a bit large with a
relative signal strength of $\mu=2.3$ and a combined observed significance of 
4.4$\sigma$ (for 2.0$\sigma$ expected). With respect to the signal strengths shown 
in Fig.~\ref{fig:ATLASCMSHiggsCouplings-Run1} we note that the least 
model-dependent observables at the LHC are coupling ratios rather than absolute 
coupling measurements~\cite{ATLASCMSHcouplings-Run1}.

The Higgs boson has been suggested to possibly act as a ``portal'' to new physics 
responsible for dark matter. In such models, a massive dark matter particle
couples only weakly (or not at all) with the SM particles, except for the Higgs boson.\footnote{For 
example in a Dirac neutrino case the massive right-handed neutrinos would transform
as singlets under the SM gauge interactions, but would couple to the Higgs boson.} 
If the dark matter particle is not too heavy, the Higgs decays 
invisibly to it and is searched for via, eg., a VBF topology where the forward jets are used 
to trigger and select the events~\cite{ATLAS-VBFHinv-Run1,CMS-VBFHinv-Run1}. 
Limits of about 25\% are currently set for an invisible Higgs boson decay. 
In general, owing to 
its low mass and consequently narrow width of 4.1\;MeV compared to the widths 
of the $W$, $Z$ or top quark of 2.1\;GeV, 2.5\;GeV and 1.3\;GeV, respectively, 
the Higgs boson has good sensitivity to new physics as even small couplings
to new  states (if light enough) can measurably impact its branching 
fractions (see~\cite{ATLAS-HBSM-Run1} for an analysis of constraints on new physics 
from the measurements of the Higgs couplings and invisible decays). It is therefore 
important to continue to measure the Higgs couplings, including the invisible one,
with highest possible precision. 

\subsection{Heavy flavour physics}

There have been beautiful flavour and low-$p_T$ physics measurements at the LHC. 
The LHCb experiment has produced a flurry of important results among which the 
observation, together with CMS~\cite{CMSLHCbBsmumu-Run1}, of the very rare 
decay $B_s\to\mu\mu$ at a branching fraction of 
$2.8^{\,+0.7}_{\,-0.6}\cdot10^{-9}$ in agreement with the SM prediction of 
$3.7 \pm 0.2\cdot 10^{-9}$~\cite{SMBsmumu}.
The left panel of Fig.~\ref{fig:Bsmumu-Run1} shows representative SM and BSM 
Feynman graphs, and the right panel shows the combined CMS and LHCb data
and the result of a simultaneous signal and background fit. The $B_s\to\mu\mu$  
decay proceeds through a loop as there is no tree-level FCNC in the SM. It is in 
addition CKM and helicity suppressed, thus the low branching fraction. The decay is sensitive 
to additional scalar bosons as, eg., predicted in supersymmetry. ATLAS recently published 
the Run-1 result giving a branching fraction value in agreement with CMS and LHCb
and approximately 2$\sigma$ below the SM prediction~\cite{ATLASBsmumu-Run1}.

\begin{figure}[t]
\centerline{\includegraphics[width=0.9\linewidth]{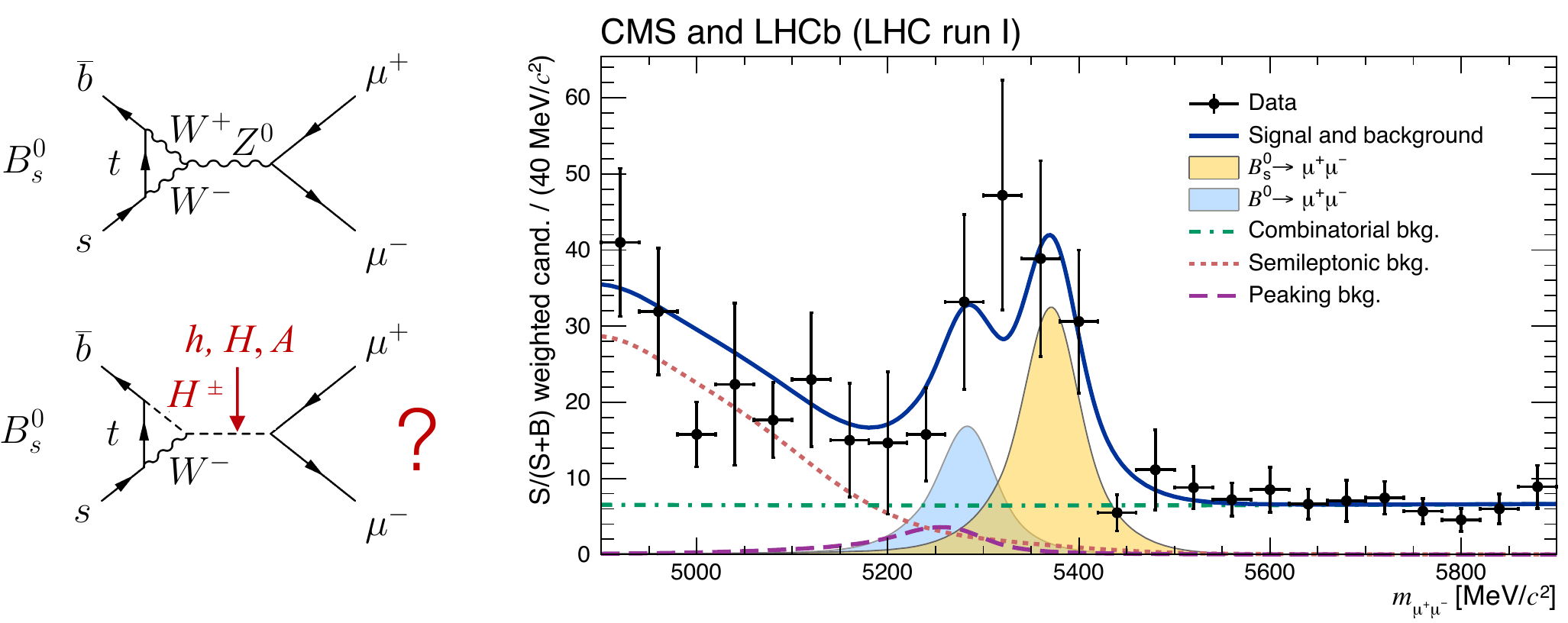}}
\vspace{0.0cm}
\caption[]{Left: Feynman graphs for the decay $B_s\to\mu\mu$ in the SM (top) and 
                beyond the SM (bottom). Right: weighted distribution of the dimuon 
                invariant mass for all CMS and LHCb measurement 
                categories. The yellow region indicates
                the fit result for $B_s\to\mu\mu$, while the blue region shows that for
                the doubly CKM suppressed decay $B_d\to\mu\mu$. The dashed and 
                dotted lines indicate the various background contributions as obtained 
                from the fit (the abundance of the peaking background is predicted) 
                and the solid line their sum~\cite{CMSLHCbBsmumu-Run1}. 
                \label{fig:Bsmumu-Run1}}
\end{figure}
Another high-priority flavour result from the LHC is the measurement of the 
mixing-induced CP violation parameter $\phi_s$ in a flavour-tagged, time-dependent 
$B_s\to J/\psi\, \phi$ analysis. That measurement represents one of the most sensitive 
CP-violation tests of the SM as $\phi_s$ is small and predicted with negligible theoretical 
uncertainty within the CKM paradigm. ATLAS~\cite{ATLASPhis-Run1}, CMS~\cite{CMSPhis-Run1} 
and LHCb~\cite{LHCbPhis-Run1} have measured simultaneously
$\phi_s$ and $\Delta\Gamma_s$, the width difference of the two $B_s$ mass eigenstates,
with LHCb exhibiting the best precision. It found the combined result 
$\phi_s=-0.010\pm0.039$\;rad in agreement with the SM. 
Figure~\ref{fig:HFAGPhis} shows the various measurements as well as their 
combination as 68\% confidence level contours in the $\Delta\Gamma_s$ versus $\phi_s$ 
plane. The SM prediction is indicated by the black vertical bar.

\begin{wrapfigure}{R}{0.57\textwidth}
\centering
\vspace{-0.39cm}
\includegraphics[width=0.57\textwidth]{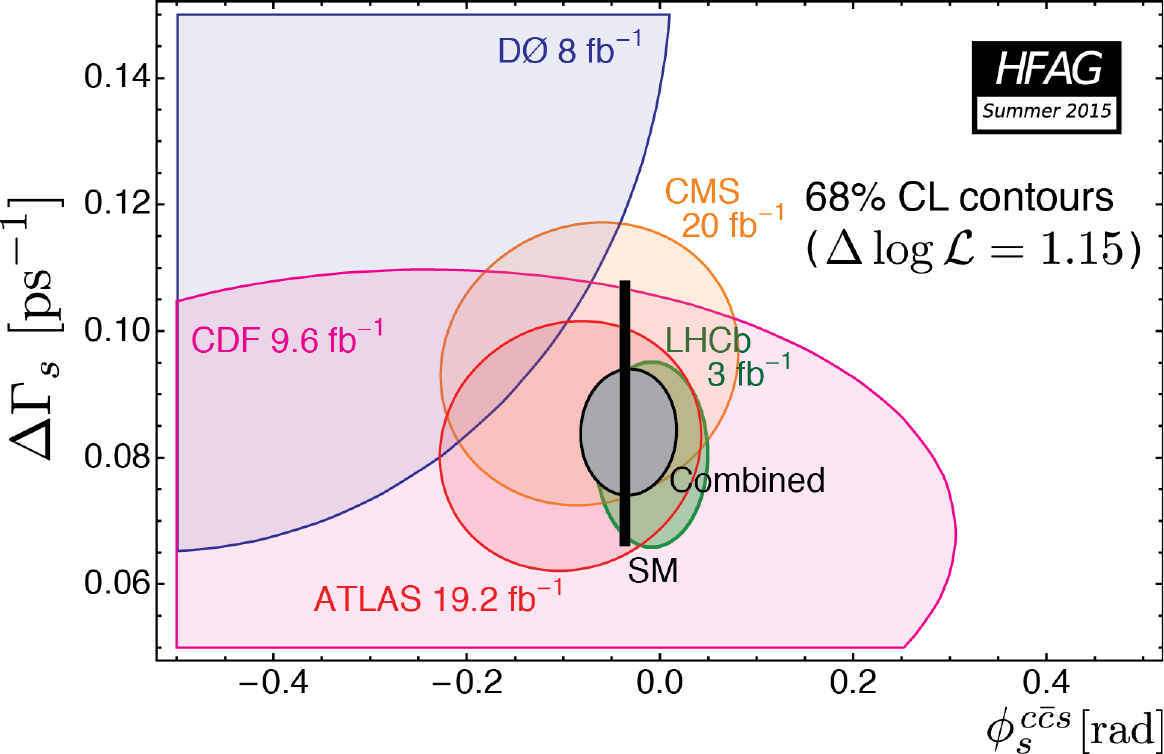}
\vspace{-0.5cm}
\caption[.]{Contours of 68\% confidence level for $\Delta\Gamma_s$ versus 
                the mixing induced CP-violation parameter 
                $\phi_s$~\cite{ATLASPhis-Run1,CMSPhis-Run1,LHCbPhis-Run1}. The SM 
                prediction is indicated by the black vertical bar.
  \label{fig:HFAGPhis}}
\vspace{-0.4cm}
\end{wrapfigure}
LHCb also contributed significantly to the long-term effort to overconstrain the CKM 
matrix in what is known as the unitarity triangle, a triangle given in the 
$\overline\rho$--$\overline\eta$ CKM parameter plane, where $\overline\eta\ne0$ 
stands for CP violation in the SM. LHCb has engaged in a vigorous programme
to determine the unitarity triangle angle $\gamma\sim {\rm arg}(-V_{ub}^\star)$. 
It can be measured through interference of $b\to u$ with $b\to c$ tree transitions 
where  hadronic amplitude parameters are determined simultaneously with $\gamma$
from the data. A combined fit~\cite{LHCb-gammaComb}, dominated by the 
measurements from charged $B^+$ to charm decays, gives 
$\gamma=70.9^{\,+7.1}_{\,-8.5}\;$deg, which is in agreement with the prediction from 
the CKM fit (not including the direct $\gamma$ measurements) of 
$68\pm2\;$deg~\cite{CKMfitter}. LHCb also measured the ratio $|V_{ub} / V_{cb}|$ 
from $\Lambda_b  \to p\mu\nu$ (a baryon decay!) with 5\% precision~\cite{LHCb-Vub}. 
The result is closer to the exclusive $B$-factory numbers for $|V_{ub}|$, 
which exhibit a tension with the larger inclusive numbers. Furthermore LHCb obtained
the world's best single $\Delta m_d$ measurement~\cite{LHCb-dmd}
 $0.5050 \pm 0.0021\pm 0.0010\;{\rm ps}^{-1}$ 
(the $B$-factories have a combined uncertainty of $0.005\;{\rm ps}^{-1}$), a
${\rm sin}(2\beta)$ measurement~\cite{LHCb-sin2b} 
of $0.731 \pm 0.035 \pm 0.020$ that approaches 
the precision of the $B$-factories, the world's best constraints on CP violation 
in $B^0_{(s)}$ mixing ($a_{\rm sl}^s$, $a_{\rm sl}^d$) in agreement with the 
SM (D0 sees a 3.6$\sigma$ deviation), and a search for CPT
violation~\cite{LHCb-CPT} (difference in mass or width) in the $B^0_{(s)}$ systems 
together with the measurement of sidereal phase dependence of the CPT violating 
parameter. 

It is interesting to speculate about the ``relevance'' of the CKM phase. So far, 
all CP violating effects measured in particle physics can be reduced to just that phase. 
On the other hand, there seems to be consensus of opinion among theorists that the CKM 
induced CP violation in the quark sector is too small by many orders of magnitude to 
generate the observed baryon asymmetry in the universe (non-zero CKM CP violation 
requires non-zero and non-degenerate quark masses, so the baryogenesis could only be 
generated during the electroweak phase transition at critical temperature 
of $T_c\sim100$\;GeV). So is the CKM phase only an ``accident of Nature''? Because 
there are three quark generations, there is a phase in the quark mixing matrix,\footnote{
There would be zero (three) phases in a two (four) generations SM, as 
$n_{\rm phases}=(n_{\rm gen}-1)(n_{\rm gen}-2)/2$.} 
and so that phase has ``some'' value? What would happen to the universe had we 
a dial to change that value~\cite{Cahn18}?

Several measurements in the flavour sector exhibit non-significant but interesting
anomalies with respect to theory predictions. A prominent example is given by 
angular coefficients describing the transition $b\to s\mu^+\mu^-$, the prediction 
of which, however, are plagued by hadronic uncertainties. Theoretically robust are
universality tests. Such tests were performed at the per-mil level at LEP 
and other $e^+ e^-$ colliders not showing any significant discrepancy from the 
expectation of universal lepton coupling. The $B$-factory experiments and LHCb 
have measured ratios of semileptonic $B$ decays among 
which~\cite{BABAR-RD,Belle-RD,LHCb-RD,Belle-RDstar}
$R_{D^{(\star)}}={\cal B}(B^0\to D^{(\star)}\tau\nu)/{\cal B}(B^0\to D^{(\star)}\ell\nu)$
and
$R_{K}={\cal B}(B^+\to K^+\mu^+\mu^-)/{\cal B}(B^+\to K^+e^+e^-)$.
The Heavy Flavour Averaging Group (HFAG) has combined the experimental results
giving~\cite{HFAG-RD} $R_{D^\star} = 0.316 \pm 0.016 \pm 0.010$, which 
is $3.3\sigma$ away from the SM prediction $0.252 \pm 0.003$~\cite{RDstar-Theory}. 
The two-dimensional combination with $R_D$ increases the deviation to $4.0\sigma$. 
For $R_{K}$ LHCb measures at low $q^2$ (given by the invariant mass of the 
dimuon or dielectron system)  the value 
$0.745^{\,+0.090}_{\,-0.074}({\rm stat})\pm0.036({\rm syst})$, 
which differs by 2.6$\sigma$ from the expected unity~\cite{LHCb-RK}. 

\begin{wrapfigure}{R}{0.65\textwidth}
\centering
\vspace{-0.55cm}
\includegraphics[width=0.65\textwidth]{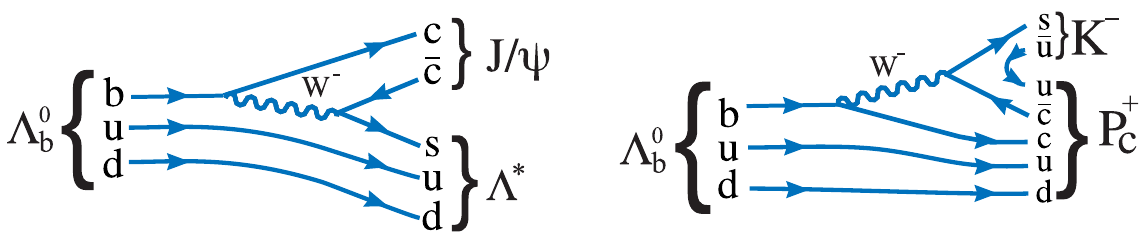}
\caption[.]{Feynman graphs for $\Lambda^0_b\to J/\psi\, \Lambda^*$ (left)
                and $\Lambda^0_b\to P_c^+ K^−$ (right).
  \label{fig:PQFeyn}}
\vspace{-0.6cm}
\end{wrapfigure}
An intriguing observation in hadron spectroscopy was announced by LHCb in summer 
2015 in a paper~\cite{LHCbPQ-Run1} that collected over 250 citations since. It is 
the observations of exotic structures in the $J/\psi\,p$ channel, consistent with 
pentaquark-charmonium states occurring in $\Lambda^0_b\to J/\psi\, K^−p$ decays
(see Fig.~\ref{fig:PQFeyn} for representative $\Lambda^0_b$ decay diagrams).
Analysing the full Run-1 data sample and performing an intricate three-body amplitude 
analysis, the observed structures could only be described by adding two resonances, 
one with mass and width of $4380\pm8\pm29$\;MeV and $205\pm18\pm86$\;MeV,
respectively, and the other (narrower) with mass and width of $ 4449.8\pm1.7\pm2.5$\;MeV and 
$39\pm5\pm19$\;MeV. LHCb dubs these two states  $P_c(4380)^+$ and $P_c(4450)^+$.
The binding mechanism for pentaquarks is not clear at present. They may consist of 
five quarks tightly bound together, but it is also possible that they are more loosely 
bound and consist of a three-quark baryon and a two-quark meson interacting relatively 
weakly in a meson-baryon molecule.

\section{Digression on electroweak precision measurements}
\label{sec:EWfit}

The global electroweak fit relating observables of the electroweak SM to each other 
by incorporating precise theoretical predictions of radiative corrections was a masterpiece 
of the LEP/SLC era. It led to the prediction of the top-quark mass prior to its discovery,
provided a strong (logarithmic) constraint on the mass of the Higgs boson, predicting
it to be light, and allowed to exclude or constrain models beyond the SM.\footnote{For 
example, it allowed to exclude the simplest technicolour 
models~\cite{TCexclusion1,TCexclusion2,TCexclusion3}. Technicolour invokes the 
existence of strong interactions at a scale of the order of a TeV and induces 
strong breaking of the electroweak symmetry. In the original form of technicolor, the 
strong interactions themselves trigger electroweak symmetry breaking without the 
need of a Higgs boson. } 
The discovery of the Higgs boson overconstrains the fit and dramatically improves its 
predictability. The fit has thus turned into a powerful test of the SM.

\begin{figure}[t]
\centerline{\includegraphics[width=1\linewidth]{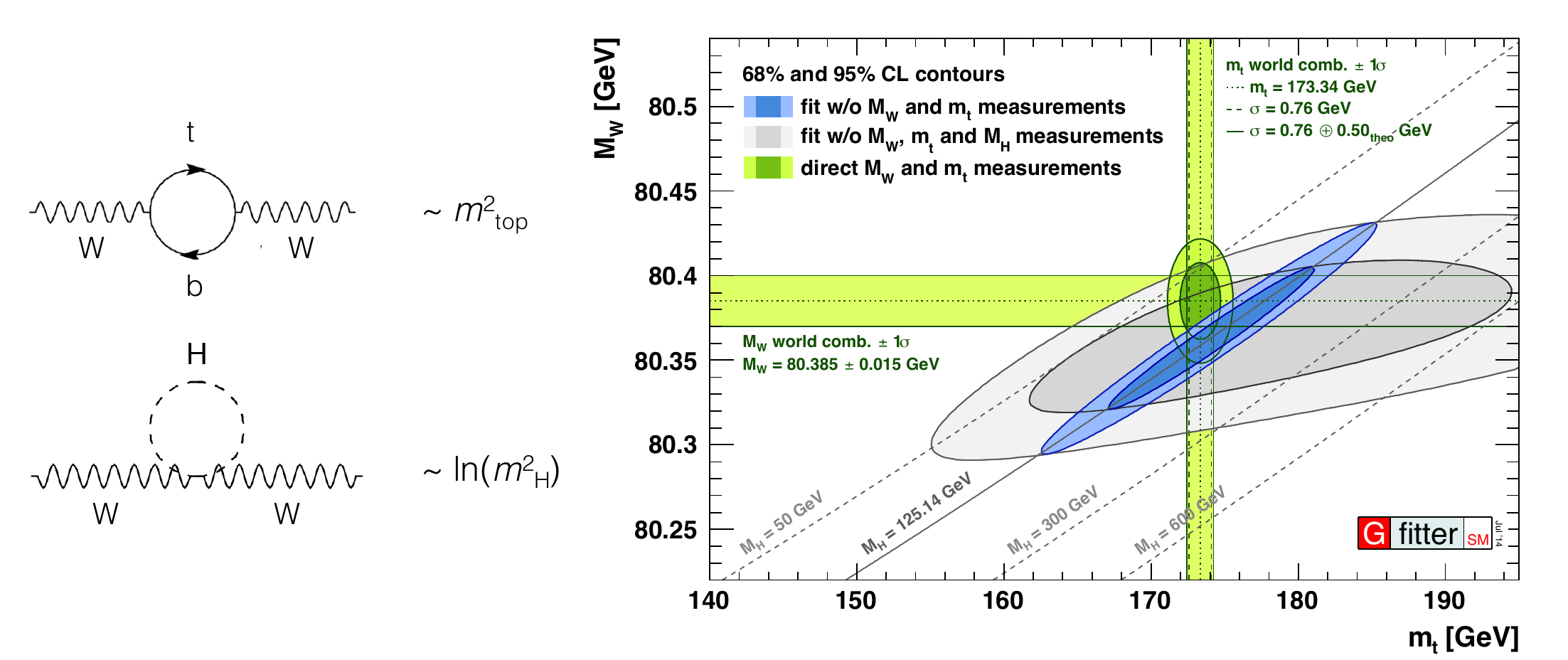}}
\vspace{0.0cm}
\caption[]{Left: Feynman graphs of radiative corrections contributing to the $W$ boson mass.
                The upper (lower) graph introduces a quadratic (logarithmic) top-quark 
                (Higgs-boson) mass dependence. The right panel shows 68\% and 95\% confidence 
                level contours obtained from scans of fits with fixed variable pairs $m_W$ versus 
                $m_t$. 
                The narrower blue and larger grey allowed regions are the results of the fit 
                including and excluding the $m_H$ measurement, respectively~\cite{Gfitter}. 
                The horizontal 
                bands indicate the 1$\sigma$ regions of the $m_W$ and $m_t$ measurements
                (world averages).
                \label{fig:Gfitter}}
\end{figure}
Figure~\ref{fig:Gfitter} shows the Feynman graphs of radiative corrections contributing 
to the $W$ boson mass. They introduce a quadratic top-quark  and logarithmic 
Higgs-boson mass dependence of the correction parameter $\Delta r$ occurring in the 
relation
\beq
    m_W^2 = \frac{m_Z^2}{2}
                   \left(1+\sqrt{1-\frac{\sqrt{8}\pi\alpha(1+\Delta r)}{G_F m_Z^2}}\right)\;,
\eeq
owing to electroweak unification. Similarly, the effective weak mixing angle, 
$\sinleff$, for lepton flavour $\ell$ depends on $m_W$ and 
$m_Z$ and, via radiative corrections and by replacing $m_W$, on the top-quark and 
Higgs-boson masses. 
The current predictions of the observables that most benefit from the known  
Higgs  mass, split into the various uncertainty terms, are~\cite{Gfitter}
\beqn
  M_W &=& 80.3584 
          \pm 0.0046_{m_t} \pm 0.0030_{\delta_{\rm theo} m_t} \pm 0.0026_{M_Z} \pm 0.0018_{\Delta\alpha_{\rm had}} \nonumber \\ 
      & & \phantom{80.3584} 
          \pm  0.0020_{\as} \pm 0.0001_{M_H} \pm 0.0040_{\delta_{\rm theo} M_W}\;{\rm GeV}\,, \nonumber \\[0.05cm]
      &=& 80.358 \pm 0.008_{\rm tot} \;{\rm GeV} \;,
\label{eq:mw}
\eeqn
and
\beqn
  \sinleff &=& 0.231488 
                 \pm 0.000024_{m_t} \pm 0.000016_{\delta_{\rm theo} m_t} \pm 0.000015_{M_Z} \pm 0.000035_{\Delta\alpha_{\rm had}} \nonumber \\
           & & \phantom{0.231496}
                 \pm 0.000010_{\as} \pm 0.000001_{M_H} \pm 0.000047_{\delta_{\rm theo} \sinfeff} \,, \nonumber \\[0.05cm]
      &=& 0.23149 \pm 0.00007_{\rm tot} \;.
\label{eq:sin2t}
\eeqn
Their total uncertainties of $8 \;{\rm MeV}$ and $7\cdot10^{-5}$, respectively, 
undercut the world average experimental errors of
15$\;$MeV and $16\cdot10^{-5}$~\cite{MWWA,sin2thW} .

The LHC experiments, as do CDF and D0 since long and continuing, are investing 
efforts into precision measurements of the electroweak observables $m_W$, 
$m_t$, and $\sinleff$. All are extremely challenging. 

\subsection{Top-quark mass}

There has been significant progress on the top-quark mass measurements
at the LHC achieving similar precision as those performed by the Tevatron 
experiments. The  currently most accurate LHC number is the CMS Run-1 
combination of measurements, based on the kinematic top mass reconstruction
and comparison with MC templates~\cite{CMS-mtopcombined}, giving
$m_t=172.44\pm0.13\pm0.47\;$GeV, where the first uncertainty 
is statistical and the second systematic. The ATLAS Run-1 combination, not yet 
including the  $\rm lepton+jets$, reads 
$m_t=172.84 \pm 0.34 \pm 0.61$\;GeV~\cite{ATLAS-mtop-Run1}.
The most recent Tevatron combination 
is $m_t=174.34\pm0.37\pm0.52\;$GeV~\cite{Tevatron-mtop} 
that shows a tension of $2.4\sigma$ or more with the CMS result.  

While these kinematic mass measurements provide the best current precision on 
$m_t$ and must be continued, it is also apparent that they approach 
a difficult systematic uncertainty regime from, mostly, the $b$-quark
fragmentation. A way to improve could be to choose
more robust observables with respect to the leading systematic effects
at the possible price of loosing statistical power. The dilepton kinematic 
endpoint is an experimentally clean observable, which has however large 
theoretical uncertainties~\cite{CMS-mtopEndpoint-Run1}. More robust could 
be the selection of charmonium states~\cite{CMS-mtopJ/psi} or charmed hadrons 
originating from a $b$-hadron produced in one of the $b$-jets. These provide
a clean but rare signature. 

ATLAS and CMS also indirectly determine the top mass  
from inclusive and differential cross-section measurements. These are 
promising approaches benefiting from theoretically well defined observables, which 
are however not yet competitive with the kinematic methods. They 
also stronger depend on the assumption that no new physics contributes to 
the measured cross sections. The currently best top pole mass determination from 
CMS~\cite{CMS-topXS} using a precise Run-1 $e\mu$-based cross-section measurement 
is $173.8^{\,+1.7}_{\,-1.8}\;$GeV in agreement with the direct 
(kinematic) measurements. 

\subsection{Weak mixing angle}

The CDF, D0~\cite{Tevatron-sinth} and 
LHC experiments~\cite{ATLAS-sinth,CMS-sinth,LHCb-sinth} 
have extracted the weak mixing angle from $Z/\gamma^\star$ polarisation 
measurements.
The total uncertainty on $\sinleff$ at the Tevatron 
is dominated by statistical effects, that of LHCb has similar statistical and 
systematic contributions, while for ATLAS and CMS parton density function 
(PDF) uncertainties are dominant.  A data-driven ``PDF replica rejection'' 
method applied by CDF allows to reduce the sensitivity to PDF and 
update the measurement when improved PDF sets are available.
Overall, these are complex measurements (in particular with respect to the 
physics modelling) that are important to pursue also in view of a better 
understanding of $Z/\gamma^\star$  production at hadron colliders. 
The precision obtained is however not yet competitive with that of LEP/SLC. 

\subsection{{\em W}-boson mass}

The $W$ boson was discovered at CERN's SPS in 1983. A first measurement of its mass 
by the UA1 experiment in 1983 at centre-of-mass energy of 546\;GeV
gave  $m_W = 81 \pm 5$\;GeV~\cite{UA1-Wmass}. In 1992, at $\sqrt{s}=630$\;GeV UA2 achieved 
$80.35 \pm 0.37$\;GeV using $m_Z$ from LEP as reference calibration~\cite{UA2-Wmass}. A factor of 
ten improvement in precision was obtained at LEP with the most recent combination giving
$80.376 \pm 0.033$\;GeV. That precision has been undercut by the Tevatron experiments 
whose latest average, using proton--antiproton collision data taken at $\sqrt{s}=1.96$\;TeV, 
is $80.387 \pm 0.016$\;GeV. The combination of the
Tevatron and LEP results leads to the present world average 
$m_W=80.385 \pm 0.015$\;GeV~\cite{MWWA,sin2thW}. While the LEP analyses are 
final, the Tevatron experiments are continuing to improve their precision and
updated results can be expected in the future. 

It likely came as a surprise to many in the particle physics community that such a precision 
measurement is now dominated by a hadron collider, which was not built with that goal in mind. 
The $W$ boson mass is arguably the hardest measurement in high-energy physics, needing 
about seven years to be accomplished. Also the LHC was not built to measure 
the $W$ boson mass, but to discover new particles. There is an unfavourable environment 
at the LHC compared to $e^+e^-$ or proton--antiproton colliders. At the Tevatron, $W$ boson
production is dominated by the valence quarks of the proton. At the LHC on the contrary, 
sea and thus heavy quarks are much more important. This difference affects all aspects 
of the measurement: detector calibration, transfer from the $Z$ to the $W$ boson, PDF 
uncertainties, $W$ polarisation, modelling of the $W$ transverse momentum. It is thus a very  
challenging undertaking, but also a very interesting one: a lot can be learned on the way!

The measurement of the $W$-boson mass at the LHC using the leptonic $W$ boson decay
relies on an excellent understanding of the final state. The observables that probe $m_W$ 
are the transverse momentum of the lepton ($p_{T,\ell}$), the transverse momentum 
of the neutrino ($p_{T,\nu}$), measured from the transverse recoil of the event, 
and the transverse mass of the lepton-neutrino system ($m_T$).
The measurement requires a high-precision momentum and energy scale calibration
(including the hadronic recoil) obtained from $Z$, $J/\psi$ and $\Upsilon$ data,
and excellent control of the signal efficiency and background modelling.
The biggest challenge is posed by the physics modelling. The production is 
governed by PDF and initial state interactions (perturbative and nonperturbative),
that can be constrained by $W^+$, $W^-$, $Z$, and $W+c$ data, and the use of 
NNLO QCD calculations including soft gluon resummation. The experimental $m_W$ 
probes are very sensitive to the $W$ polarisation (and hence to PDF, including 
its strange density). Electroweak corrections are sufficiently well known.

The experiments are thriving to address  the above issues. Many 
precision measurements (differential $Z$, $W + X$ cross sections, polarisation 
analysis, calibration performance, etc.) are produced on the way with benefits  
for the entire physics programme. Theoretical developments are also mandatory. 
Altogether this is a long-term and iterative effort.

CMS presented for the first time a $m_Z$ measurement using a $W$-like 
$Z \to \mu^+\mu^-$ analysis where one muon is replaced by a neutrino that contributes
to the missing transverse momentum in the  event~\cite{CMS-Wmass}.
It represents a proof-of-principle, although differences with the full $m_W$ analysis 
remain in the event selection, the background treatment and most of the physics modelling
uncertainties. CMS used the 7$\;$TeV dataset to take benefit from the 
lower number of pileup interactions. The momentum scale and resolution 
calibration for that  measurement relies on 
$J/\psi$ and $\Upsilon$ data. Track-based missing transverse momentum is used 
and the $W$ transverse recoil is calibrated using $Z+{\rm jets}$ events. The results
for the different probes and the positive and negative $W$-like cases are found to agree
with the LEP measurement. The uncertainties, depending on the probe used, are: 
statistical: 35--46$\;$MeV, total  systematic: 28--34$\;$MeV, 
QED radiation: $\sim$23$\;$MeV (dominant), lepton calibration: 12--15$\;$MeV.

\begin{figure}[t]
\centerline{\includegraphics[width=1\linewidth]{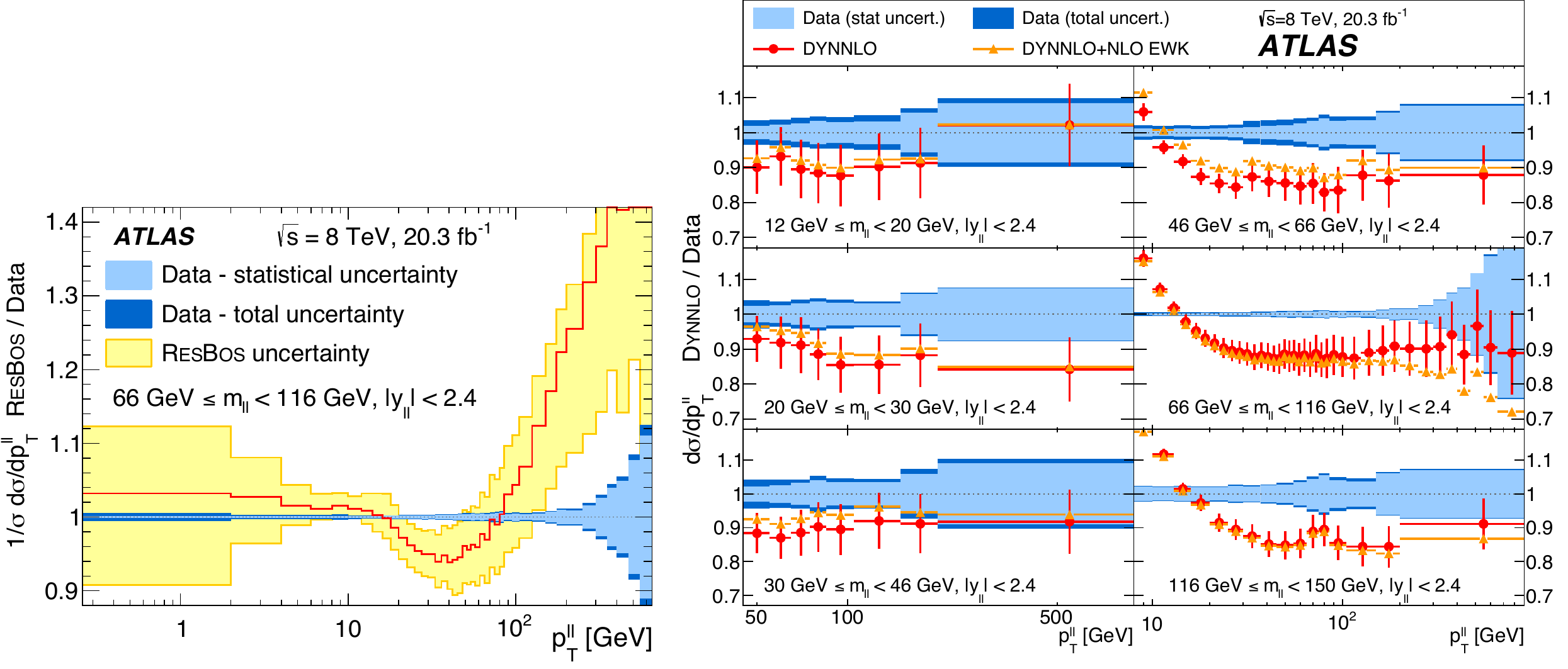}}
\vspace{0.0cm}
\caption[]{Left: ratio of ResBos predictions of the normalised differential $p_T^Z$  
                cross section to ATLAS Born-level data~\cite{ATLAS-pTZ-Run1}. 
                Right: the same ratio for different $Z$ rapidity intervals and by 
                using the DYNNLO programme for the theoretical prediction.                 
                \label{fig:ATLAS-pTZ-Run1}}
\end{figure}
ATLAS and CMS use precise measurements of the $Z$ boson $p_T$ to tune the 
$p_T$ modelling of the $W$ boson, which relies on NNLO and NNLL/resummed 
calculations. But: different generators predict different transfers from $Z$ to $W$.          
In addition, PDFs play different roles in $Z$ and $W$ production. 
Figure~\ref{fig:ATLAS-pTZ-Run1} shows normalised differential cross section 
ratios measured by ATLAS~\cite{ATLAS-pTZ-Run1} of resummed NLO predictions 
from ResBos\footnote{ResBos features ISR at approximate NNLO, $\gamma^*$--$Z$ 
interference at NLO, NNLL soft-gluon resummation, no FSR or hadronic event activity, 
CT14 PDF set.}~\cite{ResBos} to data (left) and NNLO predictions using the DYNNLO 
programme\footnote{DYNNLO features QCD production at NNLO, no soft-gluon 
resummation, CT10 PDF set.}~\cite{DYNNLO} without soft gluon resummation 
to data (right). While resummation is needed to describe the low-$p_T$ data,
NLO calculations and better are required in the high-$p_T$ regime. 

\section{The SM is complete}

Since the LHC Run-1 the SM is a complete and self-consistent theory. 
The discovery of the Higgs boson  is a triumph for the imagination and rigour 
of the scientific endeavour. It is also a triumph for the greatest experimental undertaking 
ever, at the frontier of accelerator and detector technologies, global data sharing, analysis 
and collaboration.

The Higgs  mass of 125.1\;GeV is in 
agreement with the prediction from the global electroweak fit~\cite{Gfitter2012} (cf. left 
panel of Fig.~\ref{fig:Higgs-EWfit-Stability}) and it lies marginally within the requirement 
for vacuum stability~\cite{SMFate} (right panel of Fig.~\ref{fig:Higgs-EWfit-Stability}). 
The Higgs  discovery does thus not come with a strict requirement for new 
physics below the Planck scale. 

We have now two beautiful and extremely precise theories. On one hand the SM describing 
electroweak and strong interactions (though not their unification), predicting, eg., the anomalous 
magnetic moment of the electron to a relative precision of $10^{-10}$ in agreement with 
experiment. On the other hand there is general relativity, the theory of gravitation. It 
has been tested to an accuracy of order $10^{-5}$ (Cassini probe~\cite{CassiniProbe}).
Unfortunately, the SM and general relativity do not work in regimes where both are 
important, that is at very small scales.

Indeed, many open questions not addressed by the SM remain as we have already alluded 
to in the introduction to these proceedings. We shall repeat some of them here.
\begin{itemize}\setlength{\itemsep}{0.5\baselineskip}

\item {\bf Scalar sector}. 
          Is there a single Higgs doublet or are there additional scalar states? Is the Higgs boson
          elementary or composite? What is the exact form of the scalar potential? What is the origin 
          of the Yukawa couplings?

\item {\bf Quarks and leptons}.
          What is the origin of the fermion generations, mass, mixing, CP violation?
          How was the matter--antimatter asymmetry in the universe generated?
          What is the origin of baryon and lepton number conservation and what is 
          the proton lifetime?

\item {\bf Neutrinos}.
         What is the nature of the neutrinos: Majorana or Dirac?
         Are there sterile neutrinos?
         What is the origin of neutrino mass and what are their values (and hierarchy)?          
         Is there CP violation in the neutrino mixing?

\item {\bf Strong CP problem}.
          Why is there no noticeable CP violation in strong interactions albeit predicted by the SM?

\item {\bf Dark matter}.
          What is its composition: WIMPs, axions, sterile neutrinos, hidden sector particles, 
          gravitational effect only? Is there a single or are there multiple sources?

\item {\bf Expansion of Universe}.
          Primordial expansion via inflation: which fields, and what is the role of the Higgs boson
          and of quantum gravity? Accelerated expansion today: cosmological constant problem.

\item {\bf High-scale physics}.
          Is there a solution to the hierarchy problem\footnote{The term hierarchy problem
          stands for the apparent dependence of phenomena at the electroweak scale on 
          a much higher (possibly the Planck) scale, as exemplified by the extreme ultra-violet sensitivity 
          of the Higgs potential.} and is there new physics at the TeV scale?
          Will there be grand unification of the forces? How does unification with gravity proceed?
          How is quantum gravity realised? Is everything just made of tiny strings?
\end{itemize}
\begin{figure}[t]
\centerline{\includegraphics[width=1\linewidth]{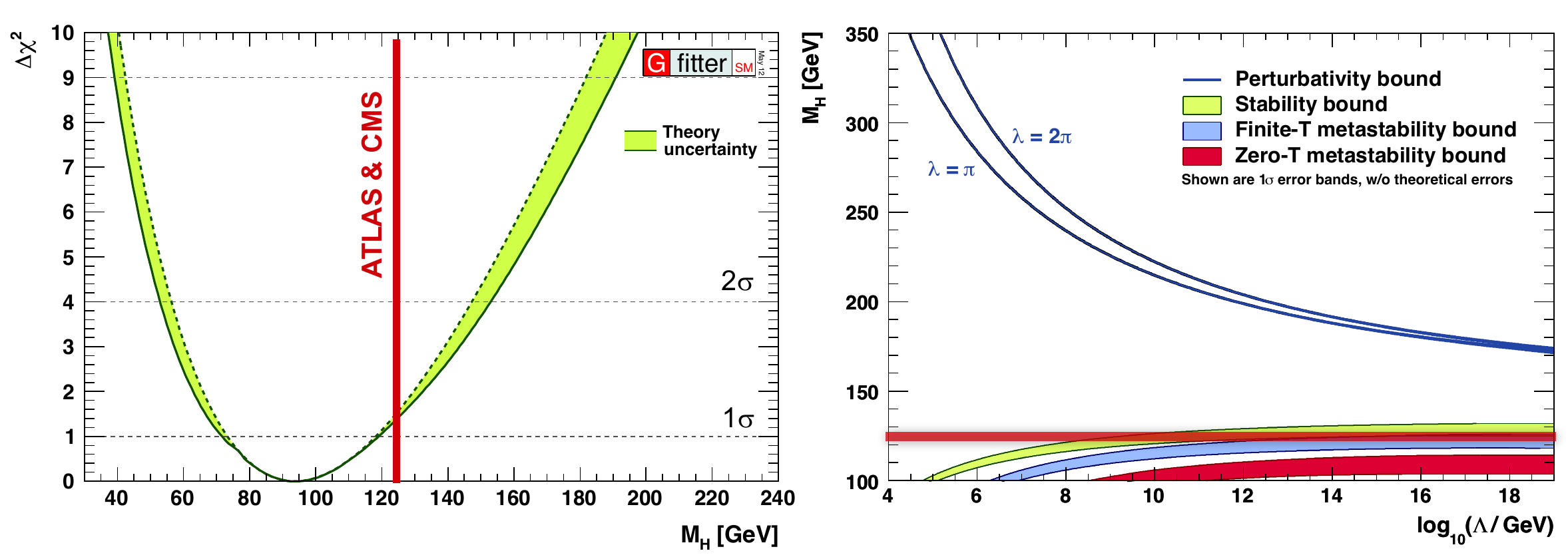}}
\vspace{0.0cm}
\caption[]{Left: $\chi^2$ curve obtained from the global electroweak fit in 2012 at the 
                moment of the Higgs boson discovery~\cite{Gfitter2012}. 
                Right: comparison of the observed Higgs  mass with the lower limits from the 
                vacuum stability constraint and the upper perturbativity limit~\cite{SMFate}. 
                \label{fig:Higgs-EWfit-Stability}}
\end{figure}

Because the SM cannot be all there is, the LHC experiments have performed a large 
number of searches for new physics during Run-1, covering a vast space of possible 
signatures as witnessed in the  exclusion plots of Fig.~\ref{fig:ATLASCMSSearches-Run1}.
Heavy resonances are excluded up to 3.5\;TeV mass in some scenarios. Gluinos up to 1.3\;TeV
are excluded for light neutralinos (supersymmetry limits are usually lower than those of 
many other new physics scenarios because $R$-parity conservation requires pair production 
of supersymmetric particles).
\begin{figure}[p]
\centerline{\includegraphics[width=0.9\linewidth]{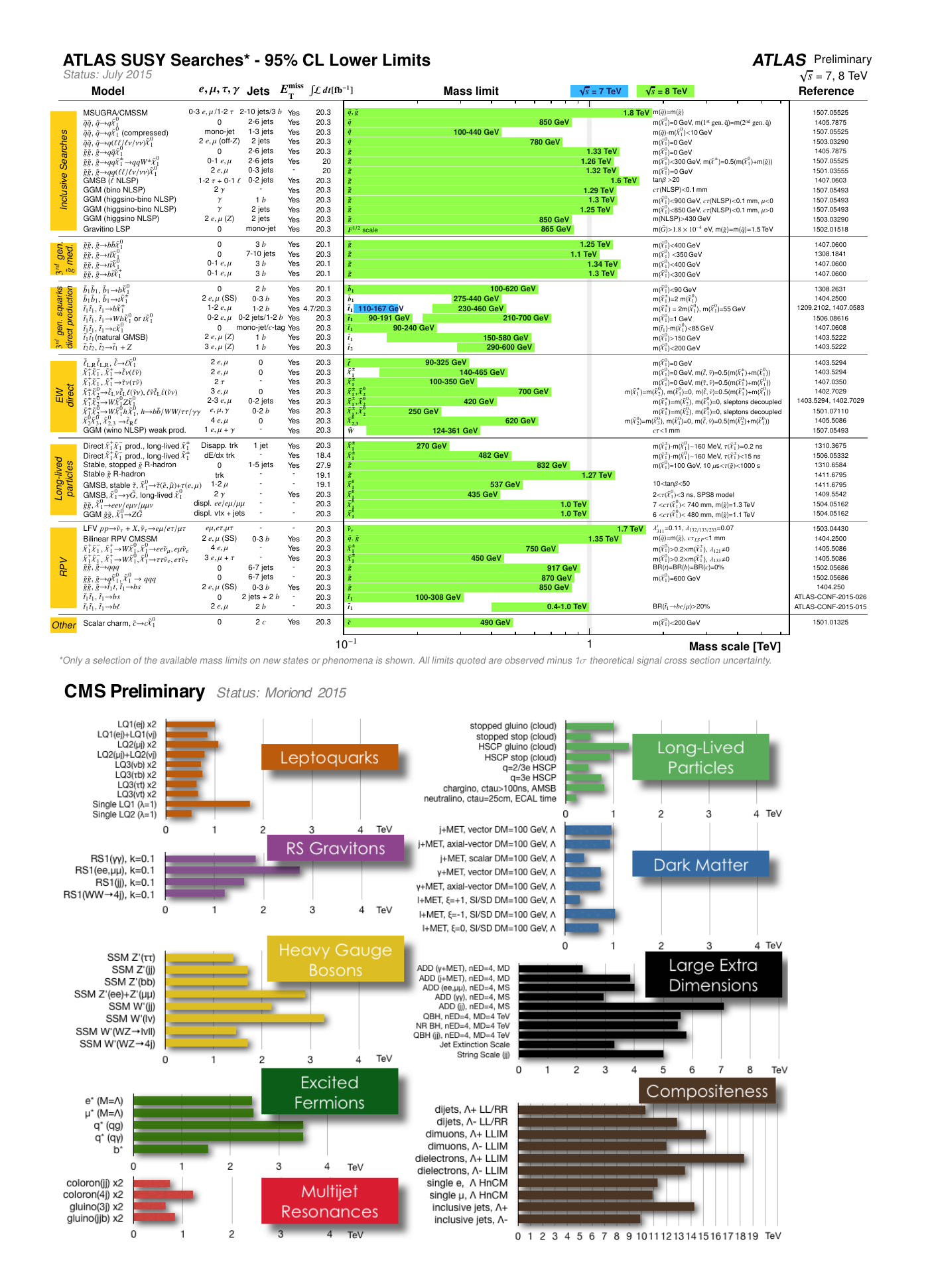}}
\vspace{0.3cm}
\caption[]{Exclusion bounds on mass scales from searches for supersymmetry (top, ATLAS) and 
               other new physics phenomena (bottom, CMS Exotica). 
                \label{fig:ATLASCMSSearches-Run1}}
\end{figure}

\vfill\pagebreak

\section{The LHC Run-2}

\begin{figure}[t]
\centerline{\includegraphics[width=1\linewidth]{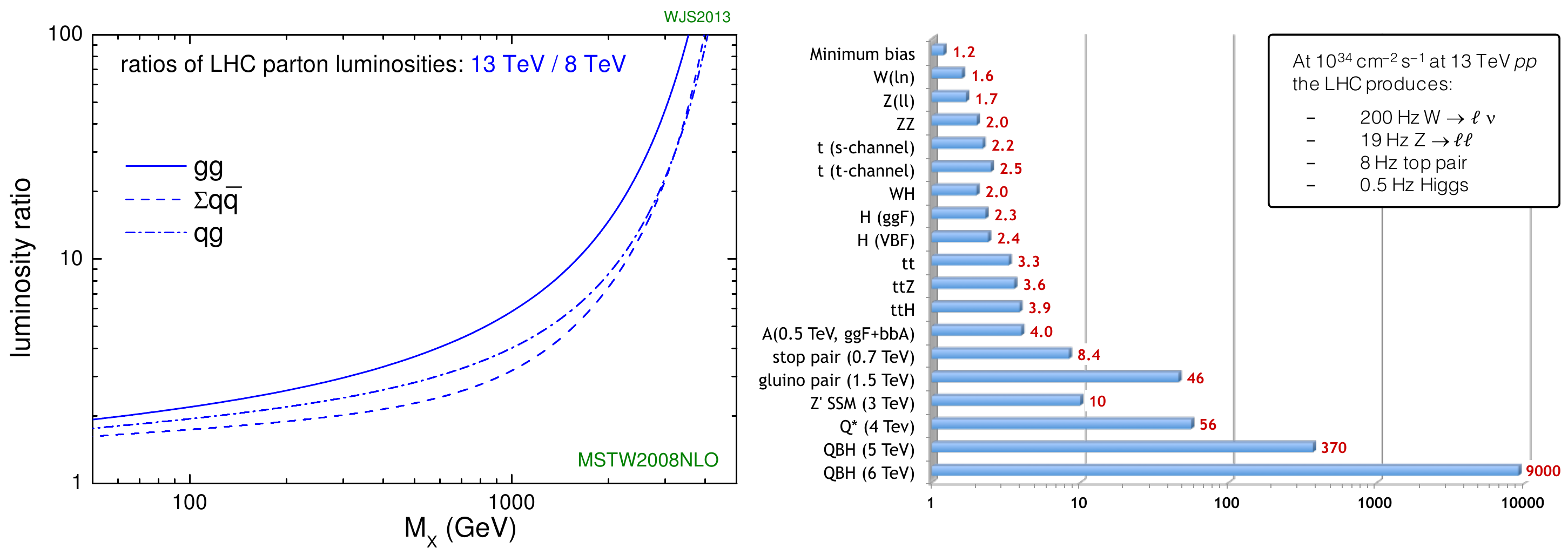}}
\vspace{0.0cm}
\caption[]{Left: parton luminosity ratio of 13\;TeV to 8\;TeV proton--proton 
               collisions~\cite{StirlingPartonLuminosities}. Right: cross section ratios
               for selected processes (the heavy flavour cross section scales roughly linearly 
               with centre-of-mass energy).
                \label{fig:PartonLuminosity13to8}}
\end{figure}
A huge milestone was achieved in 2015 when the new record proton--proton collision 
energy of 13\;TeV was reached.  After a rocky start, the LHC delivered 4.2\ifb
integrated luminosity to ATLAS and CMS. That amount of data already surpassed the Run-1 
new physics  sensitivity of many searches. During 2016 a peak luminosity of 
$1.4\cdot10^{34}\;{\rm cm}^{-2}{\rm s}^{-1}$ was reached and a total of 
$39$\ifb integrated luminosity  delivered, which exceeded expectations. 

The new centre-of-mass energy increases the cross-section of all LHC processes.
Figure~\ref{fig:PartonLuminosity13to8} gives the 13\;TeV to 8\;TeV parton luminosity 
ratios for gluon--gluon, quark--gluon and quark--quark scattering (left panel) and the 
resulting proton--proton cross-section ratios (right). The parton luminosity 
as a function of the hard scattering $Q^2=M_X^2$ (cf. Fig.~\ref{fig:ProtonCollision}) 
is defined by the convolution integral
\beq
       \frac{\partial {\cal L}_{ab}}{\partial M_X^2}
       = \frac{1}{s}\int_\tau^1 \frac{dx}{x} f_a(x,M_X^2) f_b(\tau/x,M_X^2)\;,
\eeq
where $\tau=M_X^2/s$. There is a larger parton luminosity increase with energy 
for gluon initiated processes than for quark ones. Owing to the important cross section 
rise at large $M_X$ the early Run-2 analyses put their emphasis on searches.

Most of the results from ATLAS and CMS presented at the 2016 summer 
conferences contained data up to approximately 15\ifb. CMS used different 
software releases and thus did not merge the 2015 and 2016 datasets, but
in selected cases provided a statistical combination. ATLAS performed a 
reprocessing of the 2015 data and MC allowing it to treat both years as a 
single coherent dataset. LHCb performed luminosity levelling leading to 
an approximately ten times smaller dataset in terms of integrated luminosity. 
The uncertainty on the luminosity
values from ATLAS, CMS and LHCb, for the summer 2016 results were 
2.9\%, 6.2\% and 3.8\%, respectively. The amount of pileup interactions
with an average $\mu$ (cf. Eq.~\ref{eq:pileup}) of 23 interactions was similar 
to that in 2012. LHCb observed  1.7 pileup interactions in average. 

\subsection{Standard Model and top-quark physics}

Along increasing scattering momentum transfer, SM processes  the LHC can be split the 
as follows. 
\begin{itemize}\setlength{\itemsep}{0.5\baselineskip}

\item {\bf Soft QCD}: study of particle spectra. The transverse momenta
         are typically smaller than a few GeV. More than 99.999\% of the 
         proton--proton collisions belong to that  type. 
         Measurements of soft QCD processes serve to probe LO matrix elements, 
         parton shower models, generator tunings, and for pileup modelling. 

\item {\bf Hard QCD}: study of jets. Typical jet $p_T$ greater than tens of GeV 
         up to the TeV scale; approximately $10^{-5}$ of the collisions belong to 
         that category. The measurements probe NLO QCD, the running $\as$, 
         PDFs, parton showers, etc. 

\item {\bf Hard QCD and electroweak processes}: $W$, $Z$, $H$, top 
         decaying to stable identified particles. Typical $p_T$ scale of greater than tens 
         of GeV ;  a fraction of $10^{-6}$ and less of the collisions belong to this category.
         Measurements probe NLO, NN(N)LO QCD, soft gluon resummation, 
         PDFs, electroweak physics, etc.

\end{itemize}

Standard Model and Higgs precision measurements are key to the LHC programme 
up to the High-Luminosity LHC (HL-LHC). Michelangelo Mangano at the SEARCH 
2016 workshop~\cite{Mangano@SEARCH2016} summarised the importance 
of these measurements as follows. 

\begin{itemize}\setlength{\itemsep}{0.5\baselineskip}

\item {\bf Scientific perspective}. No matter what BSM the LHC will unveil in the next 
years, improving the knowledge of Higgs properties is a must, which by itself 
requires and justifies the largest possible LHC statistics  so that stopping after 300\;\ifb 
would not be satisfying.

\item {\bf Pragmatic perspective}. Higgs and SM physics are the only guaranteed 
deliverables of the LHC programme. Need to exploit this part of the programme 
to its maximum extent!

\item {\bf Utilitarian perspective}. Elements of the SM, besides the Higgs, require further 
consolidation, control and improved precision, both in the EW and QCD sectors.
They hold a fundamental value (eg. the precise determination of 
parameters of nature and to better understand detailed scattering dynamics), 
or are critical to fully exploit the BSM search potential (eg. the knowledge 
of backgrounds, production rates and production dynamics).

\item {\bf Spinoffs}. The study of SM processes at colliders is typically  more 
complex than the search for BSM signatures and throughout the years it has been the 
main driver of fundamental theoretical innovation.

\end{itemize}

\subsubsection{Inelastic proton--proton cross section}

\begin{wrapfigure}{R}{0.53\textwidth}
\centering
\vspace{-0.40cm}
\includegraphics[width=0.53\textwidth]{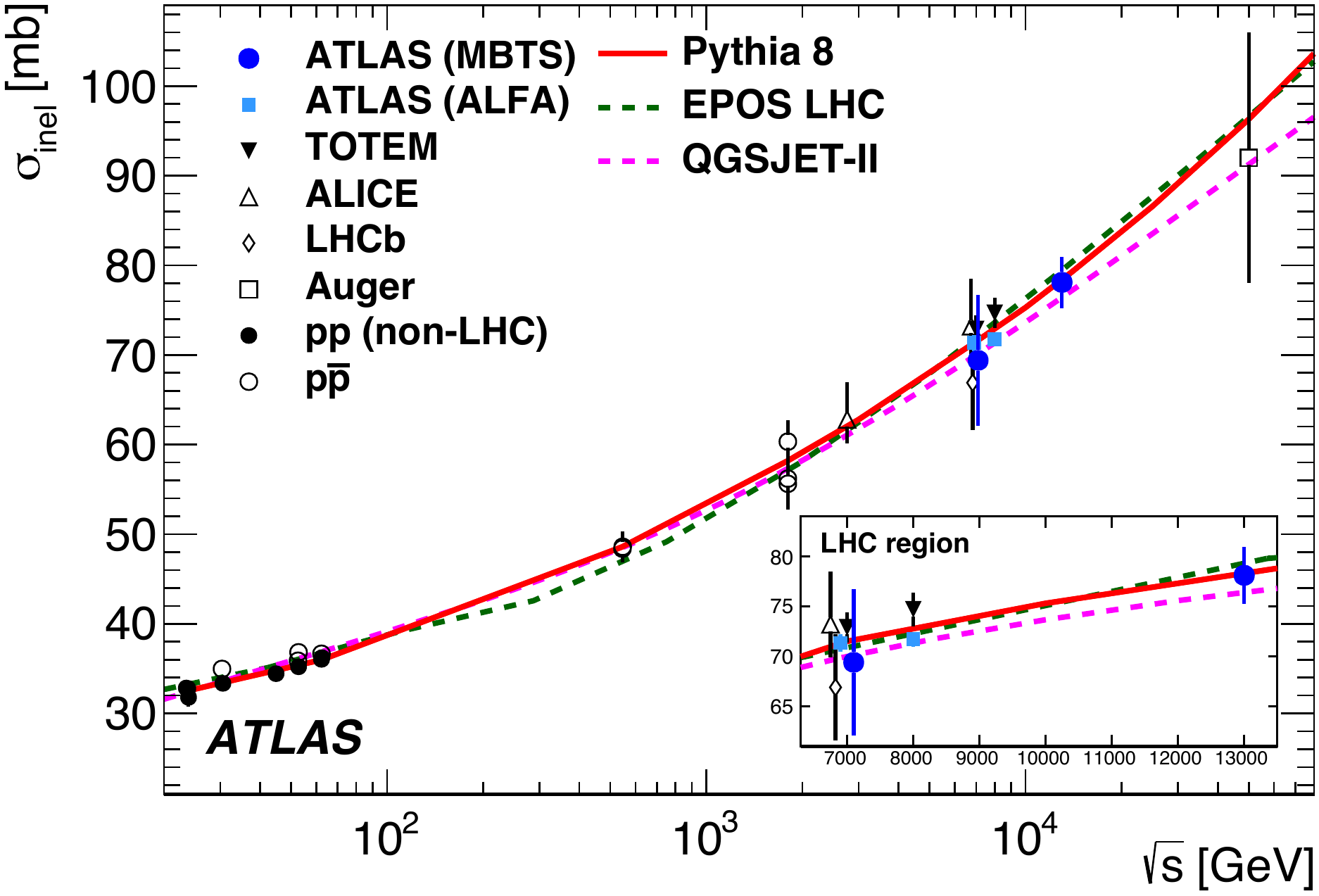}
\caption[.]{Inelastic proton--proton cross section versus centre-of-mass 
                energy~\cite{ATLAS-sigma_tot}. 
  \label{fig:ATLAS-sigma_tot}}
\vspace{-0.4cm}
\end{wrapfigure}
A key initial measurement is the inclusive inelastic cross-section at 
13\;TeV. While the most precise total cross section measurement is obtained via elastic 
scattering and the optical theorem 
($\sigma_{\rm tot}(pp\to X)\propto {\rm Im}f_{\rm elastic}({\rm t}\to 0)$,
where $f_{\rm elastic}({\rm t}\to 0)$ is the elastic scattering amplitude extrapolated 
to the forward direction, and ${\rm t}$ is the Mandelstam momentum transfer variable)
using dedicated forward devices (such measurements have achieved better than 1\% precision in 
Run-1, dominated by the luminosity uncertainty~\cite{ATLAS-ALFA-Run1,TOTEM}); 
it is also possible to determine 
$\sigma_{\rm tot}(pp\to X)$ from a measurement of inelastic scattering cross section if the 
extrapolation between fiducial to total acceptance is not too large. 
This can be achieved 
using forward detectors such as  scintillators installed in ATLAS within $2.07<|\eta|<3.86$.

The ATLAS measurement~\cite{ATLAS-sigma_tot} was performed in the fiducial 
region $\xi=M_X^2/s > 10^{-6}$, where $M_X$ is the larger invariant mass of the 
two hadronic (proton-dissociation) systems separated by 
the largest rapidity gap in the event. In this $\xi$ range the scintillators have high 
efficiency. When extrapolated to the full phase space, a cross-section of 
$\sigma_{\rm tot}(pp\to X)=78.1 \pm 0.6\pm1.3\pm2.6$\;mb 
is obtained, where the first uncertainty is experimental, the second due to the luminosity,
and the third and dominant one from the extrapolation to full phase space. The result is consistent 
with the expectation from phenomenological models (cf. Fig.~\ref{fig:ATLAS-sigma_tot},
where also measurements from other hadron collider experiments  and from the Pierre Auger 
experiment are  shown, see references in~\cite{ATLAS-sigma_tot}).

\subsubsection{Jet production}

\begin{wrapfigure}{R}{0.53\textwidth}
\centering
\vspace{-0.8cm}
\includegraphics[width=0.52\textwidth]{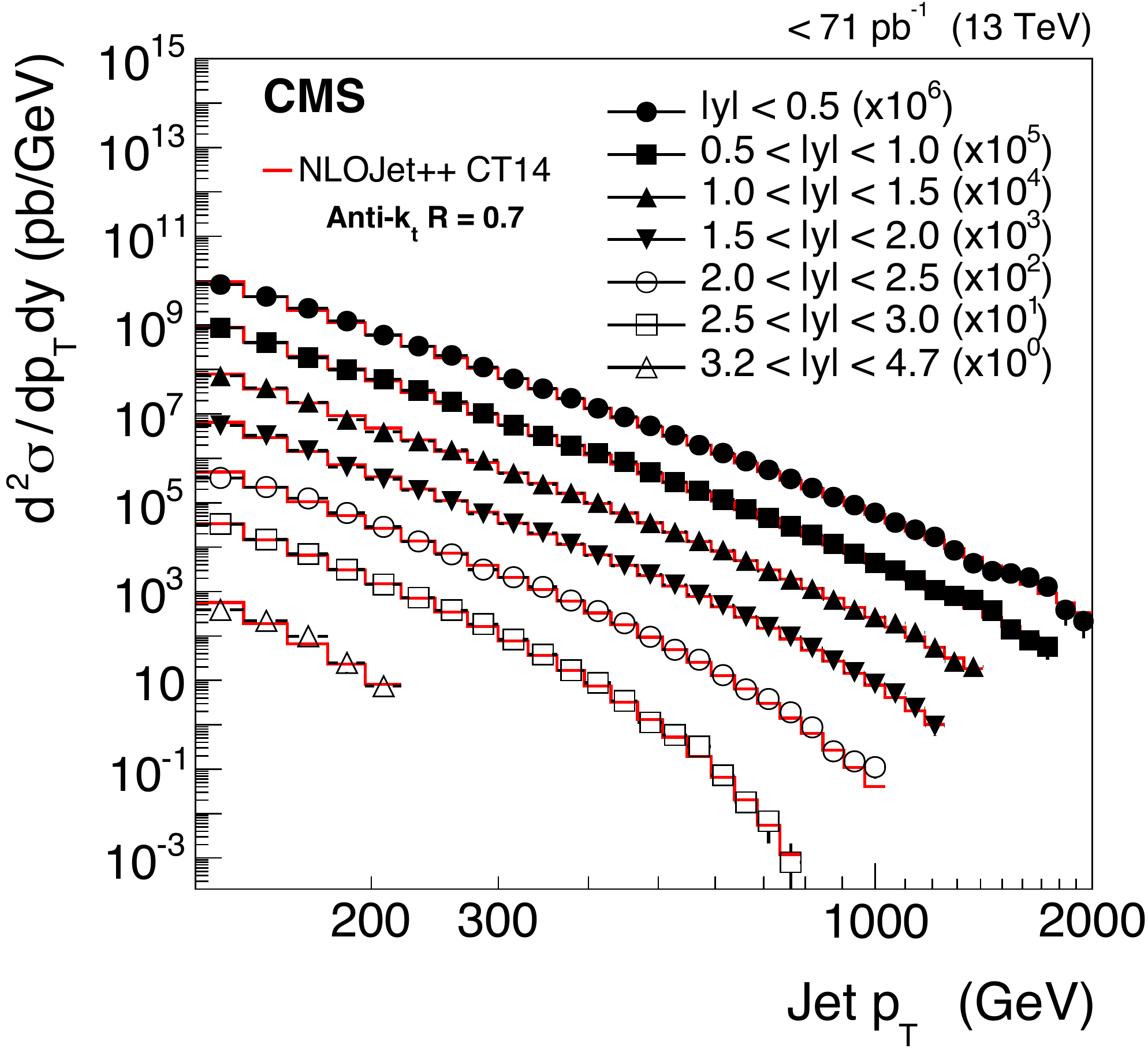}
\vspace{-0.1cm}
\caption[.]{Double-differential inclusive jet cross section versus 
                jet $p_T$ as measured by CMS at 13\;TeV~\cite{CMS-jetXS} and compared 
                to a theoretical prediction.
  \label{fig:CMS-jetXS}}
\vspace{-1.3cm}
\end{wrapfigure}
Moving up in transverse momentum, ATLAS and CMS measured jet production. 
Figure~\ref{fig:CMS-jetXS} shows the double differential inclusive jet cross section 
as measured by CMS. The unfolded data points are compared to predictions from 
NLOJet++ based on the CT14 PDF set and corrected for the nonperturbative and 
electroweak effects (line in figure). It is interesting to compare Fig.~\ref{fig:CMS-jetXS}
at 13\;TeV to the 8\;TeV result shown in Fig.~\ref{fig:CMS-jetXS-Run1} on 
page~\pageref{fig:CMS-jetXS-Run1}. For a given 
rapidity interval, the relative drop in cross section between high and low $p_T$ 
is less pronounced at 13\;TeV, as expected from the parton luminosities. Indeed, 
taking the ratio between the 13\;TeV and 8\;TeV cross sections approximately reproduces 
the left panel of Fig.~\ref{fig:PartonLuminosity13to8} for gluon--gluon scattering. 

\subsubsection{Weak boson production}

\begin{figure}[t]
\centerline{\includegraphics[width=1\linewidth]{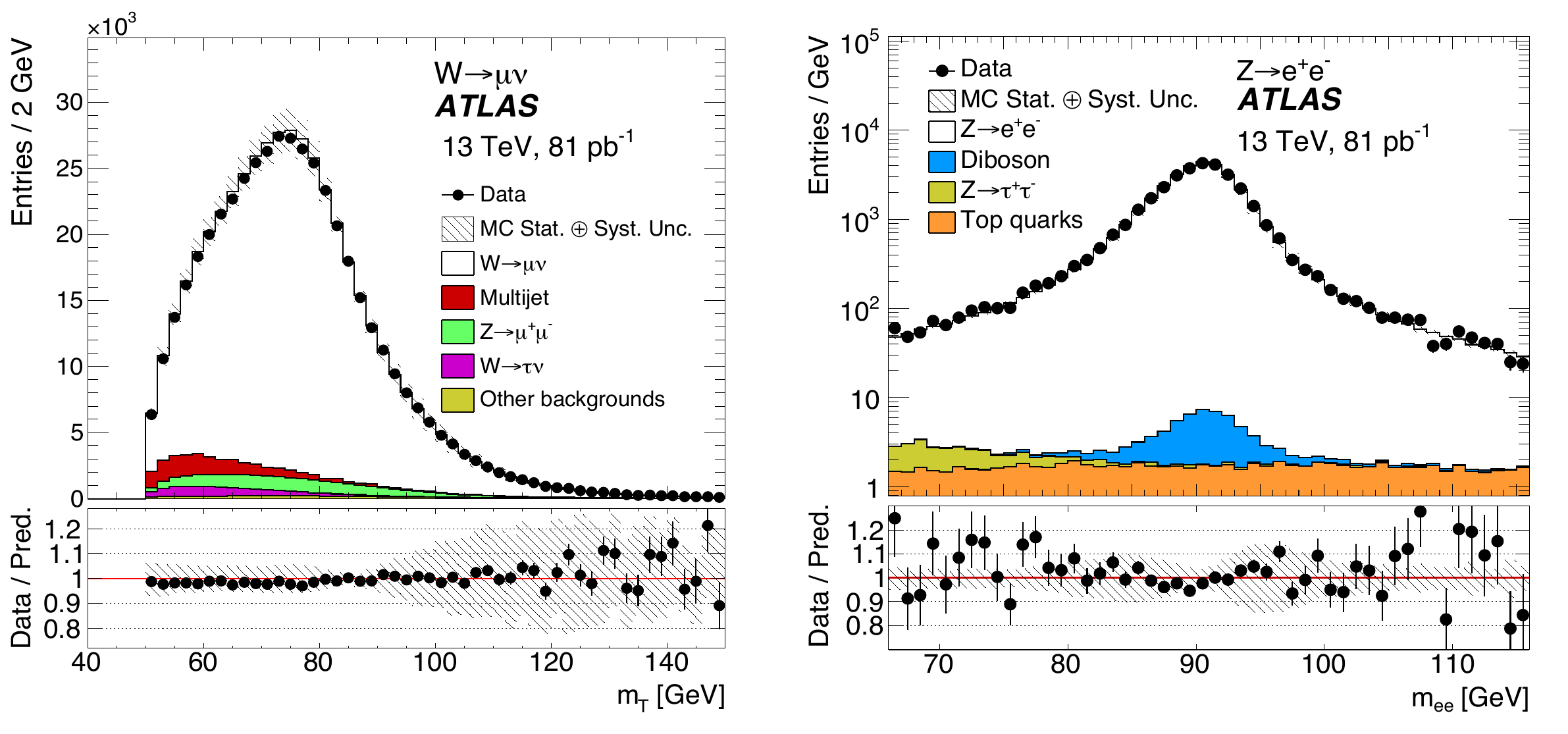}}
\vspace{0.0cm}
\caption[.]{Transverse mass (left) and invariant dilepton mass (right) for $W\to\mu\nu$
                and $Z\to ee$ candidates, respectively~\cite{ATLAS-W/Z,CMS-W/Z}.
                The predicted signal distributions are normalised to the measured cross sections.
                \label{fig:ATLAS-WZmass}}
\end{figure}
The inclusive $W$ and $Z$ boson production cross sections are expected to rise 
at 13\;TeV over 8\;TeV centre-of-mass energy by factors of 1.7 and 1.6, respectively, 
to 19.7\;nb and 1.9\;nb for the decays to muons. Leptonic $W$ and $Z$ decays 
are very pure channels as can be seen from Fig.~\ref{fig:ATLAS-WZmass}, which shows
the transverse and invariant dilepton mass distributions for 13\;TeV $W\to\mu\nu$
(left panel) and $Z\to ee$ candidates (right). The transverse mass-squared 
is defined by $m_T^2=2p_{T,\ell}\MET(1-{\rm cos}\Delta\phi_{\ell,\nu})$, where 
$\Delta\phi_{\ell,\nu}$ is the azimuthal angle difference between lepton and 
missing transverse momentum. The dilepton invariant mass-squared is given by 
$m_{\ell_1\ell_2}^2=2p_{T,\ell_1}p_{T,\ell_2}({\rm cosh}\Delta\eta_{12}-{\rm cos}\Delta\phi_{12})$.

Apart from the intrinsic interest in precise cross section measurements, leptonic 
$W$ and $Z$ decays also serve the experiments as standard candles to calibrate 
the electron and muon reconstruction performance via mass constraints and so-called 
tag-and-probe efficiency measurements. 
Tag-and-probe methods~\cite{ATLAS-electrons}
are used to select, from known resonances, unbiased samples 
of electrons or muons (probes) by using strict selection requirements on the second 
object produced from the particle's decay (tags). The efficiency of a requirement can 
then be determined by applying it directly to the probe sample after accounting for 
residual background contamination. 

\begin{wrapfigure}{R}{0.55\textwidth}
\centering
\vspace{-0.37cm}
\includegraphics[width=0.55\textwidth]{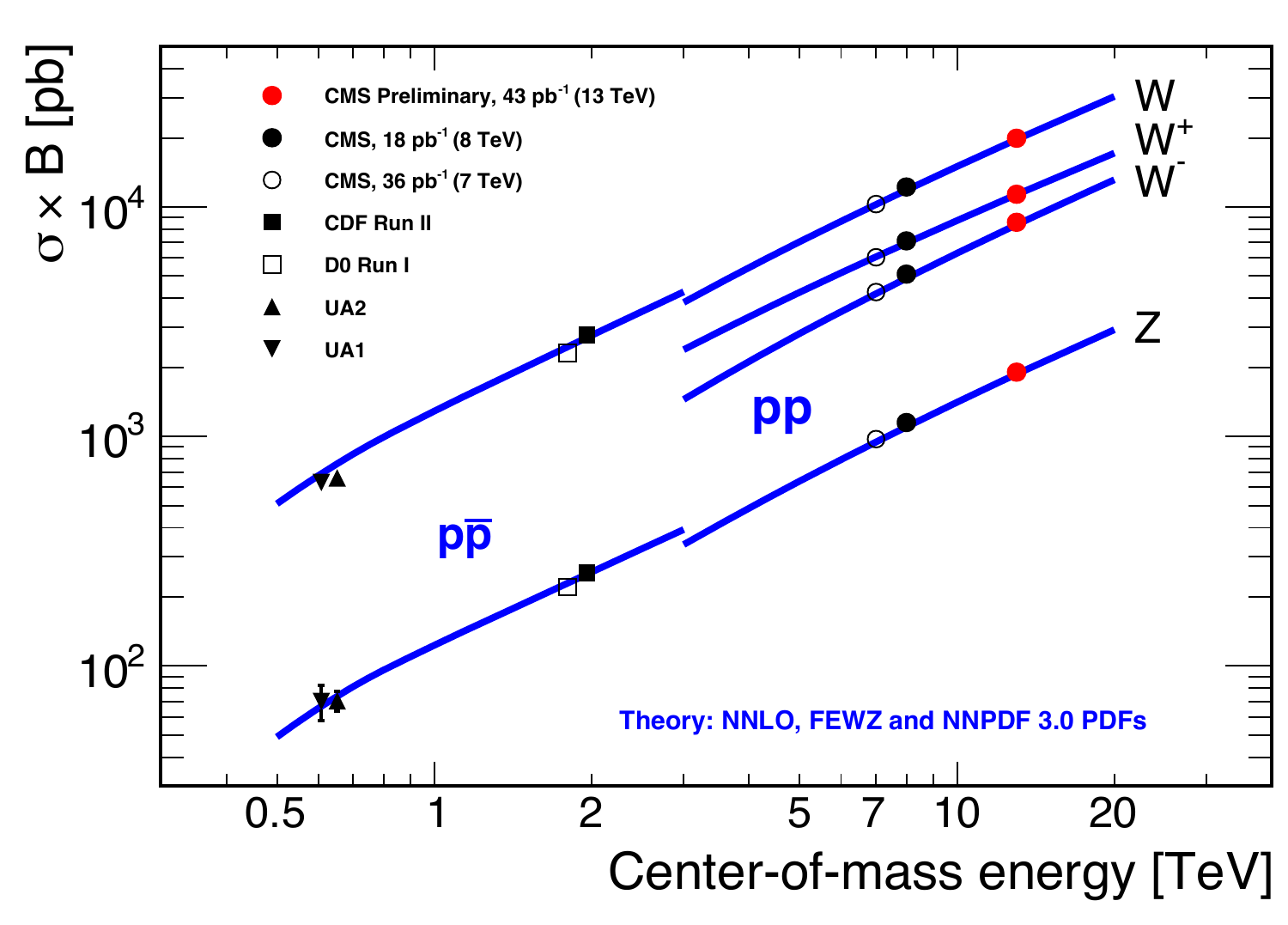}
\caption[.]{Cross sections of proton--(anti-)proton production of inclusive $W$ and $Z$ 
                 bosons versus centre-of-mass energy.
  \label{fig:CMS-WZvsSQRTs}}
\vspace{-0.1cm}
\end{wrapfigure}
Both ATLAS and CMS measured fiducial and inclusive cross sections for $W$ and $Z$ 
boson production as well as their ratios using partial 2015 datasets~\cite{ATLAS-W/Z,CMS-W/Z}. 
The fiducial 
cross sections are dominated by the luminosity uncertainty of 2.1\% (ATLAS). Comparisons of 
the measured cross-sections with NNLO QCD and NLO EW Drell-Yan predictions 
show good agreement within uncertainties. Figure~\ref{fig:CMS-WZvsSQRTs} shows the 
energy dependence of the measured inclusive $W$ and $Z$ boson cross sections 
compared to theoretical predictions. LHCb has measured the 13\;TeV $Z$ boson cross 
section in the fiducial acceptance $2.0 < \eta < 4.5$ and found agreement with the SM
prediction~\cite{LHCb-ZXS}. 

Ratios of cross sections already achieve precision of better than 1--2\% owing to a
cancellation of systematic uncertainties. They represent powerful tools to constrain 
PDFs: the $W^+ / W^–$ ratio is sensitive to the low-$x$ $u$ and  $d$ valence quarks, 
and the $W^\pm / Z$ ratio constrains the strange quark PDF, in particular when also 
using the rapidity distributions. Figure~\ref{fig:ATLAS-WZratios} shows 
the measured and predicted ratios. Fair agreement between the data and most
PDF sets is seen. An increased strange quark contribution~\cite{ATLAS-WZstrange} 
(towards SU(3) flavour symmetry of sea squarks in the proton) would likely improve 
the agreement. The right panel in Fig.~\ref{fig:ATLAS-WZratios} 
shows tests of the universality of the first and second generation leptonic 
couplings to the weak bosons.  Lepton universality in the charged 
current was measured to the  0.14\% level at LEP in $\tau$ lepton decays, however 
at low energy (off-shell), so with less sensitivity to new physics in loops.
\begin{figure}[t]
\centerline{\includegraphics[width=1\linewidth]{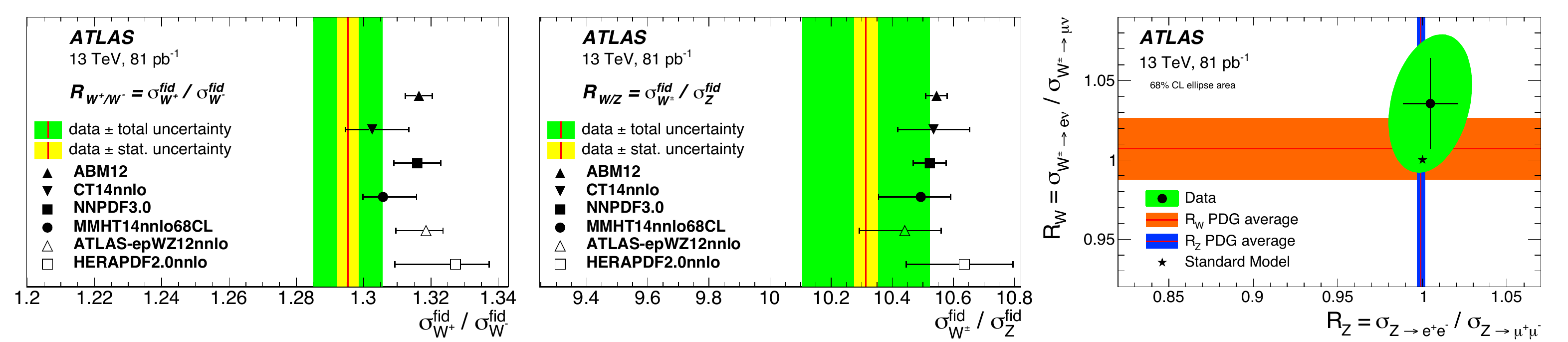}}
\vspace{0.0cm}
\caption[.]{Ratios of fiducial cross sections compared to various PDF predictions 
                (left and middle panels), and $W$ and $Z$ lepton universality tests 
                (right panel)~\cite{ATLAS-WZ}. 
                \label{fig:ATLAS-WZratios}}
\end{figure}

\begin{figure}[t]
\centerline{\includegraphics[width=1\linewidth]{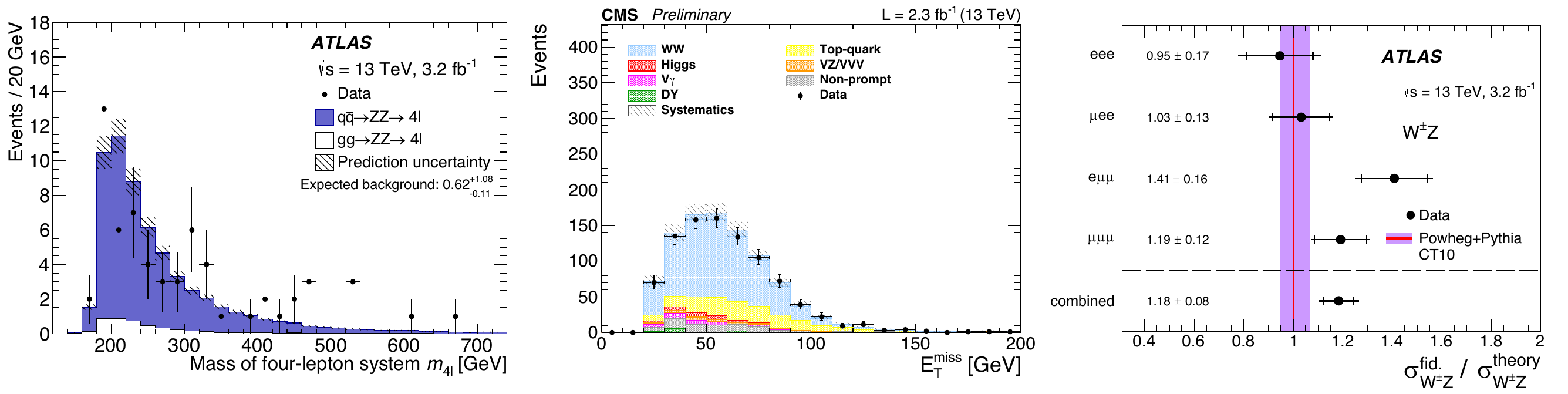}}
\vspace{0.0cm}
\caption[.]{Distributions from 13 TeV diboson selections. Left: four-lepton invariant mass
                in the $ZZ\to4\ell$ analysis~\cite{ATLAS-ZZ};  
                middle: missing transverse momentum in $WW\to 2\ell2\nu$~\cite{CMS-WW}; 
                right: ratio of measured over predicted (NLO QCD) 
                fiducial cross sections measured in $WZ\to 3\ell\nu$~\cite{ATLAS-WZ}.
                \label{fig:ATLASCMS-diboson}}
\end{figure}

\vfill\pagebreak
\subsubsection{Diboson production}

\begin{wrapfigure}{R}{0.38\textwidth}
\centering
\vspace{-0.44cm}
\includegraphics[width=0.38\textwidth]{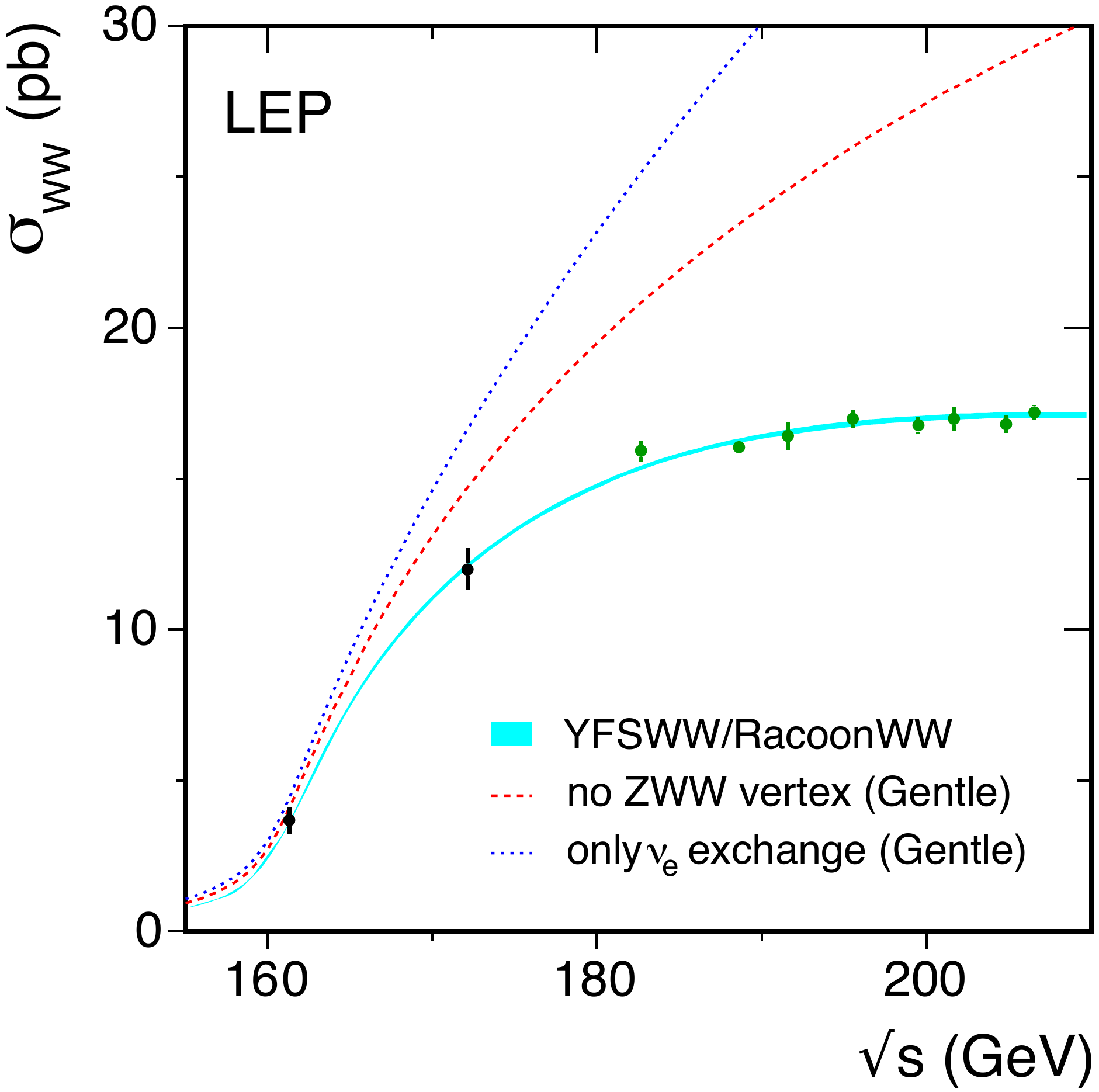}
\caption[.]{$W$ pair production cross section measured at LEP compared to the SM 
                 predictions~\cite{LEP-PhysicsReport}. 
                \label{fig:LEP-WW}}
\vspace{-0.4cm}
\end{wrapfigure}
The production of  boson pairs is a highly important sector of LHC physics that is intimately 
related to electroweak symmetry breaking. In the s-channel (Drell-Yan), via photon, $Z$ or $W$ 
exchange, diboson production is sensitive to anomalous triple gauge boson couplings (aTGC).
Triple and also quartic gauge boson couplings, the latter vertices involving the scattering among 
four gauge bosons, are predicted by the SM as the electroweak gauge bosons carry weak 
charge (non-Abelian structure of EW theory).

The production of $WW$ and $ZZ$ events was studied at LEP versus the $e^+e^-$ centre-of-mass energy
resulting in a famous plot that showed the moderation of the $WW$ cross section versus energy by 
TGC processes as predicted by the SM (see Fig.~\ref{fig:LEP-WW})~\cite{LEP-PhysicsReport}. 
The Tevatron experiments studied a multitude of 
diboson production processes. ATLAS and CMS performed inclusive, fiducial and differential 
cross-section analyses at 8 TeV. First fiducial and total cross section measurements at 13 TeV 
are also available (see Fig.~\ref{fig:ATLASCMS-diboson} for a selection of representative plots). 
Inclusive diboson $WW$, $WZ$ and $ZZ$ events are reconstructed through 
the leptonic decays of the weak bosons, leading to two-lepton, three-lepton and four-lepton 
final states, where the former two channels are accompanied by $\MET$.
Hadronic weak boson decays are not competitive for inclusive cross section measurements,
but are interesting for aTGC searches at high diboson mass where possible new physics effects 
are expected to show up first (see~\cite{Diboson-review} for a recent review).
These results show that NNLO QCD is needed to match the data, and, in case of $WW\to2\ell 2\nu$
measured in an exclusive zero-jet channel, higher-logarithmic-order (NNLL) soft gluon 
resummation.

\subsubsection{Top production}

\begin{figure}[t]
\centerline{\includegraphics[width=0.65\linewidth]{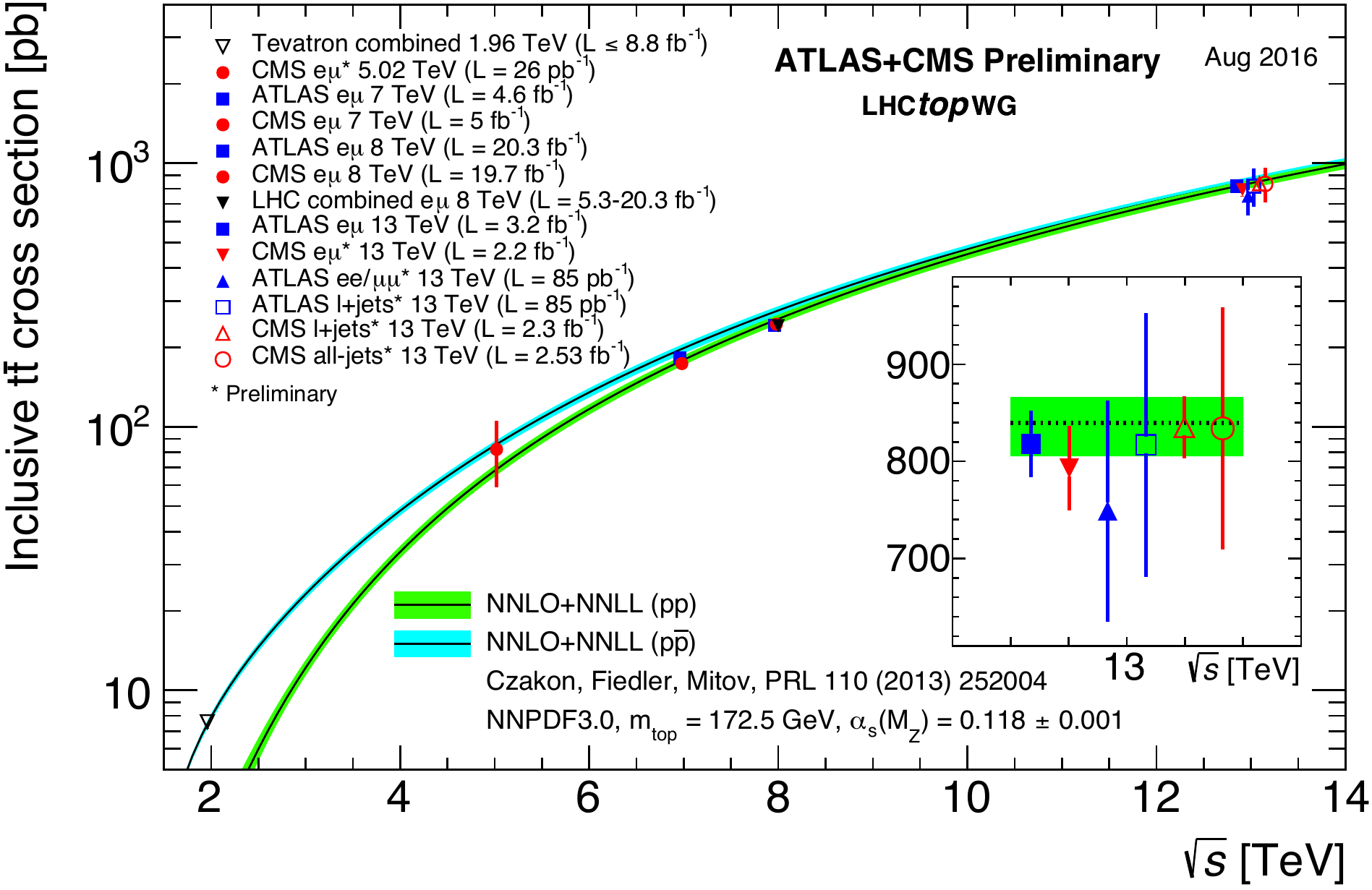}}
\vspace{0.0cm}
\caption[.]{Summary of LHC and Tevatron measurements of the inclusive top-pair production 
                cross section versus  centre-of-mass energy. The bands correspond to 
                predictions with uncertainties from NNLO QCD calculations including NNLL soft 
                gluon resummation. Measurements and theory calculations assume
                $m_t=172.5$\;GeV.
                \label{fig:ttXS}}
\end{figure}
Top--antitop production at the LHC is dominated by gluon--gluon scattering in the initial state.
A factor of 3.3 cross-section increase at 13\;TeV centre-of-mass energy compared
to 8\;TeV is expected. The inclusive $pp\to \ttbar+X$ cross section can be robustly measured 
using dilepton events selecting different lepton flavours to suppress Drell-Yan background. 
A method applied successfully during Run-1 allows to simultaneously determine the $\ttbar$
cross section and $b$-tagging efficiency from data~\cite{ATLAS-ttXSemu-Run1}. We shall briefly
discuss it here for the corresponding 13\;TeV measurement 
as it is an instructive example for a straightforward experimental
approach relying where possible on data. 

The method employs an exclusive selection of $e\mu$ events 
with one and two $b$-tags. The observed number of events is given by
$N_1=L\cdot\sigma_{\ttbar}\cdot\e_{e\mu}\cdot 2\e_b\cdot(1-C_b\cdot\e_b)+N_1^{\rm bkg}$ and
$N_2=L\cdot\sigma_{\ttbar}\cdot\e_{e\mu}\cdot C_b\cdot\e_b^2+N_2^{\rm bkg}$, where
$N_{1(2)}$ is the number of observed events with one (two) $b$-tags, $L$ the integrated 
luminosity of the analysed data sample, $\e_{e\mu}$ the combined $\ttbar\to e\mu+X$ 
selection acceptance and efficiency determined from MC, $\e_b$ the probability to $b$-tag 
$q$ from $t \to Wq$ determined from data ($\e_b$ includes the selection acceptance 
and efficiency), and $C_b=\e_{bb}/\e_b^2$ is a small non-factorisation correction 
($1.002\pm0.006$) determined from MC. The selection of $\ttbar\to e\mu+X$ events
is very pure with, for the ATLAS 2015 dataset (3.2\;\ifb), $N_1=11\,958$, $N_2=7\,069$, and
$N_1^{\rm bkg}=1\,370\pm120$, $N_2^{\rm bkg}=340\pm88$ event counts~\cite{ATLAS-ttXS}. 
The background is dominated by the $Wt$ single-top process. 
Solving the equations simultaneously for $\sigma_{\ttbar}$ and $\e_b$ gives:
$\sigma_{\ttbar}=828\pm8\pm27\pm19\pm12$\;pb, where the first error is statistical,
the second systematic, and the third (fourth) due to the luminosity (beam-energy) 
uncertainty.\footnote{The LHC beam energy during the 2012 proton--proton run was calibrated 
to be $0.30 \pm0.66$\,\% below the nominal value of 4\;TeV per beam. That estimate,
dominated by systematic uncertainties, was made 
by measuring the revolution frequency, that is, the speed difference of protons and 
lead ions during proton-lead runs in early 2013~\cite{EbeamUncertainty}, taking advantage 
of the simultaneous presence of both particle types with the same orbits in the LHC. The 
measurement result agrees with the beam energy derived from the magnetic calibration 
curves of the dipole magnets that are used to generate the current settings of the power 
converters which feed the magnets during beam operation. The magnetic calibration is expected 
to be accurate within an uncertainty of about 0.07\%, which is significantly better than 
the measurement uncertainty based on proton--lead data.}
The total relative uncertainty of 4.4\% is already comparable with the 4.3\% obtained at 8 TeV.
The measurement is in agreement with the theoretical prediction of $832^{\,+40}_{\,-46}$\;pb, based 
on NNLO QCD including NNLL soft gluon resummation and of similar precision as the measurement. 
The systematic uncertainty affecting the measurement (total 3.3\%) is dominated by theoretical 
sources in particular the modelling of nonperturbative effects related to parton showering and 
hadronisation. It is interesting (though not mandatory for the method to work) to observe that 
the resulting value for $\e_b=0.559 \pm 0.004 \pm 0.003$ is in agreement with the value of 0.549 
found in MC simulation. 
Figure~\ref{fig:ttXS} shows a summary of various  LHC and Tevatron $\ttbar$
cross section measurements versus centre-of-mass energy, and compared to theoretical predictions. 
The $e\mu$ method provides the most precise inclusive results at all LHC centre-of-mass energies. 

The experiments also performed first differential cross section measurements at 13\;TeV which 
show reasonable modelling although deviations at large jet multiplicity and top $p_T$ persist, 
similar to those seen in Run-1. 

\begin{wrapfigure}{R}{0.60\textwidth}
\centering
\vspace{-0.40cm}
\includegraphics[width=0.60\textwidth]{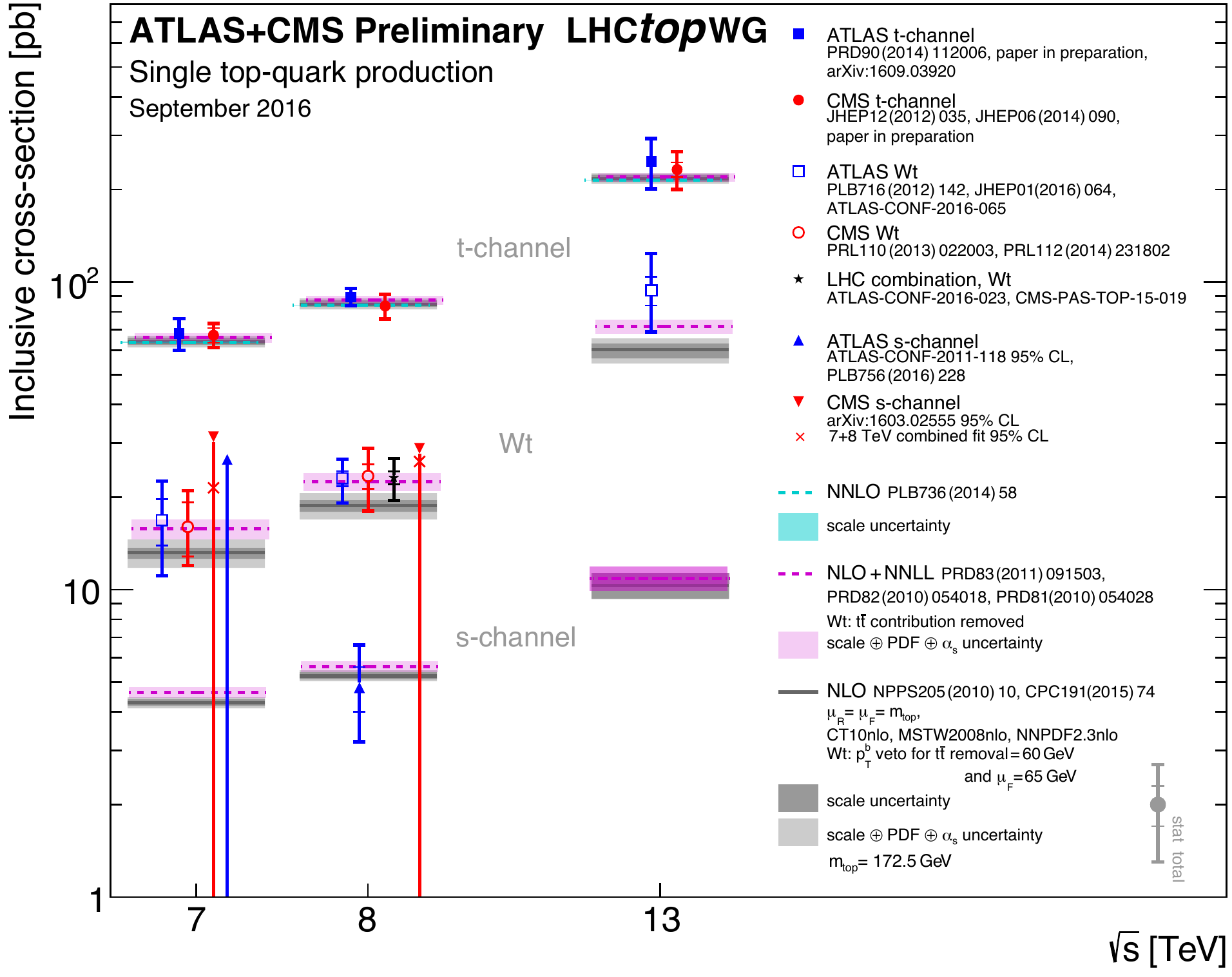}
\caption[.]{Summary of ATLAS and CMS measurements of single-top production cross sections 
                versus centre-of-mass energy. The measurements are compared to theoretical 
                calculations (see text for details).
                \label{fig:singletop-summary}}
\vspace{-0.4cm}
\end{wrapfigure}
Electroweak single-top production (cf. Feynman graphs in Fig.~\ref{fig:single-top}) amounts,
in case for the dominant t-channel, to about one third of the $\ttbar$ production cross section
and is expected to increase by a factor of 2.5 at 13\;TeV compared to 8\;TeV. A summary of 
the inclusive cross section measurements for all available centre-of-mass energies is displayed
in Fig.~\ref{fig:singletop-summary} (see~\cite{ATLAS-top-tchan,CMS-top-tchan,ATLAS-top-Wt} 
for the 13\;TeV results). Agreement with theoretical predictions based on 
pure NLO QCD, NLO QCD complemented with NNLL resummation, and NNLO QCD (t-channel only)
is observed. In the t-channel, also the charge asymmetries are measured at 13\;TeV and found 
in agreement with the SM prediction (the ratio of $t q$ to $\overline t q$ production is measured
to be $1.72\pm0.20$). The s-channel is challenging at the LHC and requires more data. 

The opening of the phase space at 13\;TeV allows to produce heavier final states, such as 
the associated production of $\ttbar$ with a $W$ or $Z$ boson (cf. Feynman graphs 
in Fig.~\ref{fig:ttV-Feyn} on page~\pageref{fig:ttV-Feyn}). 
Because of different production mechanisms (dominantly gluon s-channel 
scattering in case of $ttZ$ and t-channel quark--antiquark annihilation for $ttW$) the 
13\;TeV to 8\;TeV cross-section ratios are  different for the two channels: 3.6 for $ttZ$
compared to only 2.4 for $ttW$. Both experiments have produced first inclusive 13\;TeV
ccross section results~\cite{ATLAS-ttV,CMS-ttV} 
finding for $ttW$: $1.5\pm0.8$\;pb (ATLAS) and $0.98^{\,+0.32}_{\,-0.28}$\;pb (CMS),
and for $ttZ$: $0.9\pm0.3$\;pb (ATLAS) and $0.70^{\,+0.21}_{\,-0.19}$\;pb (CMS). The 
corresponding SM predictions are $0.60 \pm 0.08$\;pb ($ttW$) and  $0.84 \pm 0.09$\;pb 
($ttZ$). Both, ATLAS and CMS are slightly on the high side for $ttW$ reproducing
a similar pattern already observed in the Run-1 data. Improving these results is important 
in its own rights, but also because the $ttW/Z$ channels are important backgrounds to 
$ttH$ in final states with multiple leptons, where in particular $ttW$ is difficult to separate.

\pagebreak
\subsection{Reobservation of the Higgs boson at 13\;TeV}

The expected 13\;TeV to 8\;TeV cross section ratios amount to 2$\sim$2.4 for $VH$, $ggH$, 
VBF, and 3.9 for $ttH$ production. The combination of the 2015 and 2016 data available
by the 2016 summer conferences should therefore already achieve similar or better significance 
and precision on Higgs boson production than in Run-1.

\begin{wrapfigure}{R}{0.55\textwidth}
\centering
\vspace{-0.42cm}
\includegraphics[width=0.55\textwidth]{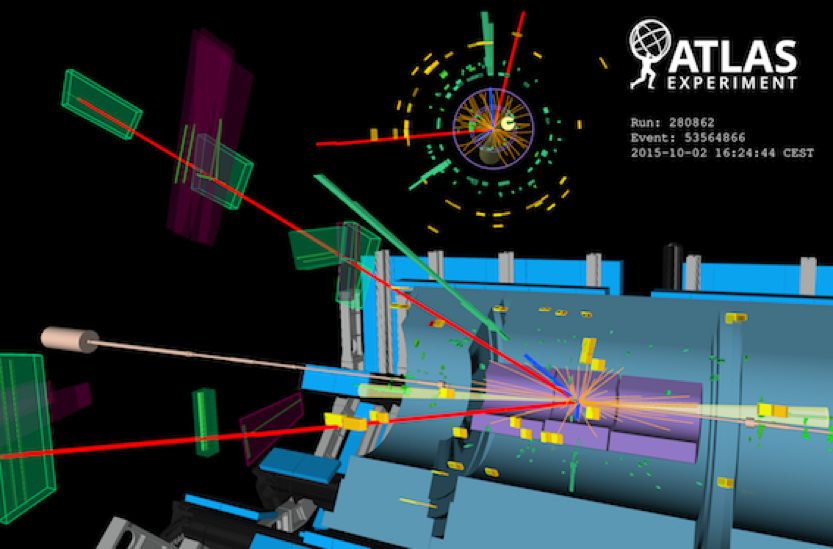}
\caption[.]{Display of $H \to ee\mu\mu$ candidate from 13\;TeV pp collisions measured by ATLAS. 
                 The event is accompanied by two forward jets with pseudorapidity difference of 6.4 
                 and invariant dijet mass of 2\;TeV. This event is consistent with VBF production of 
                 a Higgs boson decaying to four leptons.
                \label{fig:ATLASHiggsEvent}}
\vspace{-0.2cm}
\end{wrapfigure}
Figure~\ref{fig:ATLASHiggsEvent} shows a rare and beautiful VBF $H\to4\ell$ candidate event. 
Such an event has large signal to background probability. 
Preliminary results for the cleanest bosonic channels $H\to4\ell$ and $H\to\gamma\gamma$ 
were released by  ATLAS and CMS for the 2016 summer 
conferences~\cite{ATLAS-H4l,CMS-H4l,ATLAS-Hyy,CMS-Hyy}. The Higgs boson was 
reobserved with high significance at the expected mass in both channels by either experiment
(cf. Fig.~\ref{fig:ATLASCMS-H4lyy} for the corresponding diphoton and four-lepton mass spectra).
The extracted inclusive cross sections have still large uncertainties and are found in agreement 
with the SM expectations. In the four-lepton channel, ATLAS found a cross section of 
$81^{\,+18}_{\,-16}$\;pb compared to $55 \pm 4$\;pb expected. CMS measured a signal strength
of $\mu_{H\to4\ell}=0.99^{\,+0.33}_{\,-0.26}$\;pb. CMS also measured the mass to be 
$m_H=124.50^{\,+0.48}_{\,-0.44}$\;GeV in agreement with the 
Run-1 ATLAS and CMS combined value of $125.09 \pm 0.24$\;GeV. In the diphoton channel
signal strengths of $\mu_{H\to\gamma\gamma}=0.85^{\,+0.22}_{\,-0.20}$ and $0.91\pm0.21$ are
measured by ATLAS and CMS, respectively. The measured cross sections versus centre-of-mass
energy are shown in Fig.~\ref{fig:ATLASCMS-HXS} for the inclusive cases (ATLAS, left) and 
fiducial measurements (CMS, right). 
\begin{figure}[t]
\centerline{\includegraphics[width=1\linewidth]{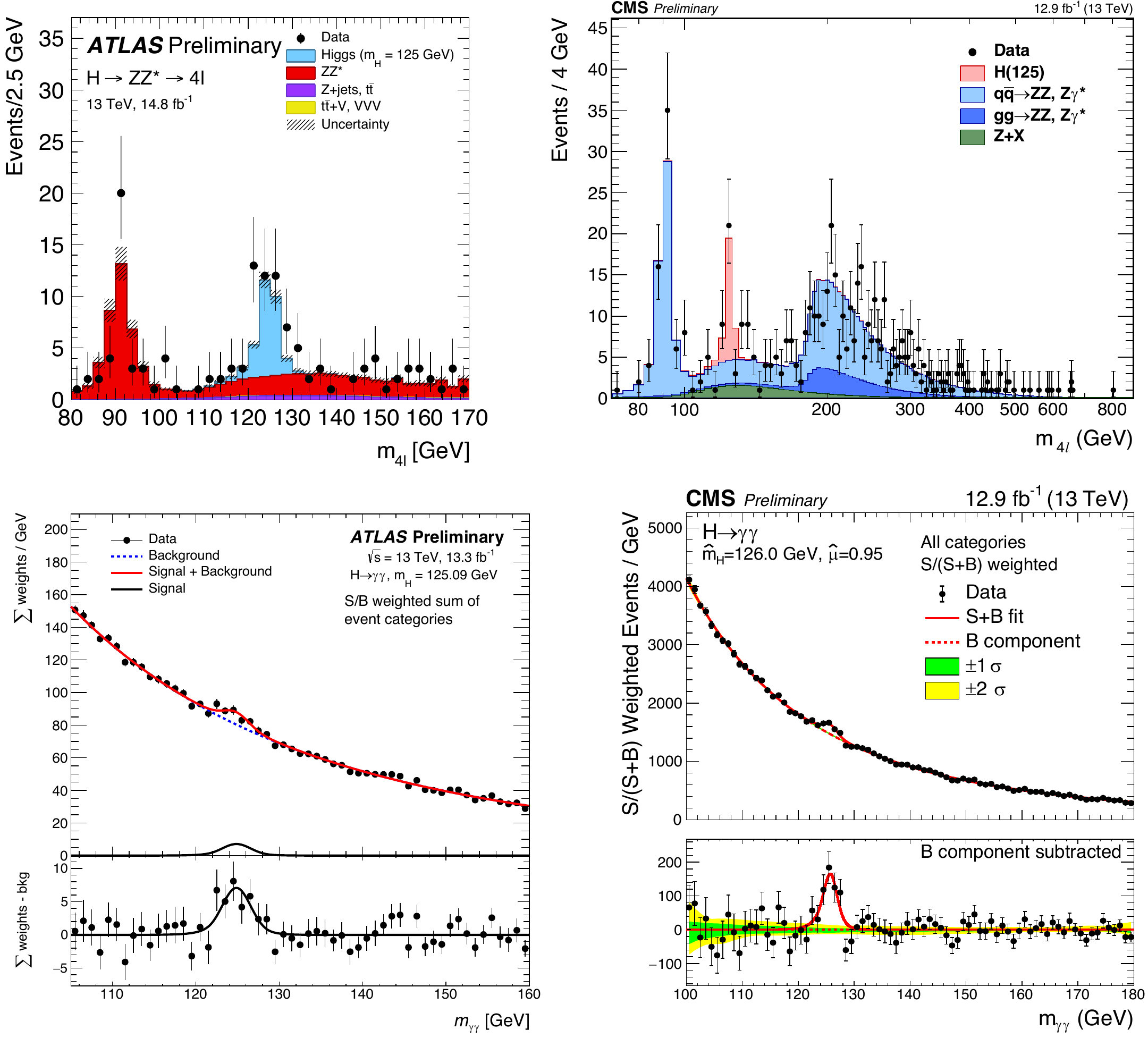}}
\vspace{0.0cm}
\caption[.]{Four-lepton (top row) and diphoton (bottom row) invariant mass distributions 
                for ATLAS (left column) and CMS (right column) for the combined 2015 and 2016
                datasets (ATLAS) and 2016 dataset (CMS) taken at 13\;TeV proton--proton 
                centre-of-mass energy~\cite{ATLAS-H4l,CMS-H4l,ATLAS-Hyy,CMS-Hyy}.
                The bottom plots show each event weighted by the 
                signal-to-background ratio of the event category it belongs to.
                \label{fig:ATLASCMS-H4lyy}}
\end{figure}
\begin{figure}[t]
\centerline{\includegraphics[width=1\linewidth]{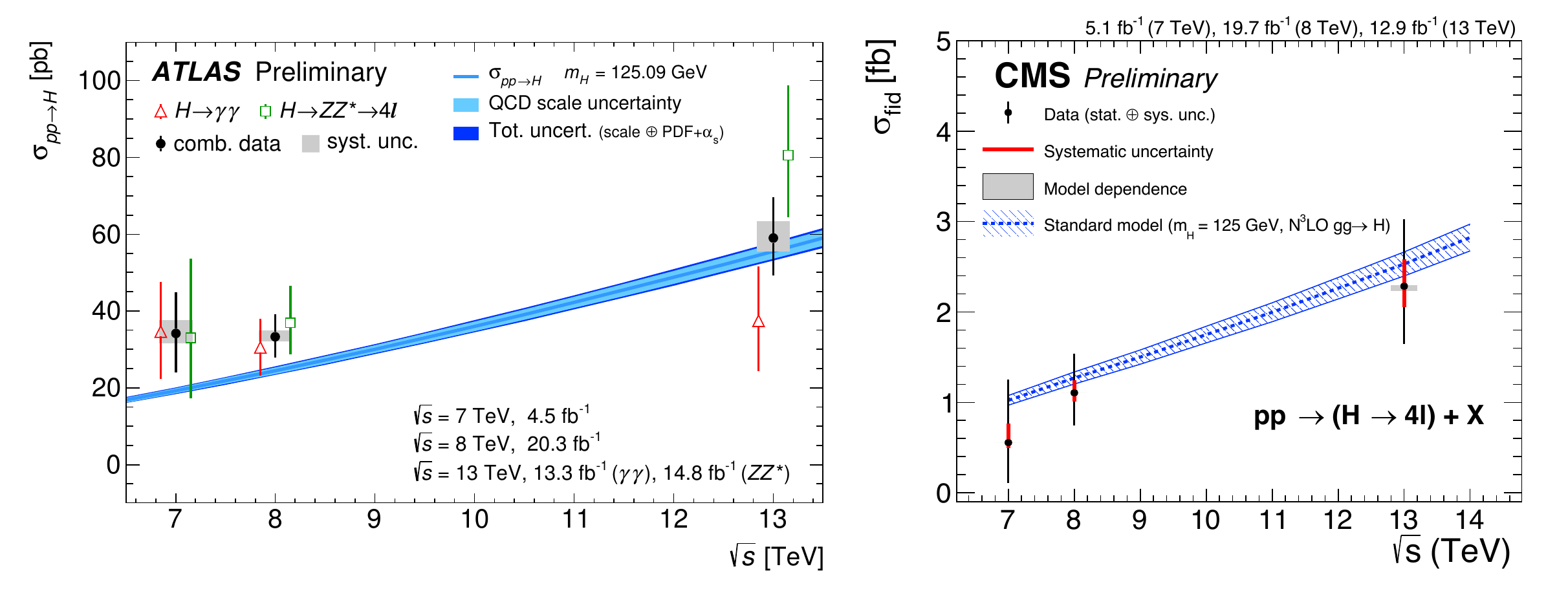}}
\vspace{0.0cm}
\caption[.]{Total (ATLAS, left) and fiducial (CMS, right) $pp \to H + X$ cross sections measured 
                at different centre-of-mass energies and compared to SM predictions at up to 3NLO 
                in QCD. The left plot shows the individual results of the $4\ell$ and $\gamma\gamma$ 
                channels and their combination~\cite{ATLAS-H4lyycomb}. 
                The right plot shows the $4\ell$ fiducial measurements. Agreement with the SM predictions
                is observed within yet large uncertainties. 
                \label{fig:ATLASCMS-HXS}}
\end{figure}
ATLAS has combined the $4\ell$ and $\gamma\gamma$ results to perform a coupling 
analysis~\cite{ATLAS-H4lyycomb}. 
The combined inclusive cross section is also shown on the left panel of Fig.~\ref{fig:ATLASCMS-HXS}.
The experiments also measured differential cross sections that are compared to NNLO plus parton 
shower predictions. No deviation from the SM prediction is found within yet large uncertainties. 

ATLAS also released first preliminary studies of associated $VH$ production with the 
decay $H\to bb$~\cite{ATLAS-VHbb}. 
The channel is very challenging due to large backgrounds that need to be controlled with high precision 
in order to extract the signal. Run-1 had provided a signal strength slightly below the SM 
expectation~\cite{ATLASCMSHcouplings-Run1,ATLAS-VHbb-Run1,CMS-VHbb-Run1}. 
The Run-2 yield was again low with $\mu_{VH(\to bb)}= 0.21^{\,+0.51}_{\,-0.50}$
after combining the zero, one, and two charged lepton final states covering the $ZH$ and $WH$
modes. 

ATLAS also looked into inclusive production of $H\to\mu\mu$~\cite{ATLAS-Hmumu} 
that has an expected branching fraction 
of 0.02\%, but might be enhanced due to new physics effects. The sensitivity to that decay  depends 
primarily on the dimuon mass resolution. The sensitivity can be improved, similarly to that of
$H\to\gamma\gamma$, by splitting the event sample into categories with different 
mass resolution and/or signal signal-to-background ratios, such as low versus high $p_T$, central versus 
forward muons, ggF versus VBF, etc. The observed 95\% confidence level limit for $H\to\mu\mu$
is found at 4.4 times the SM prediction, reducing to 3.5 when combined with Run-1. 
About 300\ifb are needed to reach SM sensitivity. These results  allow to exclude a universal 
Higgs coupling to fermions, as $H\to\mu\mu$ would have been 
observed had it the same branching fraction as $H\to\tau\tau$.

The Higgs  production mode that most benefits from the increased centre-of-mass energy 
is $ttH$ that was found a bit enhanced compared to the SM prediction in the Run-1 
Higgs couplings combination (cf. Fig.~\ref{fig:ATLASCMSHiggsCouplings-Run1}, 
page~\pageref{fig:ATLASCMSHiggsCouplings-Run1}). The motivation was thus large to look for 
that mode in Run-2. 

The associated production of $ttH$ is the only currently accessible 
channel that directly measures the top--Higgs coupling (cf. Feynman graph
in Fig.~\ref{fig:HiggsProdAndBRs} on page~\pageref{fig:HiggsProdAndBRs}). 
All major Higgs  decay channels, $\gamma\gamma$, 
multileptons, and $bb$, are analysed, where in particular the latter two channels represent
highly complex analyses. The multilepton mode targets Higgs  decays to 
$\tau\tau$, $WW\to2\ell2\nu$, and $ZZ\to 2\ell2\nu,\,4\ell$ together with at 
least one top quark decaying leptonically. It requires at 
least two leptons with the same charge, which greatly reduces SM 
backgrounds. The dominant remaining backgrounds are misidentified prompt leptons
and $ttV$ production in particular the difficult to separate $ttW$ (cf. the right panel 
in Fig.~\ref{fig:ATLASCMS-ttH} for the distribution of a boosted decision tree trained
to distinguish $ttW$ background from $ttH$ signal). The $H\to bb$ mode is analysed 
in the one and two lepton channels. Here the biggest challenge represents background 
due to $t\overline t$ production associated 
with heavy flavour quarks ($c$ or $b$) originating mostly from  gluon splitting, which 
is poorly known and needs to be constrained from data simultaneously with the 
signal. CMS released preliminary 13\;TeV results for $ttH$ in all three Higgs decay 
categories finding for 
the relative signal strengths~\cite{CMS-ttHyy,CMS-ttHmultilep,CMS-ttHbb}: 
$\mu_{ttH(\to\,\gamma\gamma)} = 3.8^{\,+4.5}_{\,-3.6}$, 
$\mu_{ttH(\to\,{\rm leptons})} = 2.0^{\,+0.8}_{\,-0.7}$, and
$\mu_{ttH(\to,bb)} = -2.0\pm1.8$, with no significant excess observed. ATLAS also 
measured all three channels~\cite{ATLAS-ttH} 
and their statistical combination~\cite{ATLAS-ttHcomb} that is shown in the left panel
of Fig.~\ref{fig:ATLASCMS-ttH}. The combined preliminary signal strength is 
$\mu_{ttH}=1.7\pm0.8$.
\begin{figure}[t]
\centerline{\includegraphics[width=1\linewidth]{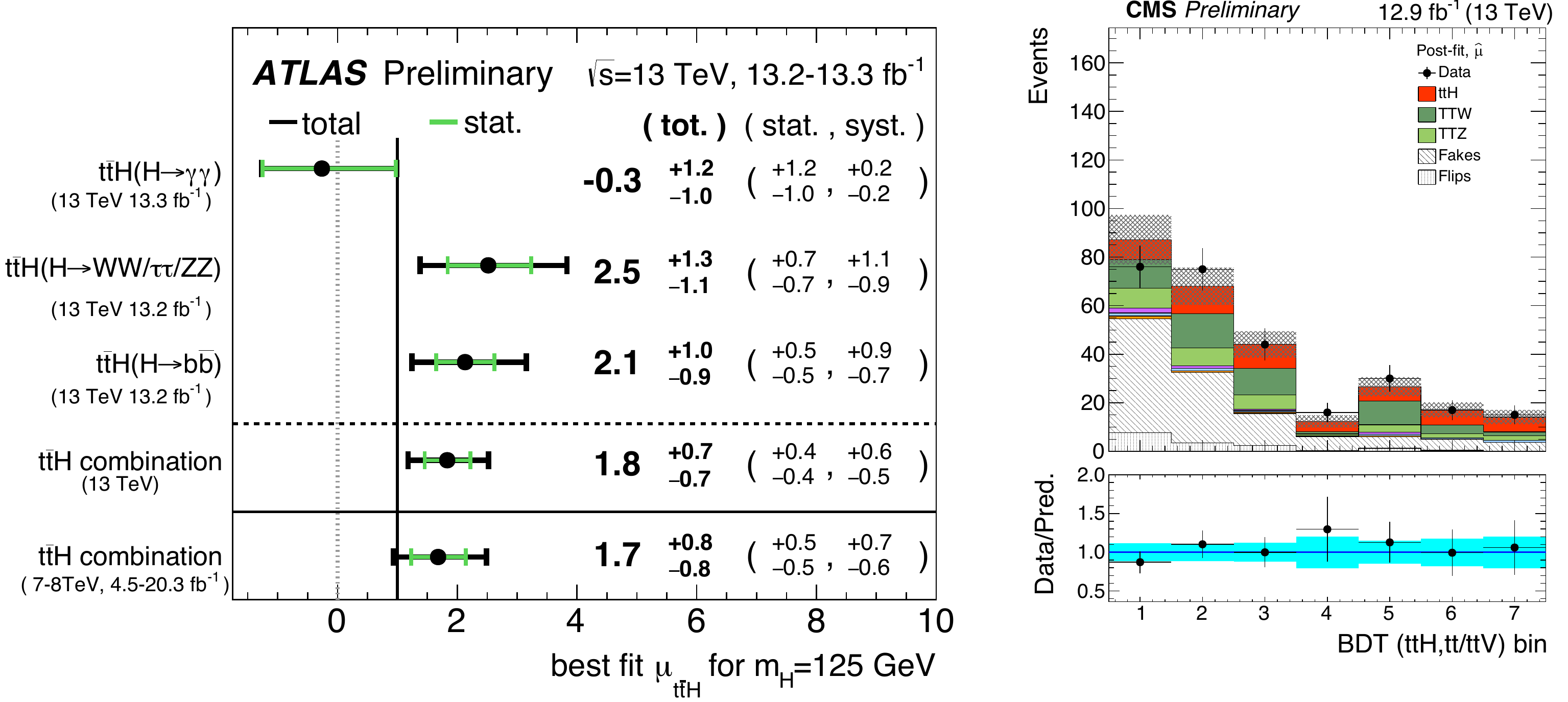}}
\vspace{0.0cm}
\caption[.]{Left: ATLAS summary of the  $\mu_{ttH}$ signal strength measurements from the 
                 individual analyses and  their combination, assuming $m_H = 125$\;GeV~\cite{ATLAS-ttHcomb}. 
                 Right: boosted decision tree output from CMS in the same-charge channel trained to 
                 separate $ttW$ background from $ttH$ signal~\cite{CMS-ttHmultilep}. 
                \label{fig:ATLASCMS-ttH}}
\end{figure}

\vfill\pagebreak

\subsection{Searches --- a fresh start}

\begin{wrapfigure}{R}{0.5\textwidth}
\centering
\vspace{-0.4cm}
\includegraphics[width=0.5\textwidth]{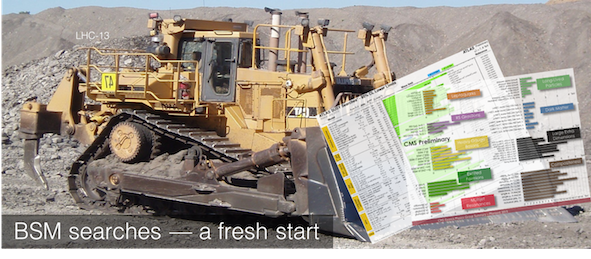}
\vspace{-0.62cm}
\caption[.]{The bulldozer (aka, LHC at 13\;TeV) moving out of the way 
                the Run-1 limits on beyond the SM searches. 
                \label{fig:BSMFreshStart}}
\end{wrapfigure}
Many of the high mass and higher cross section searches for new physics already
benefited from the 2015 13$\;$TeV data sample to extend their sensitivity, and all searches
surpass their Run-1 limits with the 2016 datasets (see Fig.~\ref{fig:BSMFreshStart}). 
Run-2 represents thus a fresh start in the quest for new physics after the negative 
searches from Run-1. The legacy of Run-1 also contained a small number of anomalies
that needed to be verified in the Run-2 data. Only 13$\;$TeV searches are
discussed in the following. 

\subsubsection{Additional Higgs bosons}

\begin{figure}[t]
\centerline{\includegraphics[width=0.95\linewidth]{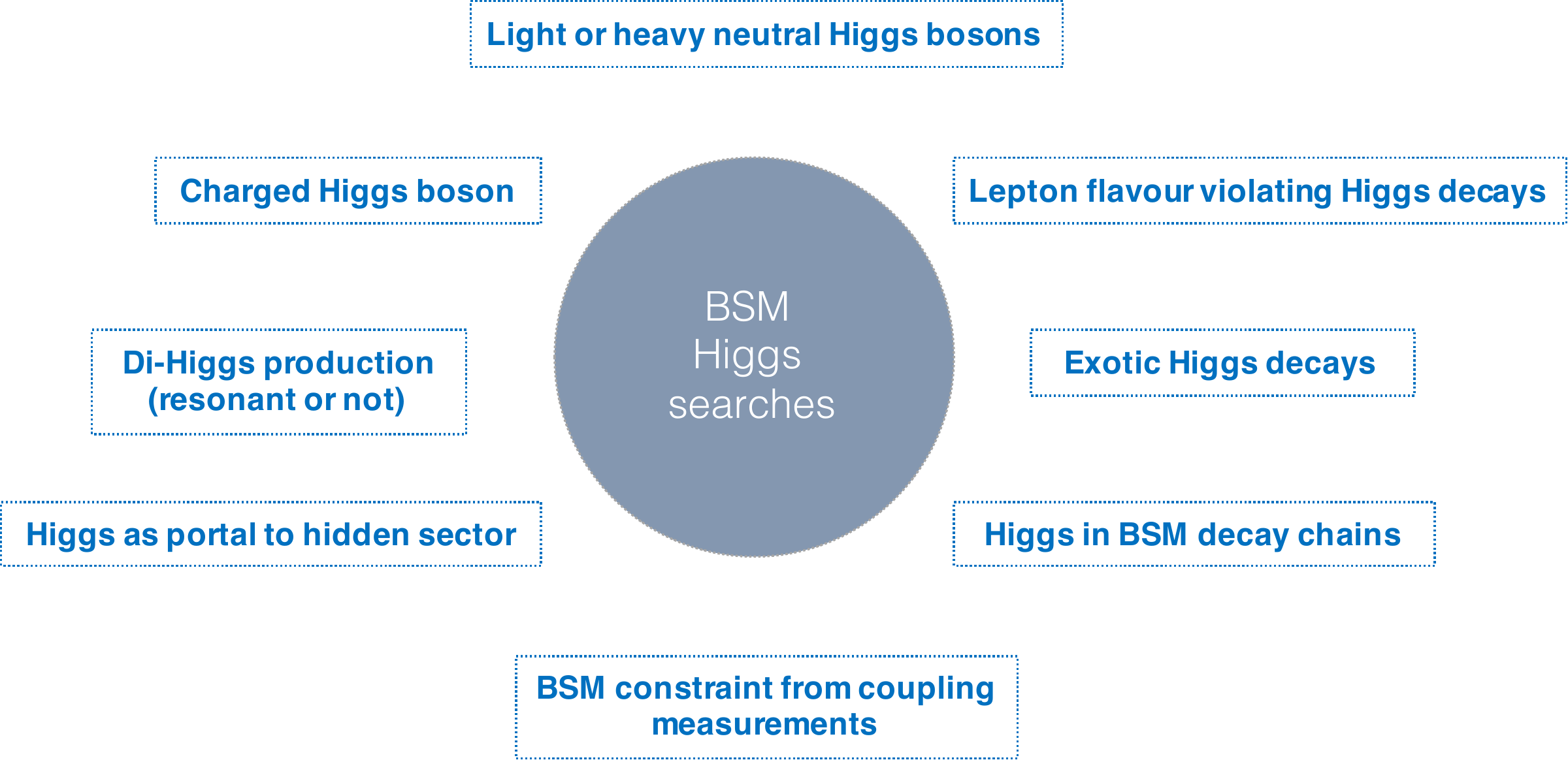}}
\vspace{0.2cm}
\caption[.]{Illustration of beyond the SM Higgs boson search areas. 
                \label{fig:BSMHiggs}}
\end{figure}
The  125$\;$GeV Higgs boson completes the four degrees of freedom of
the SM BEH doublet. Nature may have, however,  chosen a more complex scalar
sector of, eg., two BEH doublets, which extends the sector by four additional
Higgs bosons, of which two are neutral (one $CP$-even and one $CP$-odd) and 
the other two are charged. Searching for ancillary scalar bosons is thus one way 
to detect BSM physics in the scalar sector. Other ways are to look for non-SM decays
of the  Higgs boson such as decays to invisible particles where the Higgs boson
acts as a portal to new physics responsible for dark matter. The Higgs boson could
also be produced as a particle in the decay chain of new physics processes such as 
supersymmetry. New heavy resonances might decay to a pair of Higgs boson. A 
summary of BSM options around the Higgs boson is sketched in Fig.~\ref{fig:BSMHiggs}. 

\begin{figure}[t]
\centerline{\includegraphics[width=1\linewidth]{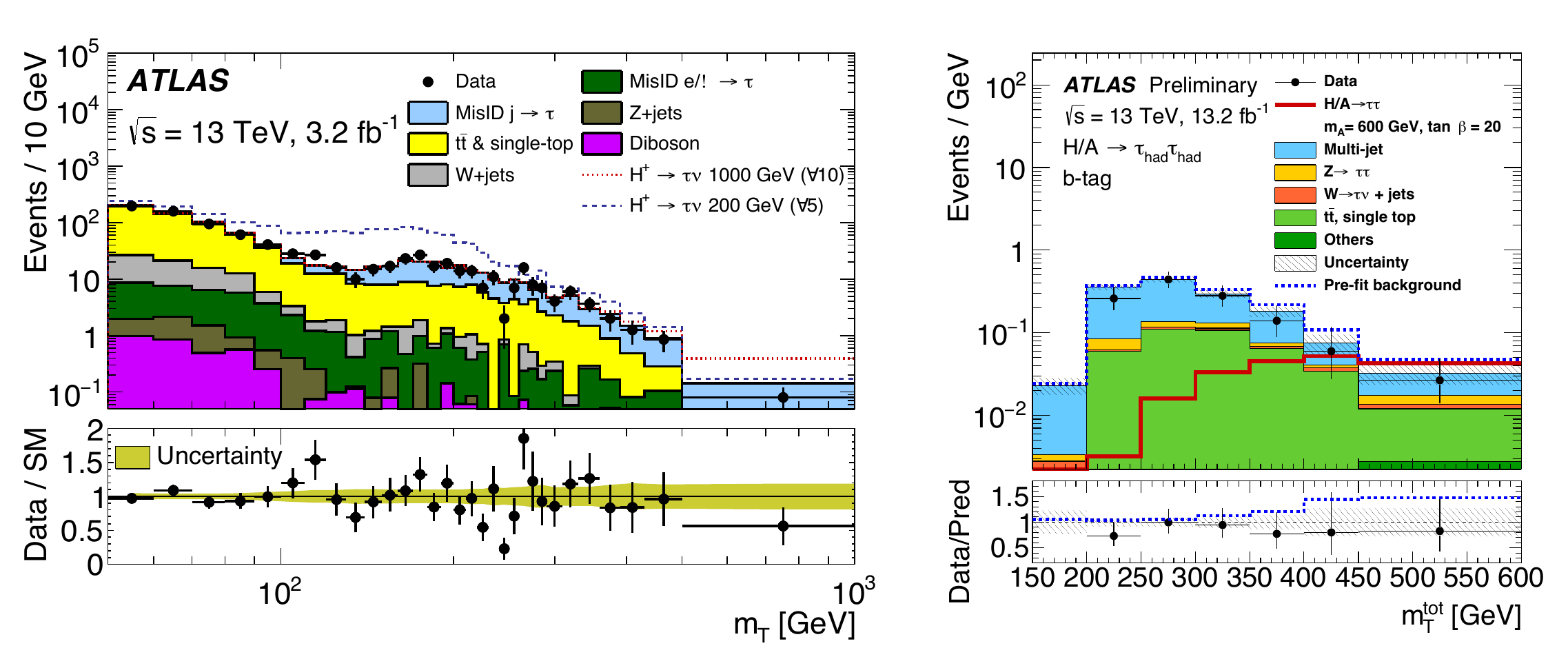}}
\vspace{0.0cm}
\caption[.]{Left: transverse mass in a search for a charged Higgs boson decaying 
                to $\tau\nu$~\cite{ATLAS-chargedH,CMS-chargedH}. Right: distribution of the 
                reconstructed transverse mass in a search for a heavy neutral Higgs boson
                decaying to a tau$\tau$ pair, where both $\tau$ leptons are reconstructed 
                via their hadronic decay modes~\cite{ATLASHtautau}. 
                \label{fig:ATLAS-BSMHiggs}}
\end{figure}
ATLAS and CMS have searched for additional Higgs bosons in Run-1 and Run-2. 
For $H^\pm\to \tau\nu$~\cite{ATLAS-chargedH,CMS-chargedH} 
($H / A \to\tau\tau$~\cite{ATLASHtautau,CMSHtautau}), the sensitivity of the 
new data exceeds that of Run-1 for masses larger than 250$\;$GeV (700$\;$GeV). 
The search for $A \to Z(\to\ell\ell,\nu\nu) h_{125}(\to bb)$ features improved sensitivity 
beyond about 800$\;$GeV~\cite{ATLAS-AtoZh}. 
Searches for $H  \to ZZ(\to \ell\ell qq,\,\nu\nu qq,\,4\ell)$ 
and $WW (\to \ell\nu qq)$ target the $>1\;$TeV mass range where the bosons are 
boosted and their hadronic decays are  reconstructed with jet substructure 
techniques.
The search for a resonance decaying to $hh_{125} (\to bb\gamma\gamma)$ had a 
small excess in Run-1 at about 300$\;$GeV~\cite{ATLAS-HHbbyy-Run1}, which has been 
excluded at 13$\;$TeV~\cite{ATLAS-HHbbyy,CMS-HHbbyy}. 
Also performed were searches for resonant and non-resonant 
$H_{125}H_{125} \to bb\tau\tau,\,bbVV_{V=Z/W},\,bbbb$
production~\cite{CMS-HHbbtautau,CMS-HHbbVV,ATLAS-HHbbbb,CMS-HHbbbb}. 
None of these many searches exhibits an anomaly so far in the 13$\;$TeV data.

A slight Run-1 excess of 2.4$\sigma$ seen by CMS in the search for the lepton-flavour violating decay 
$H\to\tau\mu$~\cite{CMS-HLFV-Run1} was not seen by ATLAS in Run-1~\cite{ATLAS-HLFV-Run1}, 
and also not confirmed by CMS in an early Run-2 analysis~\cite{CMS-HLFV}.

\subsubsection{New physics searches in events with jets}

Among the first searches performed at any  increase of collision energy
are those for heavy strongly interacting new phenomena such as excited quarks
due to quark substructure, or strong gravity effects. The signatures investigated are
a dijet resonance and angular distributions, a resonance decaying to heavy-flavour 
quarks $X\to b\overline b$ or $t\overline t$~\cite{ATLAS-hfRes,CMS-hfRes},
high-$p_T$ multijet events, high-$p_T$  lepton plus jets events, 
and a lepton--jet resonance as could occur in presence of heavy leptoquarks. 
None of these searches exhibited an anomaly. 

Figure~\ref{fig:ATLAS-dijet} shows  dijet invariant mass spectra as measured by 
ATLAS~\cite{ATLAS-dijet,ATLAS-dijetISR} (see~\cite{CMS-dijet} for the corresponding
CMS analysis).
The left panel shows the high-mass tail as obtained with standard unprescaled jet triggers. 
The right panel shows lower mass events obtained with the use of a hard ISR jet trigger 
(see Feynman graph in right panel). The measured spectra are compared to phenomenological 
fits using smoothly falling functions as expected from the QCD continuum. No significant 
deviation from these fits is seen in the data. In addition to  the ISR ``trick'', it is possible 
to reach the low mass dijet regime with high statistics by using high-rate trigger-level 
objects of evens that are prescaled for offline analysis (this technique is denoted Data 
Scouting in CMS)~\cite{ATLAS-dijetTLA,CMS-dijetScout}. 
\begin{figure}[t]
\centerline{\includegraphics[width=\linewidth]{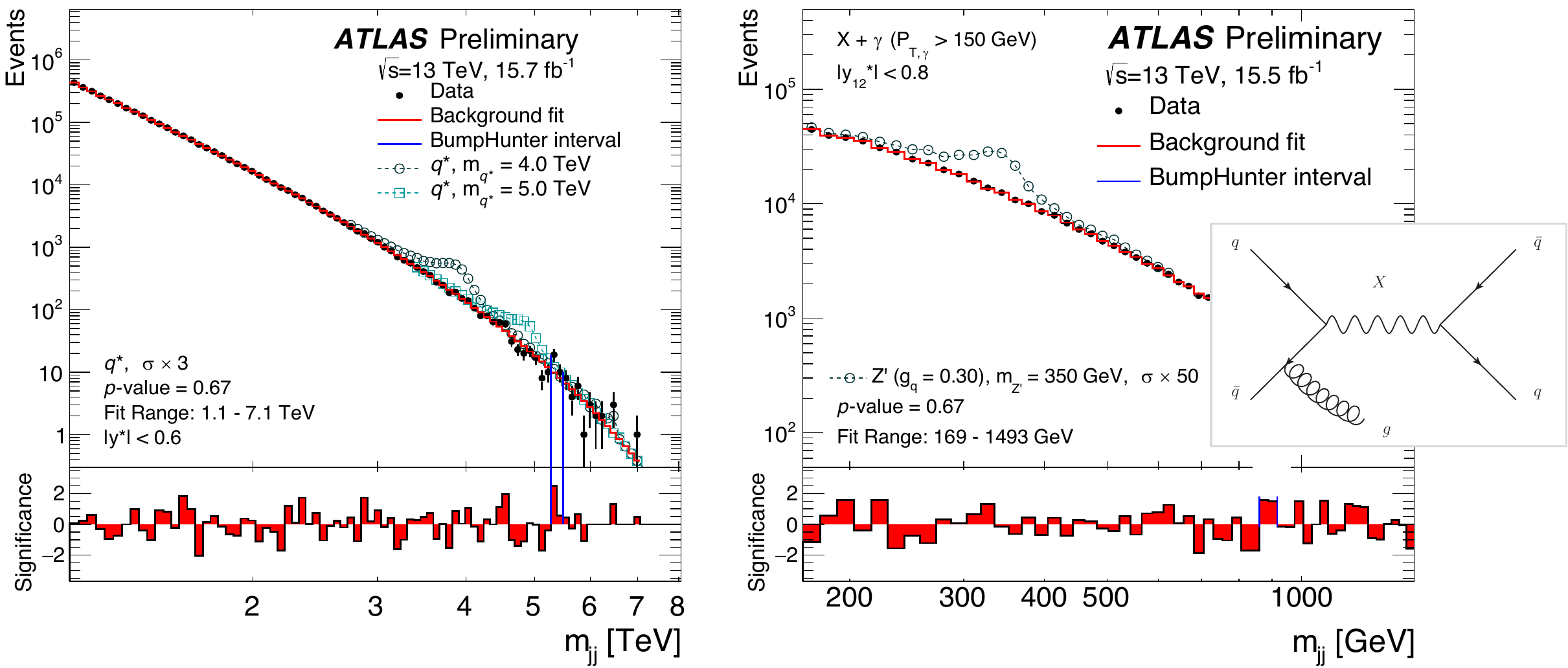}}
\vspace{-0.0cm}
\caption[.]{Dijet invariant mass distributions measured by ATLAS for the high-mass 
                resonance search~\cite{ATLAS-dijet}  
                (left panel) and the low-mass search (right) using 
                events with significant initial-state radiation (cf. Feynman graph in 
                panel)~\cite{ATLAS-dijetISR}.
\label{fig:ATLAS-dijet}}
\vspace{0.3cm}
\centerline{\includegraphics[width=0.6\linewidth]{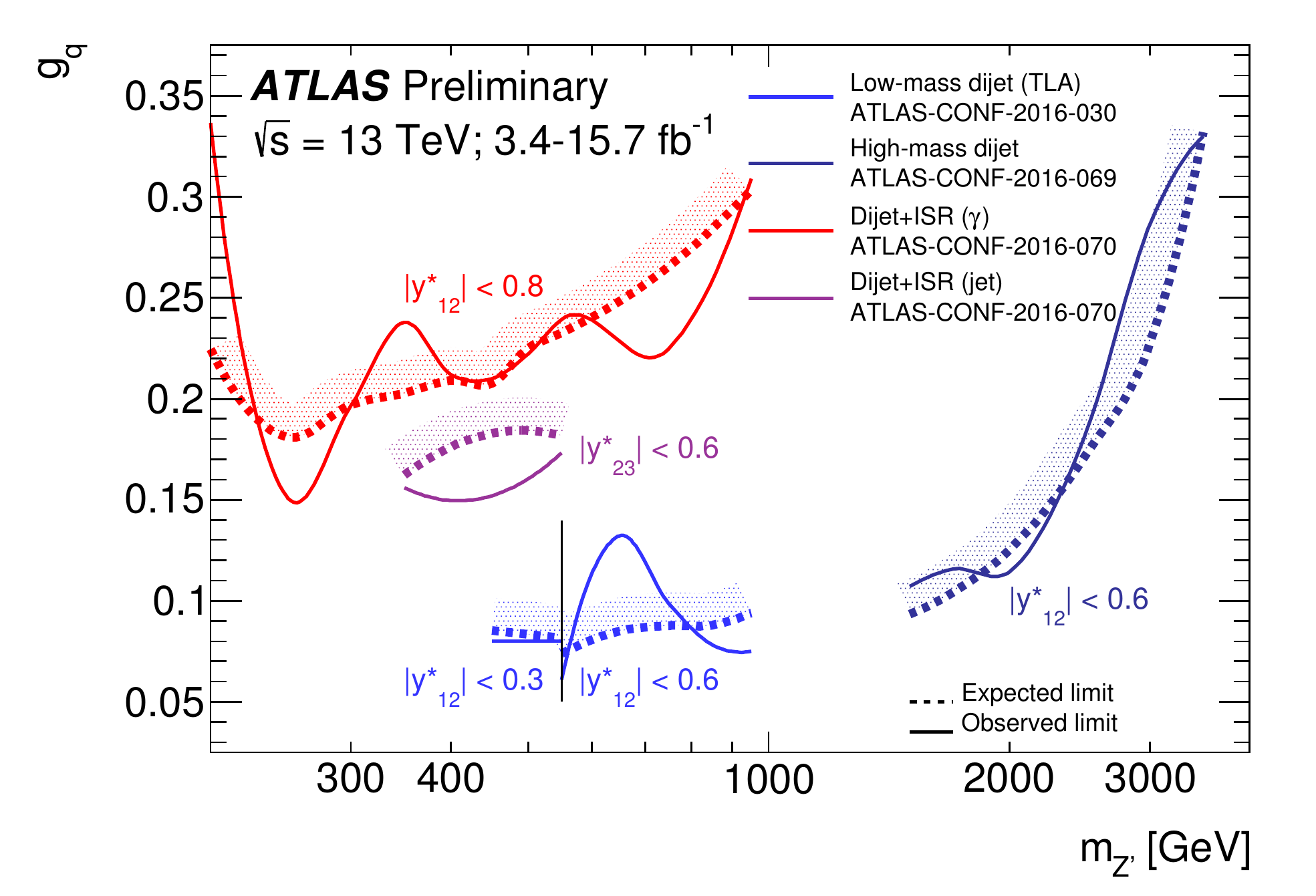}}
\vspace{-0.0cm}
\caption[.]{ATLAS bounds in the coupling-vs-mass plane on a leptophobic $Z^\prime$ 
                model obtained from dijet searches.
                \label{fig:ATLAS-dijetComb}}
\end{figure}
Figure~\ref{fig:ATLAS-dijetComb} shows the combined exclusion plot obtained by ATLAS 
in the coupling-vs-mass plane for a hypothetical leptophobic $Z^\prime$ resonance. The 
low-mass region is covered by ISR-based searches both for ISR jets and photons. 
For intermediate masses the trigger-level analysis (TLA) provides the strongest bounds, 
and for high $Z^\prime$ masses the standard dijets search takes over, smoothly extending
the TLA bound. 

\subsubsection{Searches in leptonic final states}

\begin{figure}[t]
\centerline{\includegraphics[width=1\linewidth]{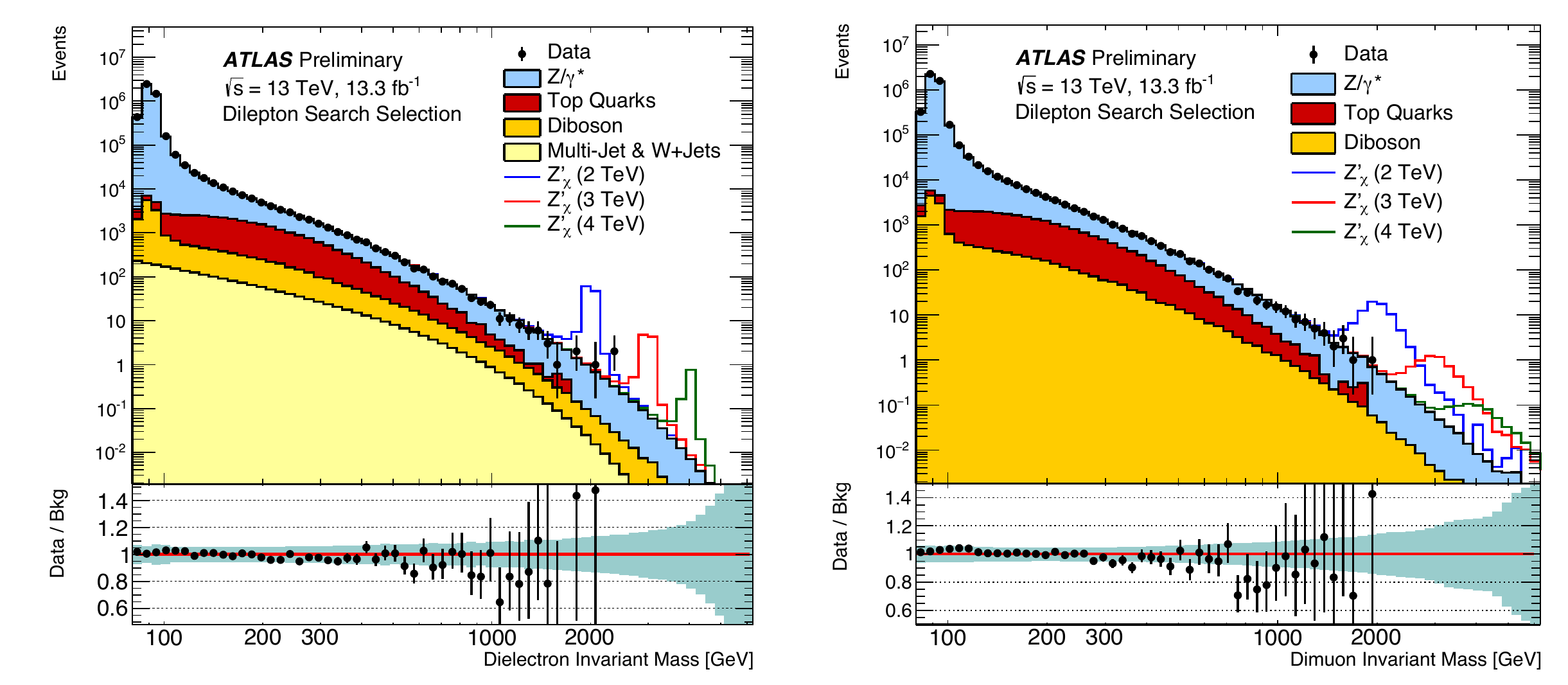}}
\vspace{0.0cm}
\caption[.]{Dielectron (left panel) and dimuon (right panel) reconstructed 
                 invariant mass distributions for data and the SM background 
                 estimates as well as their ratios. Benchmark $Z^\prime$  signals with masses of 
                 2, 3 and 4\;TeV are overlaid~\cite{ATLAS-dilepRes}.
                \label{fig:ATLAS-dilepRes}}
\end{figure}
Canonical searches for new physics are performed in high-mass Drell-Yan production 
$(Z^\prime\to\ell\ell$, $W^\prime\to\ell\nu$)~\cite{ATLAS-dilepRes,ATLAS-W',CMS-dilepRes,CMS-W'}. 
These searches require faithful SM Drell-Yan 
modelling that is tested using SM differential cross section measurements.
High transverse momentum muons represent a challenge for the detector alignment, requiring, eg.,
down to 30\;$\mu$m relative alignment precision in the ATLAS muon spectrometer. The electron
and muon channels have complementary strength: the electron energy resolution measured in the 
calorimeters
being more precise than the muon track momentum resolution, the electron channel has better
discovery sensitivity. On the other hand, there is almost no charge information from the electron
tracks, so the muon channel is needed to measure the charge of a resonance if detected
(cf. the panels in Fig~\ref{fig:ATLAS-dilepRes}). No anomaly was found in the measured spectra. 
Sequential SM $Z^\prime$ / $W^\prime$ benchmark limits are set at 4.1 / 4.7\;TeV 
(compared to 2.9 / 3.3\;TeV at 8 TeV). Figure~\ref{fig:CMS-dilepEvent} shows the 
highest-mass dielectron event measured by CMS in the early 2015 data. It has an
invariant mass of 2.9\;TeV. For comparison, the highest-mass Run-1 events have 
1.8\;TeV ($ee$) and 1.9\;TeV ($\mu\mu$).

\begin{wrapfigure}{R}{0.60\textwidth}
\centering
\vspace{0.03cm}
\includegraphics[width=0.58\textwidth]{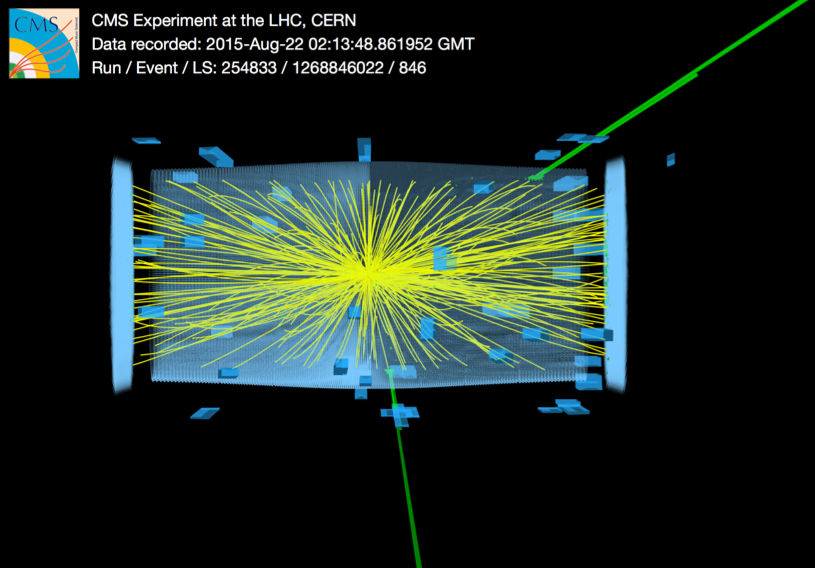}
\vspace{-0.0cm}
\caption[.]{Display of a rare, colossal $e^+e^-$ event with mass of 2.9\;TeV measured 
                by CMS. The electrons are azimuthally back-to-back.
                \label{fig:CMS-dilepEvent}}
\vspace{-0.9cm}
\end{wrapfigure}

ATLAS  and CMS also looked into high-mass $e\mu$ production not accompanied by 
neutrinos that would violate lepton flavour conservation. 
The main background here are top--antitop events that are estimated from MC simulation.

The following table~\cite{Bachacou@Search2017} gives a historical evolution of the 
95\% confidence level lower limits for 
selected leptonically and hadronically decaying benchmark resonances from Tevatron, 
via LHC up to the HL-LHC expectation (see~\cite{ATLAS-reach-HL-LHC} for the latter studies). 
The corollary from these numbers is that future improvement in reach will take more time.

\begin{table}[h]
\centerline{\includegraphics[width=1.003\linewidth]{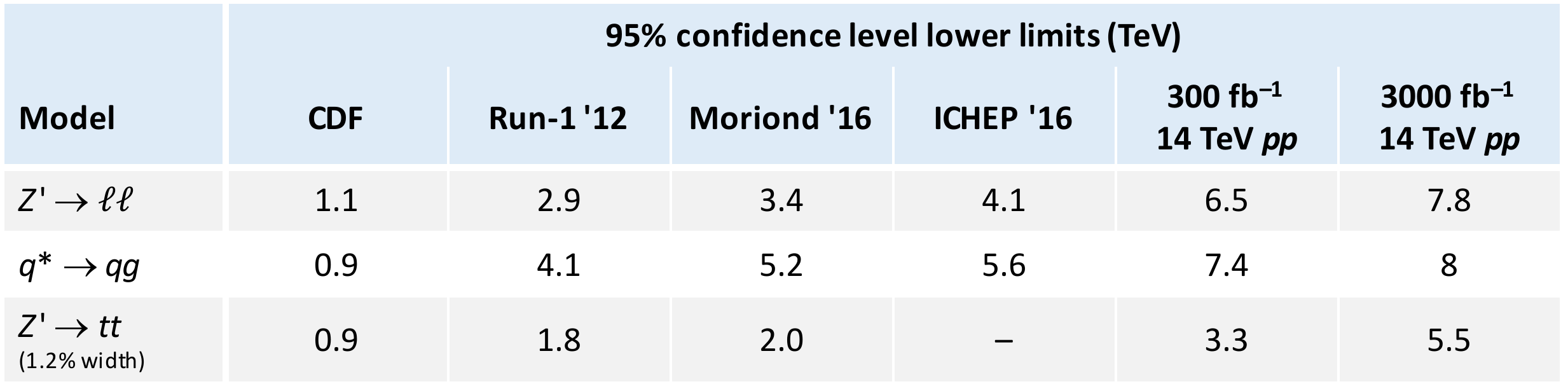}}
\end{table}

\subsubsection{Searches for diboson resonances ($VV$, $Vh$, $hh$)}

\begin{figure}[t]
\centerline{\includegraphics[width=1\linewidth]{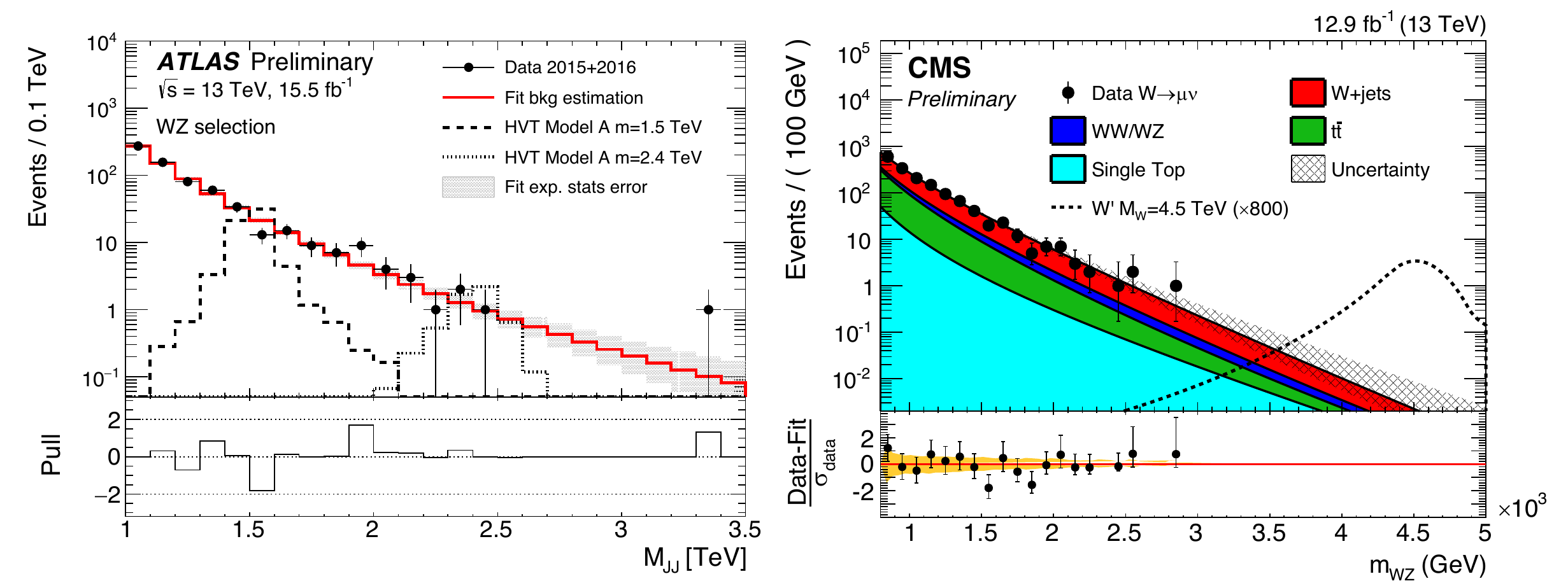}}
\vspace{0.0cm}
\caption[.]{Diboson mass in the fully hadronic channel (ATLAS~\cite{ATLAS-dibosonRes}, left panel) 
                and semileptonic  channel (CMS~\cite{CMS-dibosonRes}, right) for data and background 
                expectations.
                \label{fig:ATLASCMS-dibosonRes}}
\end{figure}
Diboson resonances occur in many new physics scenarios and also in extended scalar sector
models. If the resonances are heavy, the high transverse momentum of the decaying 
bosons boosts the hadronic decay products into merged jets. Jet substructure analysis 
is used to reconstruct hadronically decaying bosons and to suppress strong interaction 
continuum backgrounds. Some excess of events with a (global) significance of 
2.5$\sigma$ was seen by ATLAS in Run-1 around a mass of 2\;TeV in
fully hadronically decaying $VV$ events (mostly $WZ$)~\cite{ATLAS-WZRes8,CMS-WZRes8},
which was however not observed in the other weak gauge boson decay channels of 
similar sensitivity. The excess is not confirmed in Run-2~\cite{ATLAS-dibosonRes,CMS-dibosonRes}
(cf. Fig.~\ref{fig:ATLASCMS-dibosonRes}).

Searches for a new resonance in the diphoton mass spectrum were performed 
by ATLAS~\cite{ATLAS-diphotonRes8-exo,ATLAS-diphotonRes8-higgs}
and CMS~\cite{CMS-diphotonRes8} in Run-1 looking for a low to medium 
mass scalar resonance, or a medium to high mass spin-two resonance 
motivated by strong gravity models. Diphoton spectra were also analysed 
in view of high-mass tail anomalies due to new nonresonant phenomena.
Searches involving at least three photons were used during Run-1 to look for 
new physics in Higgs or putative $Z^\prime$ decays~\cite{ATLAS-multiphoton}.

Preliminary analyses of the 13\;TeV diphoton data presented at the 
2015 end-of-year seminars showed an excess of events at around 750$\;$GeV invariant
diphoton mass in  ATLAS and, albeit weaker, also in CMS. In spring 2016, reanalyses of the 
2015 data were published by ATLAS~\cite{ATLAS-diphotonRes} and CMS~\cite{CMS-diphotonRes} 
confirming the preliminary results.
CMS also included 0.6\ifb of data taken without magnetic field requiring a dedicated reconstruction. 
The photons are tightly identified and isolated and have a typical purity of 94\%.
The background modelling uses empirical functions fit to the full invariant mass
spectra (ATLAS uses a theoretical model to describe the background shape in the spin-2 
case). ATLAS observed the lowest background-only p-value for a resonance 
at around 750\;GeV with a natural width 
of about 45 GeV (6\% with respect to the mass). The local and global significance 
was found to be 3.9$\sigma$ and 2.1$\sigma$, respectively. The global significance
was derived by running background-only pseudo-experiments, modelled according to the 
fit to data, and by evaluating for each experiment the mass and width that leads to 
the largest excess, that is, the lowest p-value. One then counts the fraction of 
experiments with a p-value lower than that in data. 
This procedure corrects the local p-value for the trials factor (also called 
``look-elsewhere effect''). Indeed, the local p-value
corresponds to a non-normalised probability that does not have a well-defined interpretation. 
Only the global p-value defines a proper probability and is thus the correct reference
value. CMS also found its lowest p-value at around 750\;GeV at, however, a narrow width.
Combining 8\;TeV and 13\;TeV data a global (local) significance of 
1.6$\sigma$ (3.4$\sigma$) was seen. These results have prompted intense theoretical activity.

\begin{figure}[t]
\centerline{\includegraphics[width=1\linewidth]{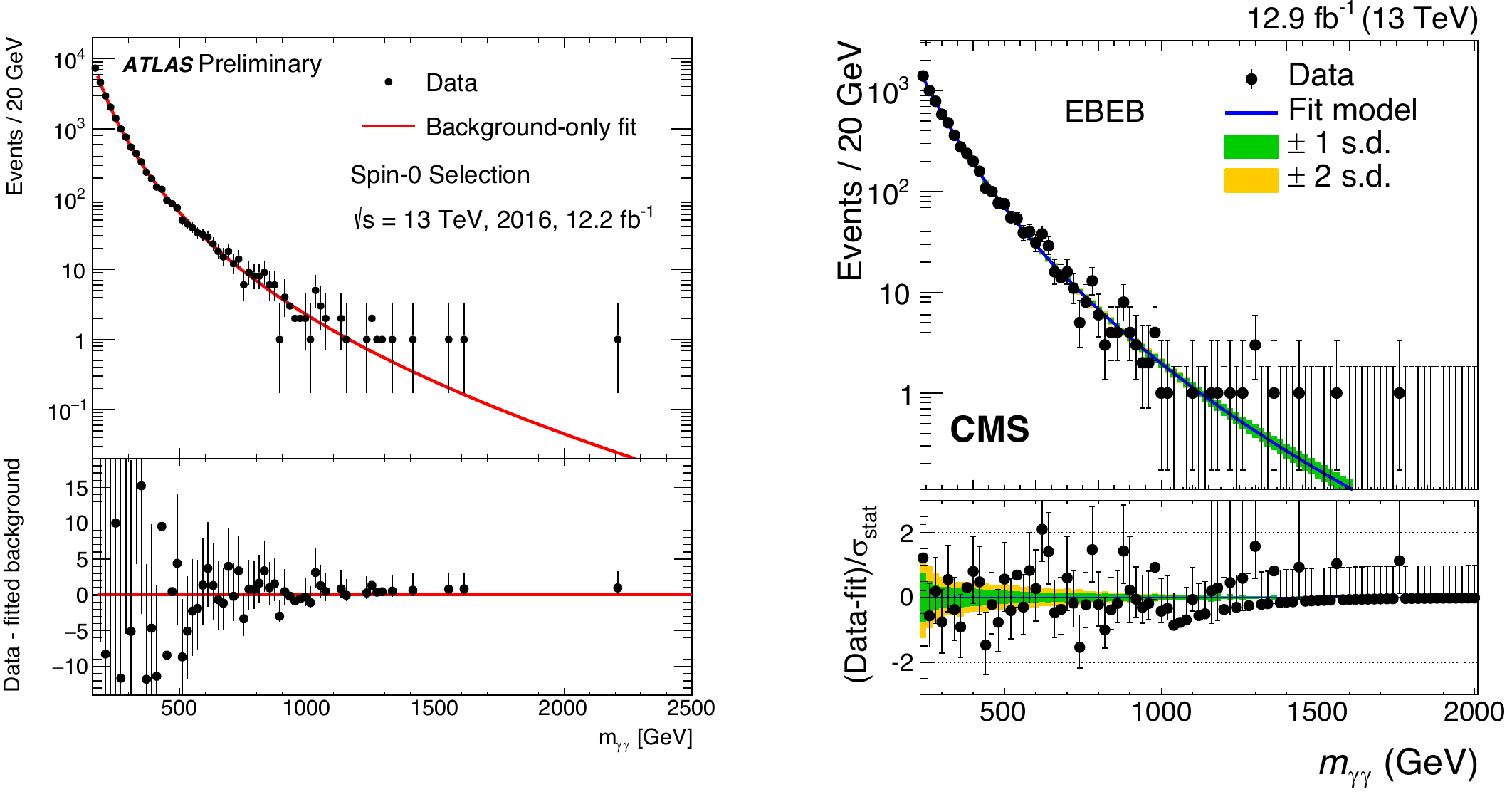}}
\vspace{0.0cm}
\caption[.]{Diphoton mass spectrum in the summer 2016 dataset from ATLAS (left panel) 
                and CMS (right). There is no noticeable excess at around 750\;GeV.
                \label{fig:ATLAS-diphotonResICHEP}}
\end{figure}
As is well known the first 12--13\ifb data taken in 2016 did not reproduce 
the excess in neither experiment~\cite{ATLAS-diphotonResICHEP,CMS-diphotonResICHEP} (cf. 
Fig.~\ref{fig:ATLAS-diphotonResICHEP}). The excesses in the 2015 data were thus the result
of a statistical fluctuation which, given the global significance, is not that unlikely to occur. 
One should also take into consideration that the actual trials factor is larger than the 
global factor quoted for these analyses
as there are many signatures probed by the experiments. This truly global significance of a local
excess is hard or impossible to estimate in a thorough manner, but the additional trials factor 
should be kept in mind.
In that respect, having a second experiment with a similar non-significant excess does 
not remove the trials factor if  the results from both experiments are retained. Removing the 2015 data
and looking solely at 750\;GeV in the 2016 data does, however, properly remove any trials factor.  

\subsubsection{Supersymmetry}

Supersymmetry (SUSY) is still among the most popular SM extensions owing to the elegance 
of the theoretical ansatz, and its phenomenological appeal by offering potential solutions to
the hierarchy problem,\footnote{As the SM, SUSY is a weakly coupled approach to electroweak 
symmetry breaking in which the Higgs boson remains elementary.}
grand unification of the gauge couplings, and dark matter. However,
if the SM is included in a supersymmetric theory with SUSY particles (sparticles)  that differ
by half-a-unit of spin from their SM partners, how is it possible that more than half the particles 
in the superworld have escaped our observations?

\begin{figure}[t]
\centerline{\includegraphics[width=0.85\linewidth]{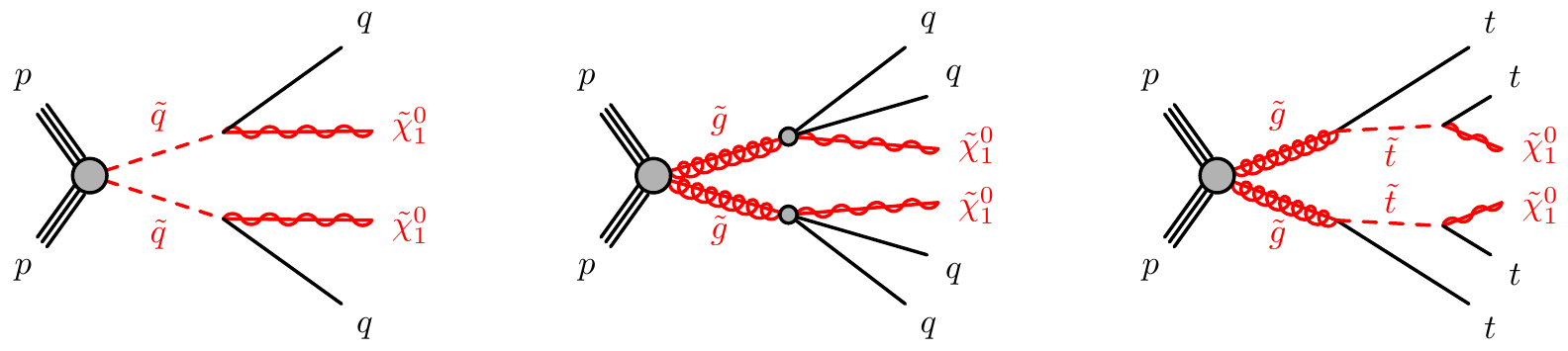}}
\vspace{0.0cm}
\caption[.]{Graphs for simplified models describing the pair production and decay of supersymmetric 
                particles. Left: squark pair production and decay to two quark-jets and two neutralinos;
                middle: gluino pair production and decay to four quark-jets and  two neutralinos;
                right: gluino pair production and decay to four top quarks and two neutralinos.
                The top quarks will each further decay to a $W$ boson and $b$ quark.
                \label{fig:SUSY-Feyn}}
\end{figure}
\begin{figure}[t]
\centerline{\includegraphics[width=1\linewidth]{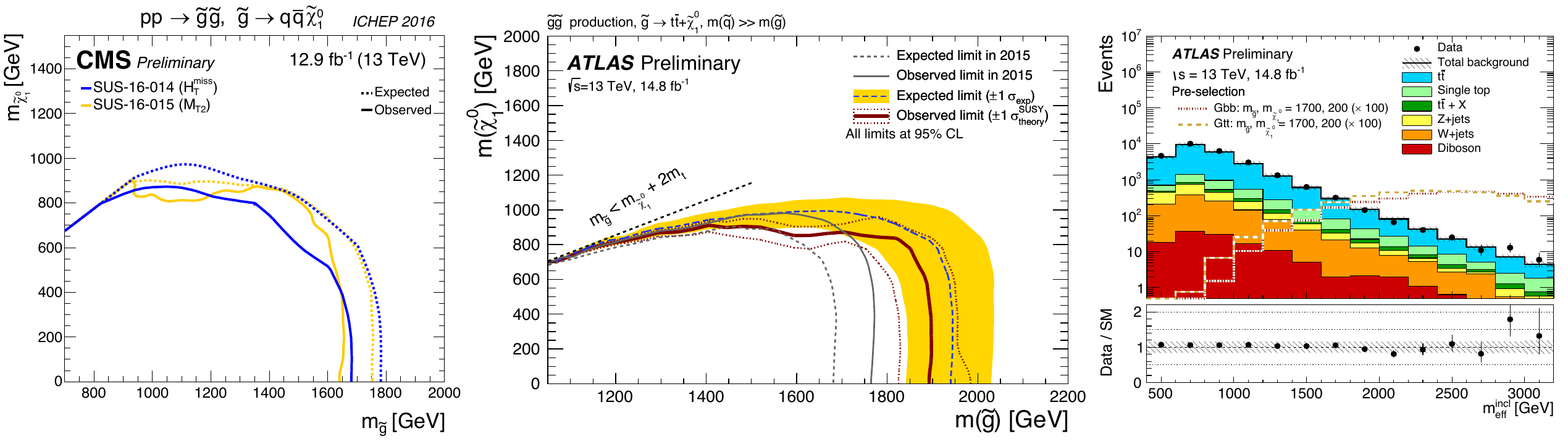}}
\vspace{0.0cm}
\caption[.]{Left: exclusion limits for strong gluino pair production and decay to four jets
                and two lightest neutralinos as obtained by CMS~\cite{CMS-gluinos1,CMS-gluinos2}.
                Observed limits are shown 
                with solid lines and expected with dashed lines. Middle: exclusion limits 
                obtained for strong gluino pair production and decay via stop squarks 
                into a four top quark final state and two neutralinos. Right: effective mass 
                distribution obtained in the search for gluino mediated stop 
                production~\cite{ATLAS-Gtt} 
                (cf. Footnote~\ref{ftn:effectivemass}, page~\pageref{ftn:effectivemass}
                for the definition of the effective mass).
                \label{fig:ATLASCMS-SUSY-strong}}
\end{figure}
Due to SUSY breaking, allowing the sparticles to acquire large masses,\footnote{In unbroken SUSY,
  fermionic $|f\rangle$ and bosonic $|b\rangle$ partner states, 
  transformed into each other via the SUSY generator $Q$,  have the same mass, as 
  $P^2Q|b\rangle=P^2|f\rangle=m_f^2|f\rangle$,
  $P^2Q|b\rangle=QP^2|b\rangle=Qm_b^2|b\rangle=m_b^2|f\rangle$, 
  and hence $m_f^2=m_b^2$.  $P^2$ is the square of the energy-momentum operator
  which commutes with $Q$.}
SUSY comes with very diverse signatures. Highest cross-section events produce gluino 
or squark pairs with decays to jets and missing transverse momentum if $R$-parity is 
conserved.\footnote{$R$-parity, defined by $R=(-1)^{3(B-L)+2S}$, where $B$, $L$, $S$ are
  the baryon, lepton numbers and spin, respectively, is assumed to be 
  conserved in most SUSY models to avoid baryon and/or lepton number violation (and 
  thus proton decay if both occur together). $R$-parity conservation is arbitrarily imposed 
  and not enforced by any known symmetry. Its consequence is that SUSY particles must be 
  produced in pairs and the lightest SUSY particle is stable.} Naturalness
suggests not too heavy SUSY top, weak and Higgs boson partners to effectively cancel the 
radiative corrections to the Higgs mass at high scale and hence provide a solution 
to the hierarchy problem.\footnote{The top quark gives the largest contribution to the 
  radiative corrections of the Higgs mass, $\delta m_H^2$, in presence of a high new physics 
  scale $\Lambda$. If the stop is heavier than the top residual
  logarithmic contributions   $\delta m_H^2\propto\ln(\Lambda^2/m^2_{\tilde t})$ remain. }
It might thus occur that stop pair production, or gluino pair 
production and decay via stop and top  to a four-top final state are the dominant SUSY 
processes at the LHC. If all strongly interacting SUSY particles are too heavy to 
be directly produced at the LHC, it could still be that the electroweak partners of the photon, 
weak bosons and five physical Higgs  states are light enough so that SUSY 
would manifest itself through ``electroweak-ino'' production featuring 
final states with leptons (and/or photons) and $\MET$. Finally, SUSY could also give rise to 
the existence of long-lived heavy particles, and, if $R$-parity is nonconserved, 
the lightest SUSY particle could decay to jets or leptons depending on the 
$R$-parity violating couplings. 

To approach the search for SUSY in a systematic 
manner, a bottom-up approach through so-called simplified models is 
used by the experiments. These models correspond to simple signatures as those
depicted in Fig.~\ref{fig:SUSY-Feyn}. While a  simplified model cannot 
encompass the full SUSY phenomenology, an ensemble of simplified models
and the corresponding searches have been shown to cover signatures of 
complete models such as the phenomenological minimal supersymmetric 
standard model (pMSSM)~\cite{ATLAS-pMSSM-Run1,CMS-pMSSM-Run1}. 

Searches for strong SUSY production study events with jets and $\MET$
with or without leptons, photons, and $b$-jets. Up to ten jets are exclusively 
selected, which requires a data-driven background determination as MC cannot 
be trusted to reliably predict such large jet multiplicity. None of the searches 
have revealed a significant anomaly. Figure~\ref{fig:ATLASCMS-SUSY-strong} 
shows on the left and middle plots exclusion limits in the lightest neutralino mass 
versus gluino mass planes. The analyses have the sensitivity to exclude gluinos of 
up to 1.8\;TeV for low-mass neutralinos depending on the scenarios. 
In case of heavy neutralinos, the final states
exhibit softer jets and less $\MET$, which leads to reduced trigger efficiency and a more 
difficult background discrimination thus reducing the sensitivity. The right panel 
in Fig.~\ref{fig:ATLASCMS-SUSY-strong} shows the effective mass distribution 
obtained in the search for gluino mediated stop production (four top quark final
state). The distribution reaches beyond 3\;TeV with the dominant background from
top-quark production. No excess of events is seen in data compared to the background 
estimation. 

\begin{figure}[t]
\centerline{\includegraphics[width=0.95\linewidth]{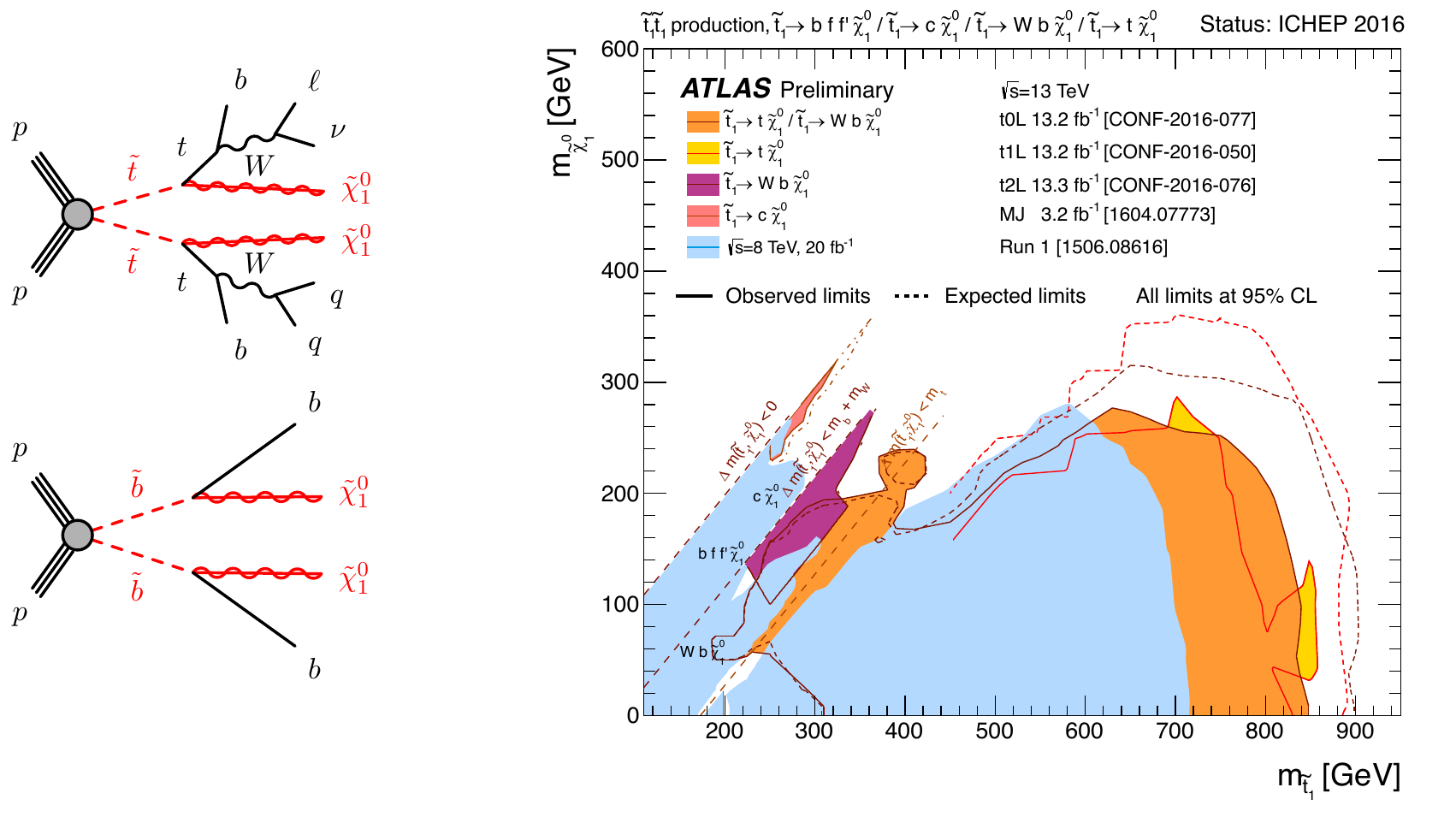}}
\vspace{0.0cm}
\caption[.]{Graphs for simplified models of stop and sbottom pair production and decay. 
                The right panel shows exclusion limits in the neutralino versus stop mass 
                plane as obtained by ATLAS with several dedicated analyses (see references in figure).
                \label{fig:ATLAS-SUSY-stop}}
\end{figure}
If gluinos are too heavy to be produced in significant quantities, squark mixing could 
make third generation squarks lighter than the first and second generation squarks. 
Direct searches for stop and sbottom squark production have been the topic of intense
efforts in both ATLAS and CMS since Run-1. The analyses are distinguished according to the 
number of identified leptons (0, 1, 2) and differently optimised signal regions target 
different stop/sbottom and neutralino mass regimes. In the stop case, the signatures
also depend on whether the stop decays in a two-body signature to an on-shell top quark
and the lightest neutralino, or off-shell via three or four body decays to the top decay products. 
The right panel in Fig.~\ref{fig:ATLAS-SUSY-stop} shows the exclusion limits for simplified models 
of stop and sbottom pair production and decay obtained by ATLAS with several dedicated 
analyses. The analyses have sensitivity to exclude stop masses up to 900\;GeV. As
in the gluino and first-generation squark cases, the limits for heavy neutralinos are 
significantly worse. 

\begin{figure}[t]
\centerline{\includegraphics[width=0.98\linewidth]{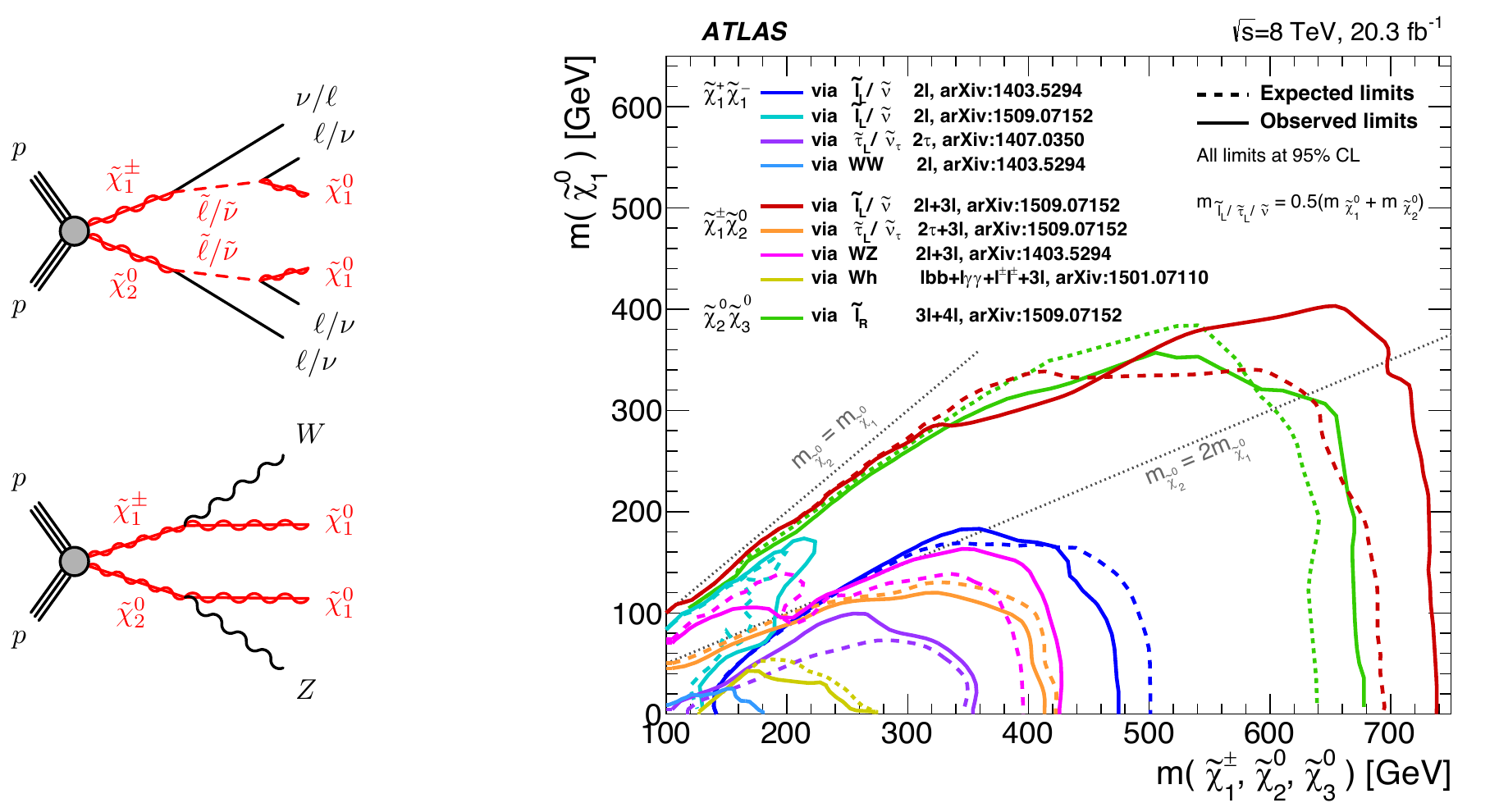}}
\vspace{0.0cm}
\caption[.]{Graphs for simplified models of associated lightest chargino and next-to-lightest
                neutralino production and decay through sleptons (top graph) or, if sleptons
                are too heavy, through $W$ and $Z$ bosons. The right plot shows the exclusion
                bounds obtained by ATLAS for various electroweakino production scenarios
                (see references in figure). 
                \label{fig:ATLAS-SUSY-EWk}}
\end{figure}
Alternative models for new heavy quark partners introduce, for example, vector-like 
quarks, which are hypothetical fermions that transform as triplets under 
colour and who have left-handed and right-handed components with same colour and 
electroweak quantum numbers. Vector-like quarks can be singly or pair produced and 
decay to $bW$, $tZ$ or $tH$. Also exotic $X_{5/3} \to tW$ processes may exist. 

It could also be that all squarks and gluinos are beyond reach of the current LHC 
sensitivity and electroweakinos are  the lightest fermions. They have low cross-sections, 
so that the present Run-2 luminosity just suffices to surpass the Run-1 sensitivity.
Figure~\ref{fig:ATLAS-SUSY-EWk} shows graphs for simplified models of associated 
lightest chargino and next-to-lightest neutralino production and decay through 
sleptons (top graph) or, if sleptons are too heavy, through $W$ and $Z$ bosons. 
The right plot shows the exclusion bounds obtained by ATLAS for various 
electroweakino production scenarios. Electroweakino decays via sleptons are a 
favourable case due to the larger leptonic rate than in weak boson decays. 
In the models considered, chargino pair production has lower cross section than 
$\tilde\chi_1^+\tilde\chi_2^0$ production. The cross section depend on the 
mixing properties of the states: neutralinos can be bino, wino or higgsino like; 
charginos wino or higgsino like, depending on the dominant contribution.\footnote{There 
are a total of eight spin-half partners of the electroweak gauge and Higgs bosons: 
the neutral bino (superpartner of the U(1) gauge field), the winos, which are a charged 
pair and a neutral particle (superpartners of the $W$ bosons of the SU(2)$_L$ gauge fields), 
and the higgsinos, which are two neutral particles and a charged pair (superpartners of 
the Higgs field's degrees of freedom). The bino, winos and higgsinos mix to form four 
charged states called charginos ($\tilde\chi^\pm_i$) 
and four neutral states denoted neutralinos ($\tilde\chi^0_i$) . Their 
indices $i$ are ordered according to the increasing mass of the $\tilde\chi_i$ state. }

\subsubsection{Search for massive long-lived massive particles}

Massive long-lived heavy particles are predicted in many new physics models. They can occur due to 
large virtuality (such as predicted in split supersymmetry), low couplings (such as predicted
in some gauge mediated SUSY breaking scenarios where the gravitino is the lightest SUSY 
particle), and mass degeneracy in a cascade decay, eg., via a scale-suppressed colour 
triplet scalar from unnaturalness~\cite{Unnaturalness} or anomaly-mediated SUSY breaking
scenarios with a wino-like lightest chargino~\cite{ATLAS-disapperingTrack}. The search 
for massive long-lived particles is a key part of the LHC search programme. 

The LHC experiments search for massive long-lived particles using measurements of  
specific ionisation loss in the tracking detectors, the time-of-flight in the calorimeters
and muon systems, and by reconstructing displaced vertices, kinked or disappearing tracks. 
Looking for calorimeter deposits outside of the colliding proton bunches makes it 
possible to look for very long-lived strongly interacting massive particles that were
stopped in the calorimeter layers~\cite{ATLAS-stoppedGluinos,CMS-stoppedGluinos}. 
Some signatures need dedicated triggers, most require novel analysis strategies
to determine backgrounds from data. 
\begin{figure}[t]
\centerline{\includegraphics[width=1\linewidth]{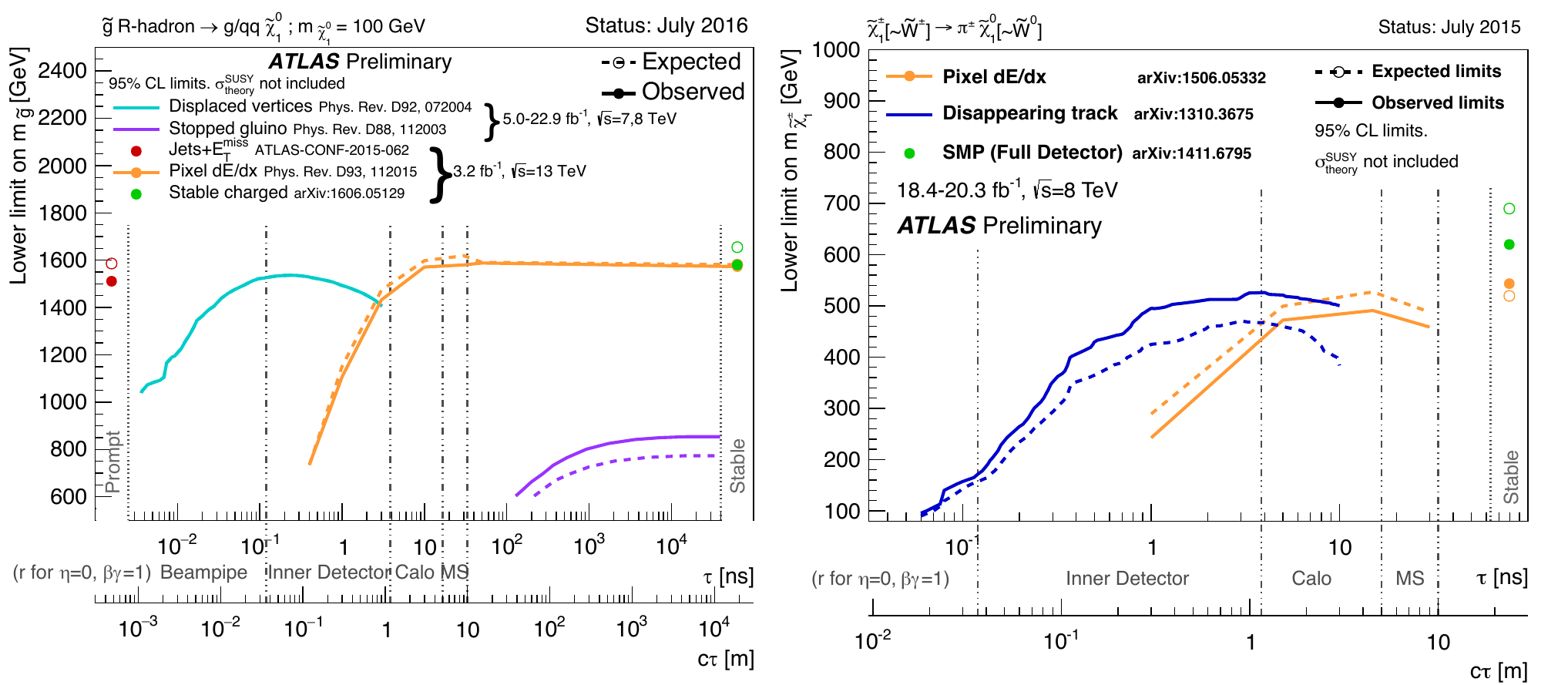}}
\vspace{0.0cm}
\caption[.]{Exclusion limits on the gluino mass versus lifetime (left panel) and 
                 chargino mass versus lifetime (right) as obtained by ATLAS (see references in figure). 
                 The dots on the left (right) of the plots indicate the limits obtained on promptly 
                 decaying (stable) gluinos/charginos. Varying searches cover the full lifetime spectrum.
                \label{fig:ATLAS-SUSY-LLP}}
\end{figure}
Figure~\ref{fig:ATLAS-SUSY-LLP} shows exclusion limits on the gluino mass versus 
lifetime (left panel) and chargino mass versus lifetime (right) as obtained by ATLAS. 
The dots on the left (right) of the plots indicate the limits obtained on promptly 
decaying (stable) gluinos/charginos. Varying searches cover the full lifetime spectrum.
It is interesting to observe that the standard SUSY searches are not blind to scenarios
with long-lived sparticles if their lifetime is short enough to still decay before the 
calorimeter. 

\subsubsection{Searches for dark matter production}

\begin{wrapfigure}{R}{0.33\textwidth}
\centering
\vspace{-0.55cm}
\includegraphics[width=0.3\textwidth]{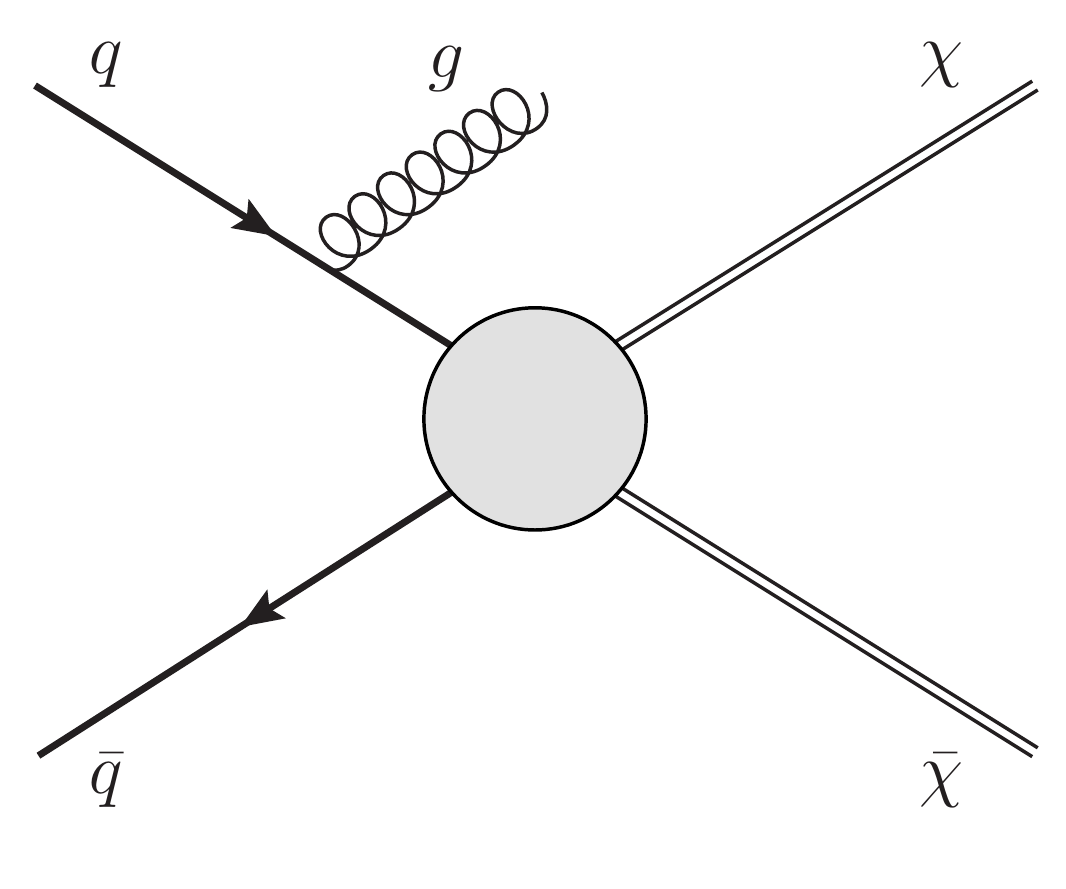}
\vspace{-0.15cm}
\caption[.]{Graph for WIMP pair production with initial-state radiation jet.
                \label{fig:Monojet-Feyn}}
\vspace{-0.4cm}
\end{wrapfigure}
\begin{figure}[p]
\centerline{\includegraphics[width=\linewidth]{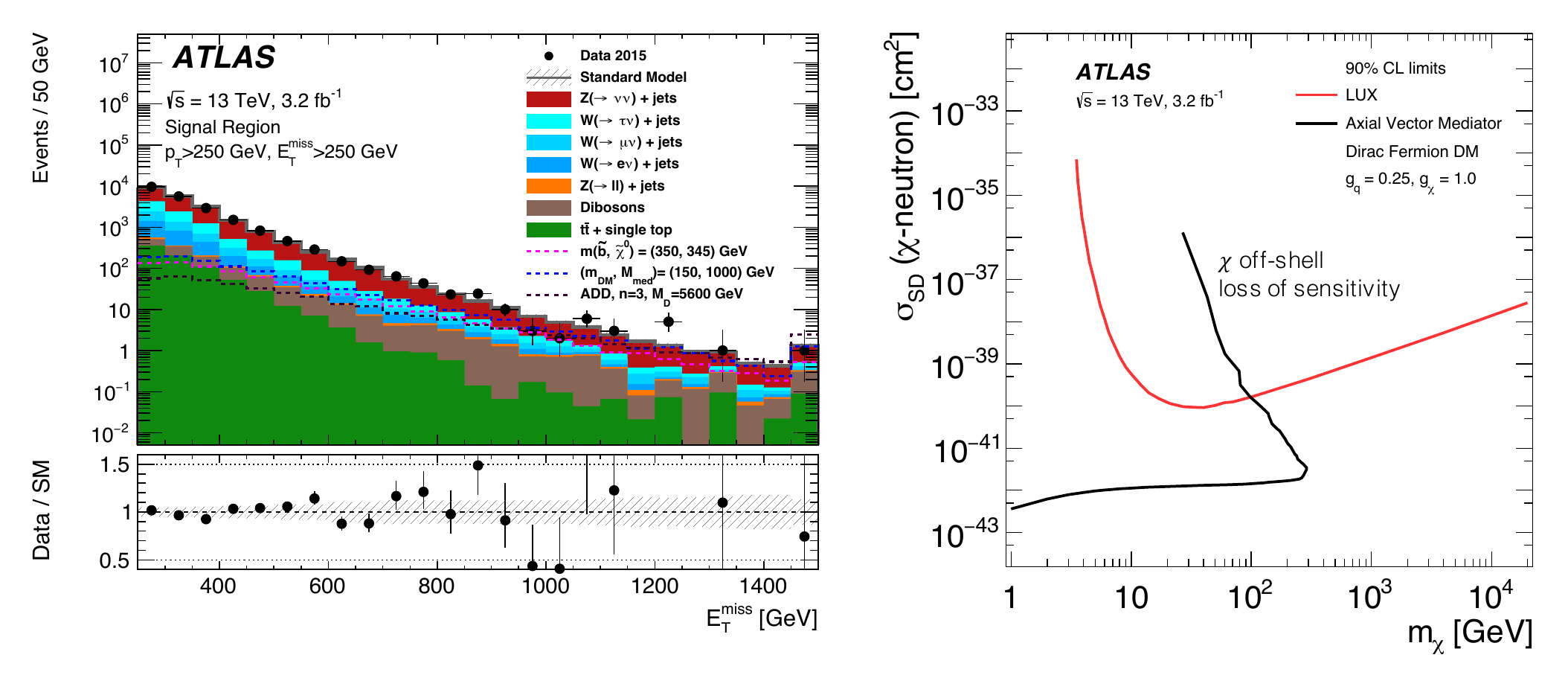}}
\vspace{-0.0cm}
\caption[.]{Left: distribution of missing transverse momentum measured
                 by ATLAS at 13\;TeV in a ``mono-jet'' search~\cite{ATLAS-monojet}.
                 The dominant backgrounds stem from leptonic $Z$ and $W$ plus jets events. 
                 Also shown are distributions for new physics  benchmark models. 
                Right: exclusion limit on the spin-dependent WIMP--neutron
                scattering cross section versus the WIMP mass  in the context of a 
                $Z^\prime$-like simplified model with axial-vector couplings.
                The result is compared with limits from the LUX experiment~\cite{LUX-darkmatter}. 
                All limits are shown at 90\% confidence level, which is the standard benchmark
                in direct dark matter detection experiments. 
                The comparison to LUX is valid solely in the context 
                of this model, assuming minimal mediator width and the coupling values 
                $g_q = 1/4$ and $g_\chi = 1$~\cite{ATLAS-monojet}.

  \label{fig:ATLAS-monojet}}
\vspace{0.7cm}
\centerline{\includegraphics[width=1\linewidth]{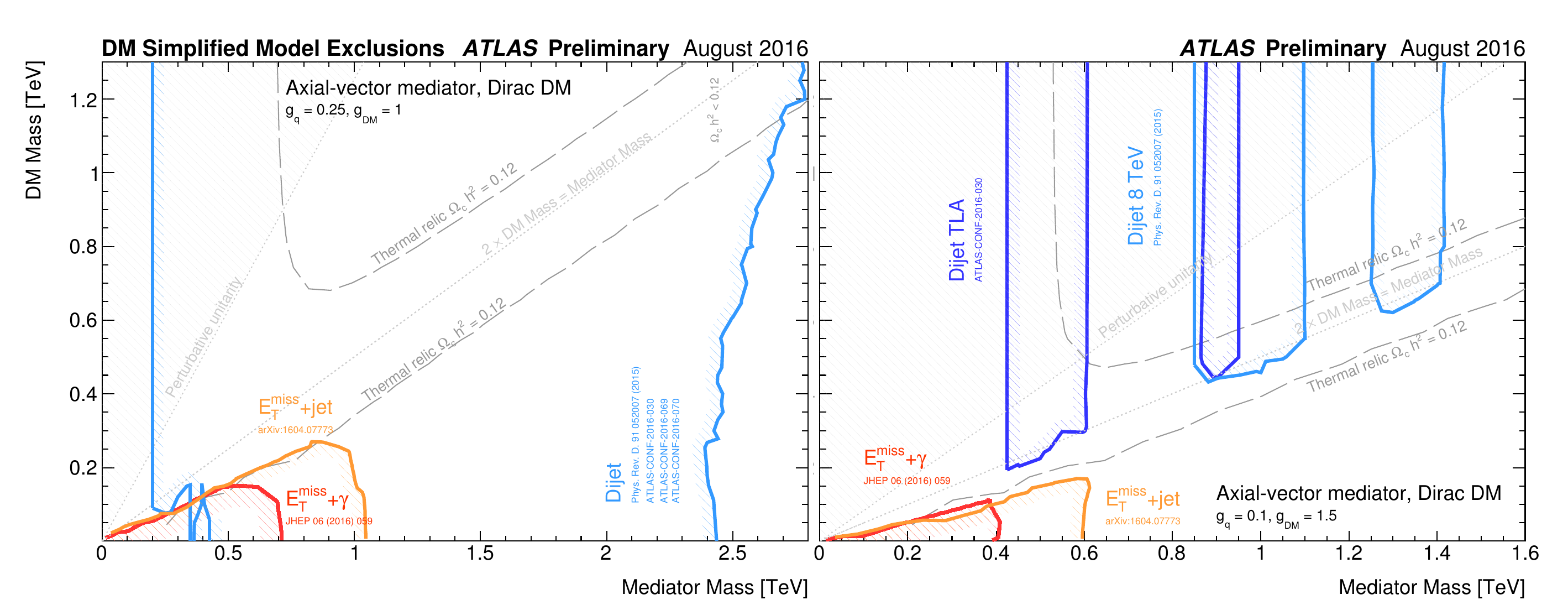}}
\vspace{0.1cm}
\caption[.]{Regions in a dark matter (DM) versus mediator mass planes excluded at 
  95\% CL by a selection of ATLAS DM searches, for a possible 
  interaction between the SM and DM, the lepto-phobic 
  axial-vector mediator described in~\cite{LHC-DMforum}. The left panel shows exclusion 
  bounds for quark coupling $g_q=1/4$, universal to all flavors, and dark matter coupling 
  $g_{\rm DM}=1$. On the right panel $g_q=1/10$ and $g_{\rm DM}=3/2$ are assumed. 
  Shown are the results from the monojet, monophoton and dijet resonance searches. 
  Dashed curves labelled ``thermal relic'' indicate combinations of DM 
  and mediator mass that are consistent with the cosmological 
  DM density  and a standard thermal history. Between the 
  two curves, annihilation processes described by the simplified model 
  deplete the relic density. A dotted curve indicates the kinematic 
  threshold where the mediator can decay on-shell into DM. 
  Points in the plane where the model is in tension with  perturbative 
  unitary considerations are indicated by the shaded triangle at the 
  upper left. The exclusion regions, relic density contours, and unitarity 
  curve are not applicable to other choices of coupling values or model. 
  See~\cite{ATLAS-DM-plot} for more information.
  \label{fig:ATLAS-DM-plot}}
\end{figure}
If dark matter particles (assumed to be weakly interacting and massive, WIMPs) 
interact with quarks and/or gluons they can be directly pair produced in the proton
collisions at the LHC~\cite{LHC-DMforum}.
Since the WIMPs remain undetected, to trigger the events a large boost via initial 
state jet or photon radiation (or other recoiling particles) is needed leading to 
large missing transverse momentum from the recoiling WIMP pair. 
The final state signature depends on the unknown details of the proton--WIMP 
coupling. A set of ``$X$ + $\MET$'' searches is therefore needed for full experimental coverage. 
The most prominent and among the most sensitive of these is the so-called ``mono-jet'' 
search, which extends to a couple of high-$p_T$ jets recoiling against the $\MET$
(cf. Fig.~\ref{fig:Monojet-Feyn}). 
Large irreducible SM backgrounds in this channel stem from 
$Z(\to\nu\nu)+{\rm jets}$  and $W(\to\ell\nu)+{\rm jets}$ events, 
where in the latter case the charged lepton is either undetected or a hadronically 
decaying tau lepton. These backgrounds are determined in data control 
regions requiring accurate input from theory to transfer the measured normalisation 
scale factors to the signal regions. 

Numerous 13$\;$TeV results have been released by ATLAS and CMS, including 
jets + $\MET$~\cite{ATLAS-monojet,CMS-monojet}, 
photon + $\MET$~\cite{ATLAS-gammaMET,CMS-gammaMET}, 
$Z/W$ + $\MET$~\cite{ATLAS-VMET,CMS-ZMET}, and 
$bb/tt$ + $\MET$~\cite{ATLAS-bMET,CMS-bbMET,CMS-ttMET} signatures. 
None of these has so far shown an anomaly.
Figure~\ref{fig:ATLAS-monojet} shows the missing transverse momentum 
distributions measured by ATLAS in the monojet jets + $\MET$ search. 

Since the mediator is produced via quark annihilation ($g_q$) it can also decay to quarks
and hence the dijet resonance search is sensitive to it. Figure~\ref{fig:ATLAS-DM-plot} 
shows for a specific benchmark model and two different coupling sets (see figure caption) 
ATLAS exclusion regions in the 
DM versus the model's mediator mass plane as obtained from the jets + $\MET$
and photon + $\MET$ analyses as well as from the dijet resonance search. These 
searches have complementary sensitivity.

Finally, we note that even in the case of a signal in one of the LHC WIMP searches
the LHC may not be able to prove that a signal is indeed dark matter because of insufficient
constraints on the lifetime of the detected WIMPs. 

\section{The road to the future}

\begin{figure}[b]
\vspace{-0.1cm}
\centerline{\includegraphics[width=1\linewidth]{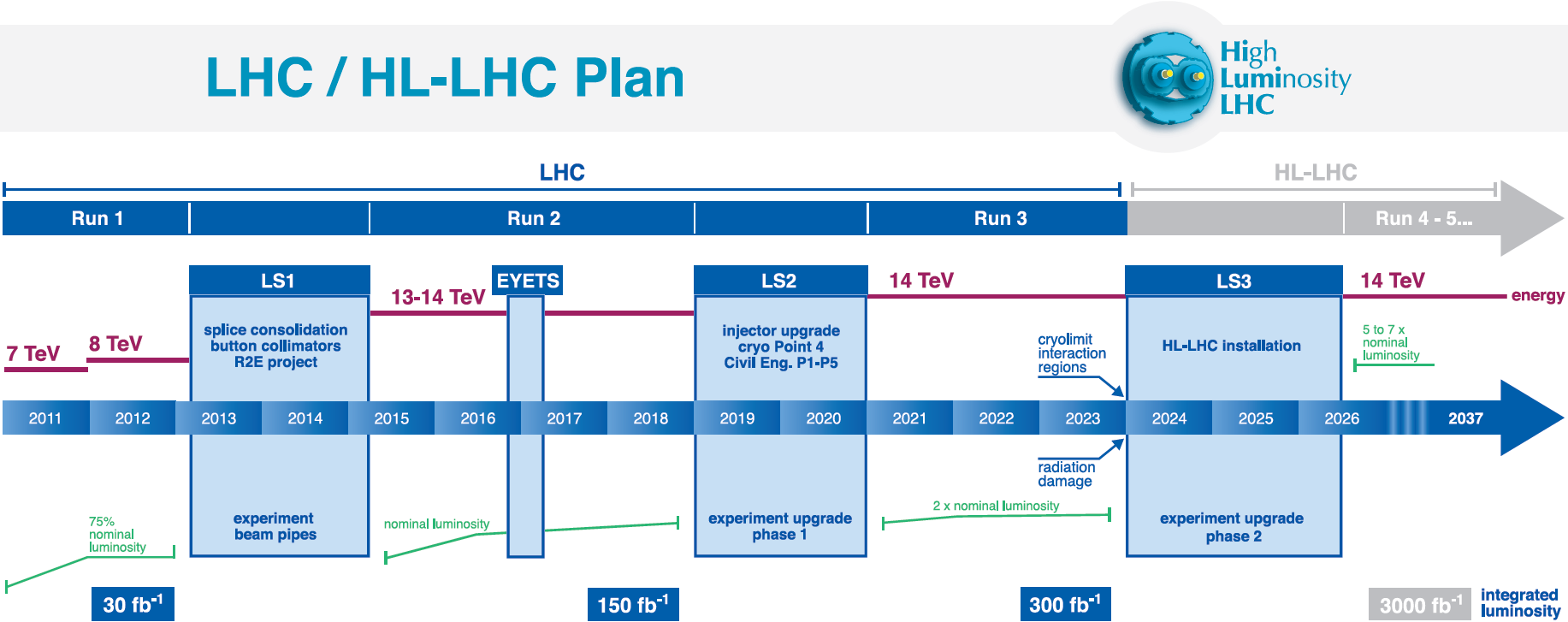}}
\vspace{0.0cm}
\caption[.]{Timeline of the LHC programme up to the high-luminosity LHC (HL-LHC). 
                \label{fig:LHC-timeline}}
\end{figure}
The LHC experimental programme follows a well-defined suit of data taking periods followed 
by longer technical stops used to repair and upgrade the accelerator and experiments. With the 
approval of the HL-LHC project by the CERN Council in 2016 a roadmap for twenty more exciting 
years of physics with the LHC has been established. That roadmap is sketched in 
Fig.~\ref{fig:LHC-timeline}. The current Run-2 will continue until end of 2018 with a
delivered integrated luminosity at 13\;TeV (or higher) that may reach 120\ifb. The following 
two year long shutdown (LS2) will be used to upgrade the injector for an increased 
beam brightness (batch compression in the PS, new optics in the SPS, collimator upgrades). Also 
the experiments upgrade their detectors to prepare for the increased Run-3 luminosity. 
The following data taking period between 2021 and 2023 should allow the LHC to deliver a 
total of 300\ifb at 13--14\;TeV proton--proton centre-of-mass energy. This is followed by 
the major HL-LHC upgrade during 2024 until 2026, featuring a new LHC triplet design (low-$\beta^*$ 
quadrupoles, crab cavities), and injector upgrades for luminosity levelling~\cite{HL-LHC}. 
Here, also the experiments will undergo major upgrades to prepare for the high-luminosity 
phase~\cite{ATLAS-scoping,CMS-scoping}. Collisions are expected to resume in 2026
allowing to deliver to each experiment (ATLAS and CMS) 300\ifb per year.  
The following table summarises some of the LHC beam parameters during Run-1, Run-2, 
and as expected for Run-3 and the HL-LHC.\footnote{Recall that 
$L\propto (\sigma_x\sigma_y)^{-1}=(\e_n\beta^*/\gamma)^{-1}$} \\

\centerline{\includegraphics[width=1\linewidth]{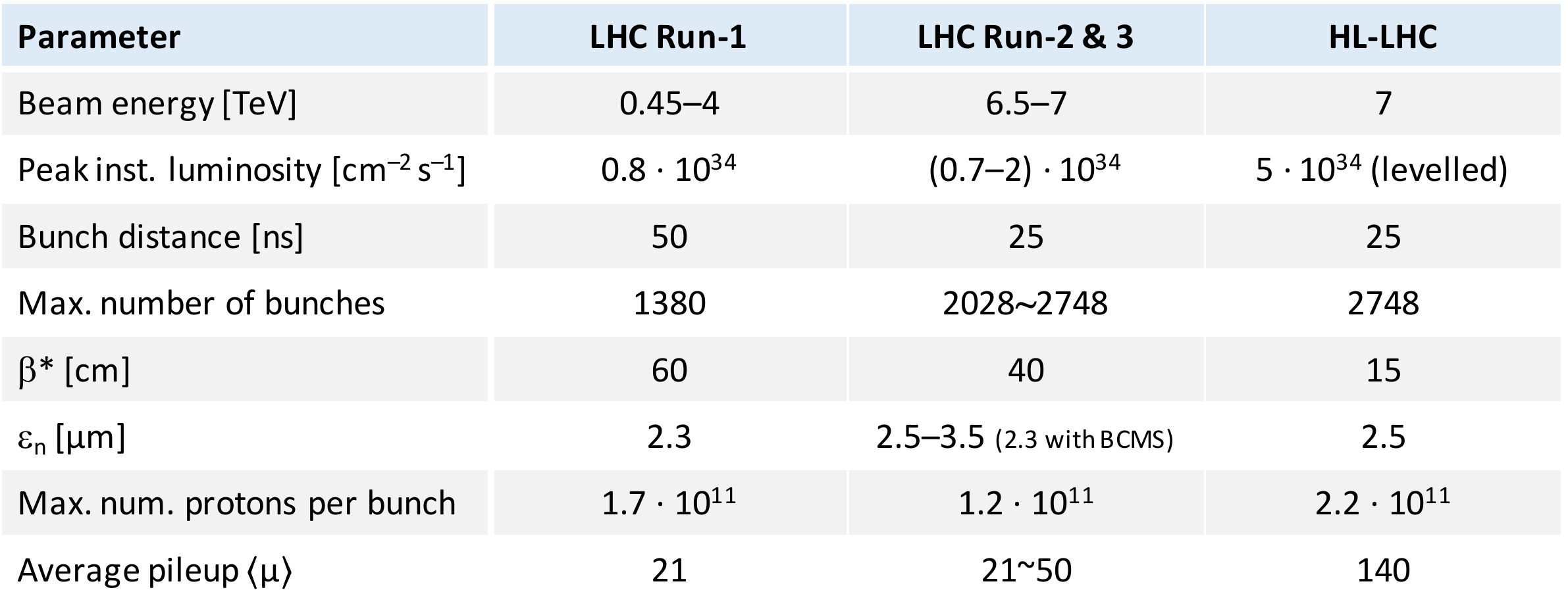}}

\begin{wrapfigure}{R}{0.6\textwidth}
\centering
\vspace{-0.5cm}
\includegraphics[width=0.6\textwidth]{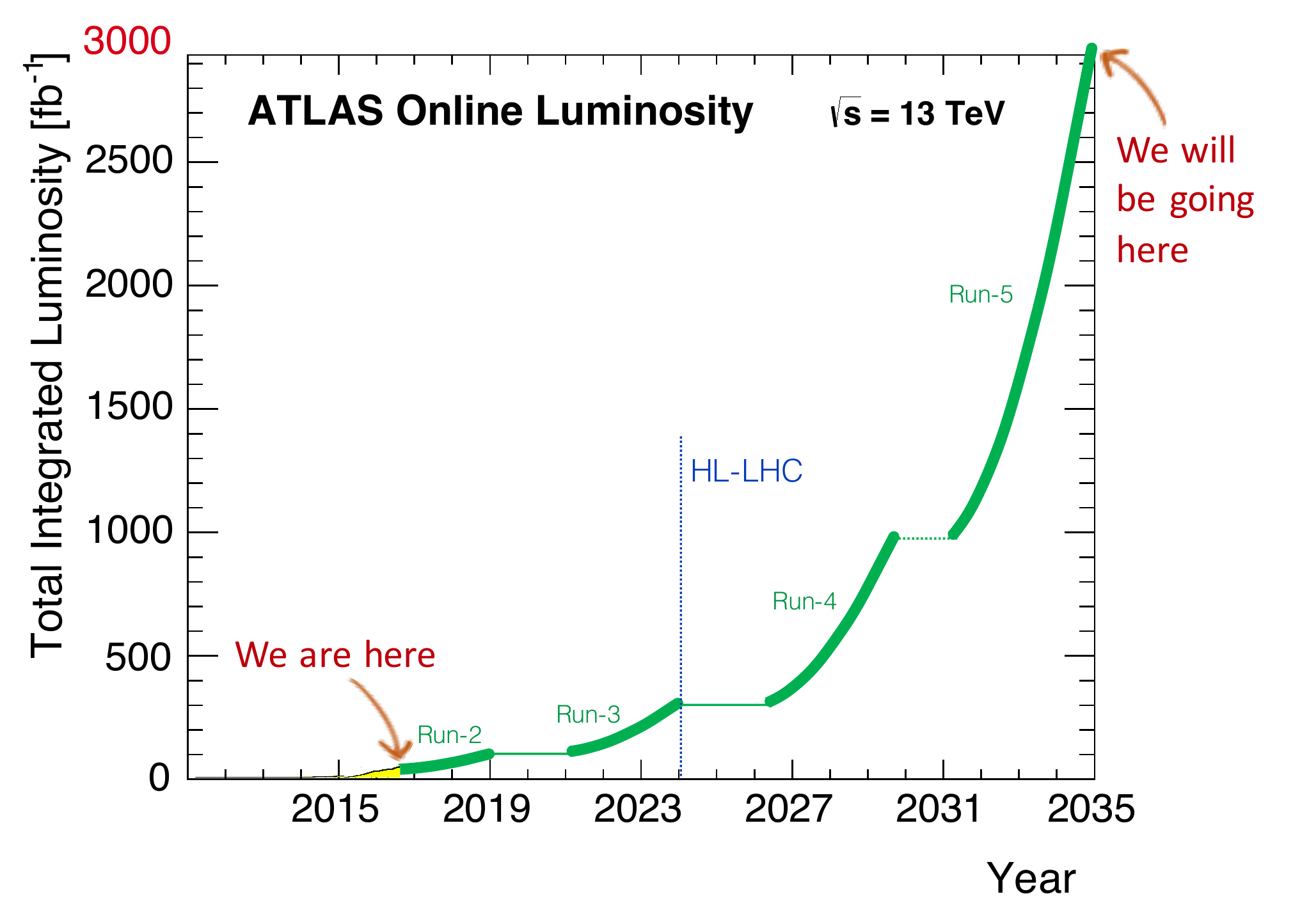}
\vspace{-0.6cm}
\caption[.]{Sketch illustrating the integrated luminosity evolution during the various LHC 
                phases~\cite{LHC-lumi-outlook}. LHC physics will hardly look the same again.
                \label{fig:LHC-lumi-outlook}}
\end{wrapfigure}
If one wants to succinctly highlight the main physics results of the LHC proton--proton 
programme during Run-1, one should emphasise the discovery of the Higgs boson, 
searches for additional new physics (all negative), multiple SM measurements, 
the observation of rare processes such 
as $B_s \to\mu\mu$, precision measurements of SM processes and parameters, and the 
study of CP asymmetries in the $B_s$ sector. For Run-2 and Run-3, the focus lies on searches for 
new physics at the energy frontier, improved measurements of Higgs couplings in the main 
Higgs boson channels, consolidation and observation of the remaining Higgs decay and 
production modes,  measurements of rare 
SM processes and more precision, improved measurements of rare $B$ decays and 
CP asymmetries. Finally, the HL-LHC will serve for precision measurements of Higgs 
couplings, the search for and observation of very rare Higgs modes (among these di-Higgs 
production), the ultimate new physics search reach 
(on mass and forbidden decays such as FCNC), and ultimate SM and heavy flavour physics precision 
for rare processes (VBS, aT/QGC, etc.). Although any new physics found along the way would likely 
be a game changer in this planning process, these physics goals are ``must do'' 
topics for the HL-LHC. 

The substantial increase in luminosity will pose major technical challenges for the experiments.
The average pileup will rise to $\langle\mu\rangle=140$ inelastic collisions per bunch crossing
at (levelled) $5\cdot10^{34}\;{\rm cm}^{-s}{\rm s}^{-1}$, which will increase the background levels, 
the average event size and the time it takes to reconstruct the events (dominated by the 
track reconstruction). Faster detectors and readout electronics, as well as more sophisticated
trigger systems will be required to efficiently identify physics signatures while keeping 
the transverse momentum thresholds at the current level. Finally, the detectors will need 
to withstand substantial radiation dose. Ambitious and costly upgrade programmes of the
experiments address these challenges by improving the trigger and data acquisition systems, 
the front-end electronics, entirely replacing the inner tracking system (thereby increasing 
the tracker acceptance), and, in case of CMS, the endcap calorimeter, and more. 

Among the large amount of prospective studies for the physics potential of the HL-LHC (and 
compared to the Run-3 integrated luminosity of 300\ifb) I would like to mention here the 
prospects for Higgs  coupling measurements, the constraint on the Higgs  width from
Higgs off-shell coupling measurements,  and a precision measurement of $B_{s}\to\mu\mu$. 

\begin{figure}[t]
\centerline{\includegraphics[width=0.85\linewidth]{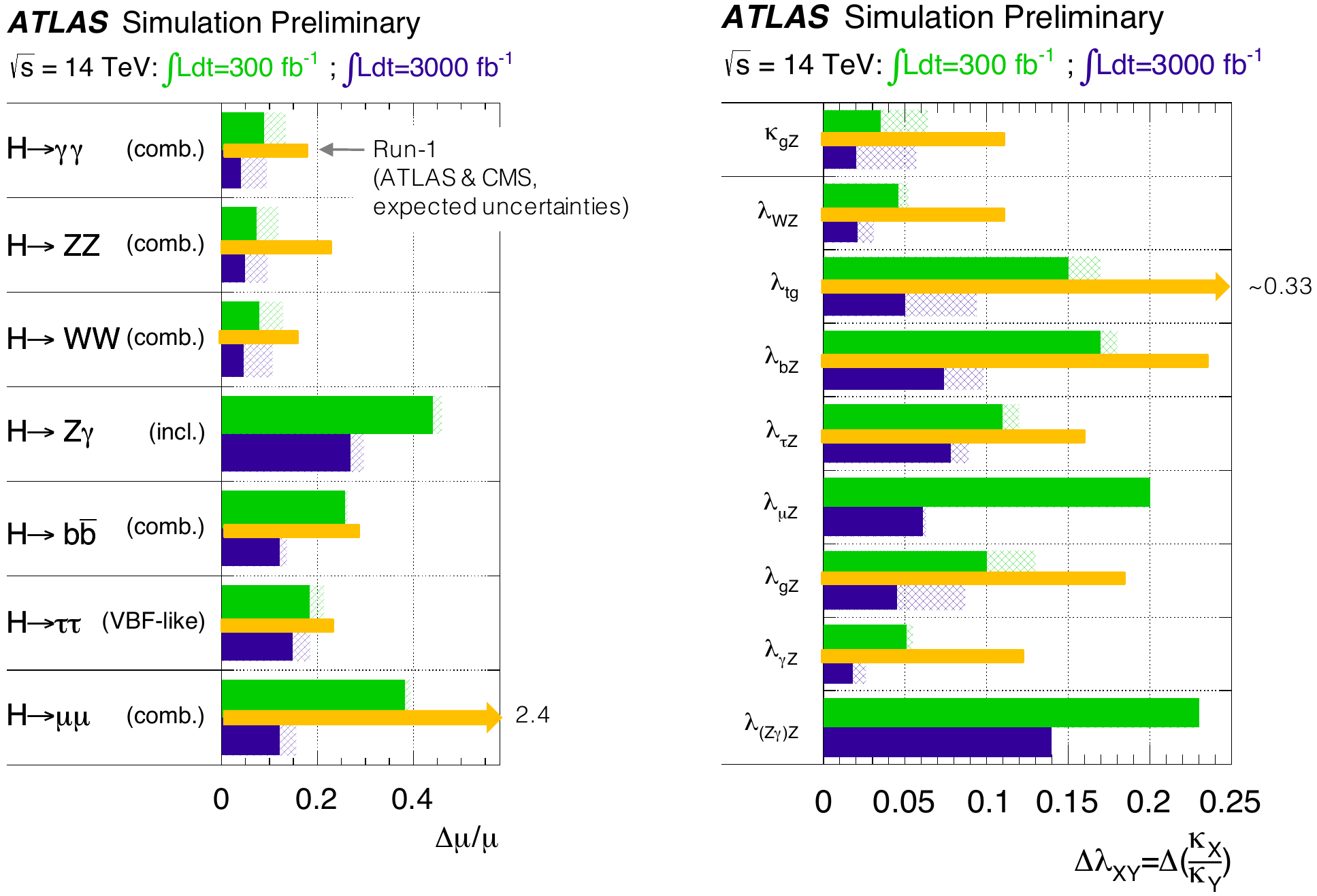}}
\vspace{0.0cm}
\caption[.]{Current (orange) and prospects for future precision (green for Run-3, blue for HL-LHC) 
                 on the measurements of the Higgs  signal strengths (left panel) and the coupling 
                 modifier ratios (right). Hatched areas indicate the impact of theoretical uncertainties on 
                 expected cross-sections~\cite{ATLAS-HiggsCouplings-HL-LHC}. 
                 (The original figures have been modified.)
                \label{fig:ATLAS-HiggsCouplings-HL-LHC}}
\end{figure}
From a rather conservative extrapolation of the Run-1 Higgs coupling measurements 
ATLAS has derived the prospects shown in Fig.~\ref{fig:ATLAS-HiggsCouplings-HL-LHC}
assuming SM central values for the couplings.
As a reminder, the coupling modifiers are defined by $\kappa_i^2=\sigma_i/\sigma^{\rm SM}_i$
and $\lambda_{ij}=\kappa_i/\kappa_j$. The best precision of a few percent on the relative Higgs signal 
strengths is obtained for the diphoton, four-lepton and $2\ell2\nu$ decay channels. The decays
to $\tau\tau$ and $bb$ are challenging and will be limited by systematic uncertainties. The rare 
decays to $Z\gamma$ and $\mu\mu$ will have been observed and be statistically limited. 
The coupling modifier ratios (cf. right panel of Fig.~\ref{fig:ATLAS-HiggsCouplings-HL-LHC}) 
show a similar pattern. The important Higgs--top to Higgs--gluon coupling ratios 
are expected to be measured with a precision reaching 5\% at the HL-LHC. Theory uncertainties 
are limiting the achievable precision in several cases. Some of the uncertainties 
cancel in the coupling modifier ratios. 

Both CMS and ATLAS have constrained the Higgs off-shell coupling in Run-1 analyses and through 
this obtained upper limits on the Higgs total width 
$\Gamma_H$~\cite{ATLAS-Higgs-width,CMS-Higgs-width}.
The method uses the independence of the off-shell cross section on $\Gamma_H$ and relies on 
the assumption of identical on-shell and off-shell Higgs couplings.\footnote{The denominator
of a relativistic Breit-Wigner resonance lineshape has the form $(s - m^2)^2 + s^2\Gamma^2/m^2$.
For $s\sim m^2$ (on-shell) the first term in the denominator vanishes so that the coupling depends
reciprocally on the width $\Gamma$. In the off-shell regime $s\gg m^2$ the first term dominates and the 
$\Gamma$ dependence becomes negligible.}
One can then determine $\Gamma_H$ ($=$$4.1$\;MeV in SM~\cite{HiggsXS}) 
from the measurements of the off-shell and on-shell
signal strengths $\mu_{\textrm{off-shell}}$ and $\mu_{\textrm{on-shell}}$ as follows:
\beqns
   \mu_{\textrm{off-shell}}(\hat s) &=& \frac{\sigma^{gg\to H^*\to VV}_{\textrm{off-shell}}(\hat s)}
                                                     {\sigma^{gg\to H^*\to VV}_{\textrm{off-shell,SM}}(\hat s)}
                             = \kappa^2_{g,\textrm{off-shell}}(\hat s)\cdot \kappa^2_{V,\textrm{off-shell}}(\hat s)\;, \\[0.2cm]
      \mu_{\textrm{on-shell}} &=& \frac{\sigma^{gg\to H\to VV^*}_{\textrm{on-shell}}}
                                             {\sigma^{gg\to H\to VV^*}_{\textrm{on-shell,SM}}}
                             = \frac{\kappa^2_{g,\textrm{off-shell}}(\hat s)\cdot \kappa^2_{V,\textrm{off-shell}}(\hat s)}
                                       {\Gamma_H/\Gamma_{H,{\rm SM}}}\;.
\eeqns
With the Run-1 datasets, limits of the order of 5 times $\Gamma_{H,\rm SM}$ were obtained by ATLAS and 
CMS. An ATLAS HL-LHC study~\cite{ATLAS-Hwidth-HL-LHC} derived prospects for integrated 
luminosities of 300\ifb and 3000\ifb giving $\mu_{\textrm{off-shell}}=1^{\,+0.80}_{\,-0.97}$ and 
$1^{\,+0.43}_{\,-0.50}$, respectively. The latter precision allows to constrain $\Gamma_H$ to 
remarkable $4.1^{\,+1.5}_{\,-2.1}$\;MeV.

In the area of new physics searches, the emphasis will gradually move towards rare and difficult 
channels, such as low cross-section electroweak production and compressed scenarios in SUSY.
Searches for WIMPs will require improvements in the data-driven determination of the backgrounds
to take full benefit from the increased data sample. 

\begin{figure}[t]
\centerline{\includegraphics[width=0.9\linewidth]{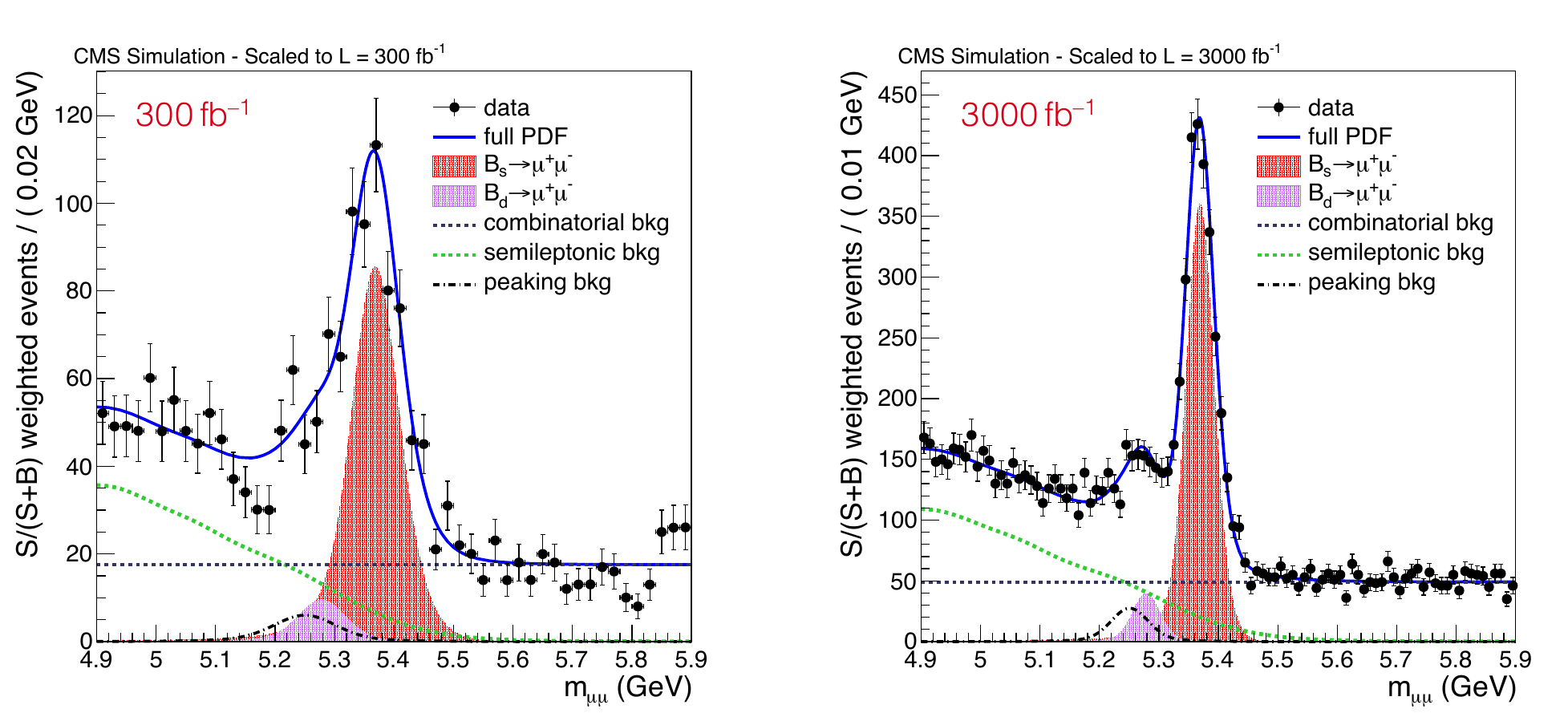}}
\vspace{0.0cm}
\caption[.]{Expected invariant mass distribution in the measurement of $B_{(s)}\to\mu\mu$
                 for 300\ifb and 3000\ifb from a prospective study by CMS~\cite{CMS-Bsmumu-HL-LHC}.
                \label{fig:CMS-Bsmumu-HL-LHC}}
\end{figure}
Among the many other interesting prospects, one should also note the continuous gain in precision 
and reach for rare or suppressed processes in the flavour sector. The rise in luminosity during Run-2 
will be slower for LHCb due to the luminosity levelling. The upgrade to 40\;MHz trigger readout 
during the long shutdown 2 in 2019 will help increase the annual muonic $B$ rate by a factor of ten.
High-profile rare decay measurements performed by LHCb, ATLAS and CMS are $B_{(s)}\to\mu\mu$ 
(and similar) as well as $b\to s$ transitions such as $B\to K^*\mu\mu$ and similar modes. 
Figure~\ref{fig:CMS-Bsmumu-HL-LHC} shows the invariant mass distribution for 
$B_{(s)}\to\mu\mu$ as expected from a prospective study by CMS~\cite{CMS-Bsmumu-HL-LHC}. 
The observation beyond $5\sigma$ significance of the loop and CKM suppressed decay $B\to\mu\mu$ 
is expected for the full HL-LHC integrated luminosity. 
CP-violation measurements of the phase $\phi_s$ will be performed by LHCb (dominant) and also 
by ATLAS and CMS, the unitarity triangle angle  $\gamma$ and other CKM parameters will be measured
by LHCb. These important measurements will benefit from any increase in integrated luminosity.
LHCb will also improve CP asymmetry  measurements in the charm sector. 
Of high importance given the current results is to pursue measurements testing lepton universality 
in $B$ decays (LHCb and Belle). Finally, further surprises and a better understanding of recently 
discovered heavy flavour spectroscopy states are expected by LHC, ATLAS and CMS. 

\section{Conclusions}

The LHC Run-2 is a key period for particle physics. The  first 100\ifb at 13\;TV centre-of-mass energy 
are critical for new physics searches in all signatures. Further consolidation of the Higgs sector with the 
observation and measurement of $H \to\tau\tau$, $H\to bb$, and associated $ttH$ production, 
as well as more precise coupling, fiducial and differential cross section measurements will
be followed up with high priority by ATLAS and CMS. The luminosity of Run-2 will hugely increase 
the amount of interesting Standard Model and flavour physics measurements that can be performed.

Throughout Run-2 it is important to stay alert. New physics does not necessarily appear at high 
mass so that one needs to continue to search everywhere. High precision measurements are key 
for a better knowledge of the Standard Model. It is thereby extremely important to measure the 
detector performance in data as precisely as possible, and this may  have priority over further 
improving the performance. Many results are dominated by theoretical 
uncertainties. The experiments need to produce measurements that allow to test theory, to 
improve PDFs, and that motivate theorists to improve calculations and event generators. 
We may cite William Thomson Kelvin, from a speech held to the British Association for 
the Advancement of Science in 2 Aug 1871: ``Accurate and minute measurement seems to the 
non-scientific imagination, a less lofty and dignified work than looking for something new. 
But [many of] the grandest discoveries of science have been but the rewards of accurate 
measurement and patient long-continued labour in the minute sifting of numerical results.''
\vspace{6pt}
\begin{details}
{\em 
I thank the organisers of the 2016 European School of High-Energy Physics for 
giving me the opportunity to lecture at this excellent school at a very pleasant 
location in Norway. 
}
\end{details}

\vfill\pagebreak
\setcounter{tocdepth}{3}
\tableofcontents
\vfill\pagebreak

\bibliographystyle{unsrt}

\end{document}